\documentclass[11pt,oneside]{article}
\usepackage[top=1in, bottom=1.2in, left=1.1in, right=1.1in]{geometry}
\usepackage{setspace}
\usepackage{amsmath, amsthm, amssymb,amsbsy,amsfonts,mathrsfs}
\usepackage{color}
\usepackage{bm}
\usepackage{natbib}
\usepackage{appendix}
\usepackage{float}
\usepackage[caption = false]{subfig}
\usepackage{graphicx}
\usepackage{enumitem}
\usepackage{enumitem}
\usepackage{authblk}
\usepackage[breaklinks=true]{hyperref}
\usepackage{booktabs}
\usepackage[flushleft]{threeparttable}
\usepackage{rotating}

\csname @openrightfalse\endcsname

\setlist{nolistsep}
\makeatletter
\newcommand*{\rom}
[1]{\expandafter\@slowromancap\romannumeral #1@}
\makeatother

\DeclareMathOperator*{\argmax}{arg\max}
\DeclareMathOperator*{\tr}{Tr}

\DeclareMathOperator*{\rank}{rank}

\newtheorem{thm}{Theorem}

\newtheorem{lemma}{Lemma}
\newtheorem{prop}{Proposition}
\newtheorem{ass}{Assumption}
\newtheorem{exm}{Example}
\newtheorem{rem}{Remark}

\setenumerate{noitemsep}

\hypersetup{
	colorlinks   = true,
	citecolor    =  blue
}

\usepackage{titlesec}

\newcommand{\Lx}{\mathcal{L}}

\newcommand{\Ex}{\mathbb{E}}

\begin{document}
	\title{\textbf{Common Correlated Effects Estimation of Nonlinear Panel Data Models}\thanks{
Address correspondence to Liang Chen, Peking University HSBC Business School, No. 2199 Lishui Road, Shenzhen, Guangdong 518055, China; Email: chenliang@phbs.pku.edu.cn.}}
	\author[1]{Liang Chen}
	\author[2]{Minyuan Zhang}
	\affil[1]{HSBC Business School, Peking Unviersity}
	\affil[2]{School of Economics, Shanghai University of Finance and Economics}
\date{First draft: October 2019\\
This draft: April 2023}	
	\maketitle
	
	\begin{abstract}
		This paper focuses on estimating the coefficients and average partial effects of observed regressors in nonlinear panel data models with interactive fixed effects, using the common correlated effects (CCE) framework. The proposed two-step estimation method involves applying principal component analysis to estimate latent factors based on cross-sectional averages of the regressors in the first step, and jointly estimating the coefficients of the regressors and factor loadings in the second step. The asymptotic distributions of the proposed estimators are derived under general conditions, assuming that the number of time-series observations is comparable to the number of cross-sectional observations.  To correct for asymptotic biases of the estimators, we introduce both analytical and split-panel jackknife methods, and confirm their good performance in finite samples using Monte Carlo simulations. An empirical application utilizes the proposed method to study the arbitrage behaviour of nonfinancial firms across different security markets. 
	
		\vspace{0.2cm}
		
	\end{abstract}
	
	\newpage
	\section{Introduction}
Nonlinear panel data models have posed a long-standing and challenging problem for identification and estimation. This difficulty is especially pronounced when the number of cross-sectional observations (denoted by $N$) is large, but the number of time-series observations (denoted by $T$) is small. The central challenge in this setting is to identify the coefficients of the observed regressors in the presence of fixed effects that have an unrestricted joint distribution with the regressors. Over the years, a considerable amount of work has been devoted to this problem, including classical results on binary choice models by \cite{manski1987semiparametric}, \cite{honore2000panel}, \cite{honore2002semiparametric}, and \cite{chamberlain2010binary}, as well as more recent developments by \cite*{davezies2020fixed}, \cite{honore2020moment}, \cite*{khan2020identification} and \cite{zhu2022sufficient}.

In \textit{long panels}, where the number of time-series observations is comparable to the number of cross-sectional observations, the identification problem can be resolved by treating the fixed effects as parameters that are jointly estimated with the coefficients of the regressors, and the primary focus of this literature is to correct the asymptotic biases caused by the estimation errors of the fixed effects, also known as the \textit{incidental parameter biases} --- see \cite{neyman1948consistent}. This line of research was pioneered by \cite{hahn2004jackknife}, and further developed by \cite{dhaene2015split}. 

Recent research on long panels has focused on models with \textit{interactive fixed effects} that use a factor model structure to characterize the unobserved errors. These models nest the conventional two-way fixed effects models as special cases, providing a more general way to capture cross-sectional dependence in panels (see \cite{chudik2015large}). For linear models with interactive fixed effects,  fundamental contributions have been made by \cite{pesaran2006estimation}, \cite{bai2009panel} and \cite{moon2015linear}. In recent years, \cite{chen2016estimation},  \cite{boneva2017discrete}, \cite*{chen2020nonlinear}, \cite*{ando2022bayesian} and \cite*{gao2023binary} have investigated the estimation of nonlinear models with interactive fixed effects.\footnote{\cite{wang2022maximum} studies the estimation of nonliner factor models, which can be viewed as a special case of nonlinear panel data models with interactive fixed effects. However, in that paper, the main focus is the estimation of the factors and factor loadings. } 

Panel data models with interactive fixed effects involve three sets of parameters: a finite-dimensional vector of coefficients (denoted by $\bm{\beta}$) for the observed regressors, a $T\times r$ matrix of latent factors (denoted by $\bm{F}$) representing the global shocks to all individuals, and an $N\times r$ matrix of factors loadings (denoted by $\bm{\Lambda}$) measuring the individual-specific responses to the factors.\footnote{Throughout the paper, $r$ is used to denote the number of factors.} While the main object of interest is $\bm{\beta}$, the latter two sets of parameters are introduced to account for individual heterogeneity and cross-sectional dependence. To estimate these parameters in nonlinear models, the papers mentioned above usually take two different approaches. The first approach, utilized by \cite{chen2016estimation}, \cite{chen2020nonlinear},  \cite{ando2022bayesian} and \cite{gao2023binary}, involves estimating $(\bm{\beta},\bm{F},\bm{\Lambda})$ jointly using some iterative algorithms. The second approach, introduced by \cite{boneva2017discrete}, extends the CCE estimation method of \cite{pesaran2006estimation} to nonlinear models.

This paper fills a gap in the literature by studying the estimation of nonlinear panel data models with interactive fixed effects using the CCE framework, where the observed regressors are assumed to be driven by the same latent factors $\bm{F}$ and the coefficients of the regressors are homogeneous across individuals (see Table 1 below). As the main contribution of this paper, a two-step estimator for this type of models is proposed. In the first step, an estimator for the latent factors is constructed using the observed regressors; in the second step, given the estimated factors, $\bm{\beta}$ and $\bm{\Lambda}$ are estimated jointly to maximize the objective function (e.g., the log likelihood function). Notably, the proposed method for estimating the factors in the first step is different from the standard CCE approach, which can suffer from the problem of \textit{degenerated regressors} (see \cite*{Karabiyik201760}). Asymptotic properties, particularly asymptotic biases, of the proposed estimators are derived under the framework of long panels and other general conditions, with a Bahadur representation established for the estimator of $\bm{\beta}$ where the leading biases are shown to be of order $1/T+1/N$. To the best of our knowledge, this is the first result of this kind for CCE estimators of nonlinear panels. In addition, both analytical and split-panel jackknife methods are introduced to correct the asymptotic biases, providing the basis for valid inference in large samples. Through Monte Carlo simulations, we find that the proposed bias-correction procedures significantly reduce the biases of the estimators and improve the empirical coverage rates of the confidence intervals in finite samples. 

	\begin{table}[H]
		\caption{Estimation of Nonlinear Panels with Interactive Fixed Effects}
		\begin{center}
			\begin{tabular}{ccc}
				\hline
				& \textbf{Joint Estimation} & \textbf{CCE Estimation}\\
				\hline
				Homogeneous Coeff. & Chen (2016), Chen et al. (2021) & \textcolor{blue}{This Paper}\\
				Heterogeneous Coeff. & Ando et al. (2022), Gao et al. (2023)& Boneva and Linton (2017)\\
				\hline				
			\end{tabular}
		\end{center}
		\label{default}
	\end{table}%

Compared with the approach that estimate $(\bm{\beta},\bm{F},\bm{\Lambda})$ jointly, the main advantage of the CCE estimation method is its much lower computational cost, because the estimation problem is decomposed into a simpler first step and a second step that can be easily implemented in standard software such as Stata and R. Additionally, for the leading examples of binary choice models, the log likelihood function is convex in $(\bm{\beta},\bm{\Lambda})$ given $\bm{F}$, simplifying the search for the global maximum of the objective function. In contrast, joint estimation of these parameters can be computationally challenging even for linear models, as the objective functions are generally not convex in $(\bm{\beta},\bm{F},\bm{\Lambda})$ (see \cite{bai2009panel}). However, the success of the CCE approach relies on the assumption that it is possible to consistently estimate the (space of) latent factors using the observed regressors, which contradicts the usual assumption of cross-sectional independence. Nonetheless, as pointed out in \cite{andrews2005cross}: \textit{``...it seems apparent that common shocks (macroeconomic, technological, legal/institutional, political, environmental, health, and sociological shocks) are a likely feature of cross-section economic data. This is true whether the population units in the cross-section regression are individuals, households, firms, industries, plants, cities, states, countries, or products.''} Thus, the seemly stronger CCE assumption that the cross-sectional dependence of observations are driven by the same latent factors could be potentially advantageous in certain applications.

Finally, in a closely related paper, \cite{boneva2017discrete} also considered the CCE estimation of nonlinear panels, but their approach assumes that the coefficients of the regressors are heterogeneous across individuals. Consequently, the estimators of these coefficients converge at the rate of $\sqrt{T}$, and their asymptotic distributions are free of asymptotic biases. In this paper, the CCE estimator of the homogeneous coefficients converges at the rate of $\sqrt{NT}$, making it far more challenging to establish its asymptotic distribution, because many higher order terms in the stochastic expansion of the estimator now become asymptotic biases whose analytical forms need to be carefully derived.

The rest of the paper is structured as follows: Section 2 introduces the model along with the CCE estimators for the coefficients and the average partial effect of the regressors. Moving on to Section 3, we establish the asymptotic properties of the estimators and demonstrate how to correct their asymptotic biases and consistently estimate their asymptotic variances. Simulation results are presented in Section 4 to evaluate the performance of the estimators in finite samples. Section 5 describes an empirical application where we use the proposed method to study the arbitrage behavior of U.S. nonfinancial firms across different security markets.  Finally, Section 6 concludes. The appendix provides the proof of Theorem 1, whereas the proofs of the other theorems are available in an online appendix to save space.

	\section{The Model and the CCE Estimator}

	\subsection{The Model}
	Let $y_{it}\in\mathbb{R}$ and $\boldsymbol{x}_{it}\in\mathbb{R}^k$ be the observed outcome and covariates respectively for individual $i$ at period $t$, and let $\boldsymbol{\lambda}_i, \boldsymbol{f}_t \in \mathbb{R}^r$ be the unobserved factor loadings (or individual effects) and factors (or time effects) respectively.  
	Suppose that we have a random sample $\{ y_{it}, \boldsymbol{x}_{it}\}$ for $i=1,\ldots,N, t=1,\ldots,T$, where the realized value of $\boldsymbol{\lambda}_i, \boldsymbol{f}_t$ are denoted as $\boldsymbol{\lambda}_{0i}, \boldsymbol{f}_{0t}$. Moreover, for some $\boldsymbol{\beta}_0 \in \mathbb{R}^k$, the likelihood function of $y_{it}$ given $(\boldsymbol{x}_{it},\boldsymbol{\lambda}_{0i}, \boldsymbol{f}_{0t} )$ can be written as 
	\[ L(y_{it}, \boldsymbol{\beta}_0'\boldsymbol{x}_{it} + \boldsymbol{\lambda}_{0i}'\boldsymbol{f}_{0t}).\]
	Through out the paper, the following examples are used to illustrate the applicability of our general theoretical results.
	
	\begin{exm}[Binary Choice Model] Define $y_{it}^{\ast} = \boldsymbol{\beta}_0'\boldsymbol{x}_{it} + \boldsymbol{\lambda}_i'\boldsymbol{f}_t -\epsilon_{it}$ and assume that the cumulative distribution function (CDF) of $\epsilon_{it}$ is $G$. We only observe a binary outcome: $y_{it} =\boldsymbol{1}\{y_{it}^{\ast} \geq 0\}$. In this case, we have
		\begin{equation*}\label{logitlikelihood}
			L(y_{it}, \boldsymbol{\beta}_0'\boldsymbol{x}_{it} + \boldsymbol{\lambda}_{0i}'\boldsymbol{f}_{0t}) =   G(z_{it} )^{y_{it}}\left[ 1-  G(z_{it} )\right]^{1-y_{it}},
		\end{equation*}
		where $z_{it}= \boldsymbol{\beta}_0'\boldsymbol{x}_{it} + \boldsymbol{\lambda}_{0i}'\boldsymbol{f}_{0t}$. Two popular choices that are widely used in practice are logit and probit models, corresponding to $G(z) = \frac{ \exp(z)}{1+ \exp(z)}$ and $G(z)=\Phi(z)$ respectively, where $\Phi(z)$ is the CDF of the standard normal distribution. 
	\end{exm} 
	
	\begin{exm}[Poisson Model] Suppose that $y_{it}$ only take non-negative integer values and that 
		\begin{equation*}\label{poissonlikelihood}
			L(y_{it}, \boldsymbol{\beta}_0'\boldsymbol{x}_{it} + \boldsymbol{\lambda}_{0i}'\boldsymbol{f}_{0t}) =  \frac{z_{it}^{y_{it}}}{y_{it}!}\cdot \exp(-z_{it}),
		\end{equation*}
		where $z_{it}= \boldsymbol{\beta}_0'\boldsymbol{x}_{it} + \boldsymbol{\lambda}_{0i}'\boldsymbol{f}_{0t}$.	
	\end{exm}

	For any $\boldsymbol{\beta}\in\mathbb{R}^k$ and $\boldsymbol{\lambda}_i,\boldsymbol{f}_t\in \mathbb{R}^r$, write $ l_{it}( \boldsymbol{\beta}, \boldsymbol{\lambda}_i'\boldsymbol{f}_t)  = \log \left[  L(y_{it}, \boldsymbol{\beta}'\boldsymbol{x}_{it} + \boldsymbol{\lambda}_i'\boldsymbol{f}_t) \right]$,
	then the objective function can be written as 
	\[      \mathcal{L}_{NT}( \boldsymbol{\beta}, \boldsymbol{\Lambda}, \boldsymbol{F} )  =  \frac{1}{NT}\sum_{i=1}^{N}\sum_{t=1}^{T}    l_{it}( \boldsymbol{\beta}, \boldsymbol{\lambda}_i'\boldsymbol{f}_t) ,  \]
	where $\boldsymbol{\Lambda}=(\bm{\lambda}_1,\ldots,\bm{\lambda}_N)'$ and $\boldsymbol{F}=(\bm{f}_1,\ldots,\boldsymbol{f}_t)'$. In the spirit of \cite{bai2009panel}, \cite{chen2020nonlinear} proposes to estimate $(\boldsymbol{\beta}, \boldsymbol{\Lambda}, \boldsymbol{F})$ jointly to maximize the above objective function. This estimation procedure leaves the relationship between the regressors, the factors and factor loadings unspecified, thus it is usually termed as the \textit{fixed-effects} estimator. The practical implementation of this estimation method usually involves iterations between $\bm{\beta}$ and $(\boldsymbol{\Lambda}, \boldsymbol{F})$. In particular, \cite{chen2016estimation} proposed an EM-type iterative algorithm that converges to local maximums of the objective function. Moreover, it is usually assumed that the number of factor $r$ is known in the asymptotic theory and in the practical implementation of the fixed-effects estimator.\footnote{\cite{chen2020nonlinear} proposed an method adapted from the eigen-ratio estimator of \cite{ahn2013eigenvalue} to estimate $r$, but the consistency of their method was not established.} The effect of overestimating $r$ in linear panel data models is analyzed by \cite{moon2015linear}, but extending their analysis to nonlinear models is much more challenging. 
	
	\begin{rem}
		Even though the function $\mathcal{L}_{NT}$ is interpreted as the log likelihood function above, our results below also apply to other extremum estimators where $\mathcal{L}_{NT}$ represents a smooth objective function. For example, when $\mathcal{L}_{NT}( \boldsymbol{\beta}, \boldsymbol{\Lambda}, \boldsymbol{F} )=  (NT)^{-1}\sum_{i=1}^{N}\sum_{t=1}^{T}-(y_{it} -\boldsymbol{\beta}_0'\boldsymbol{x}_{it} - \boldsymbol{\lambda}_{0i}'\boldsymbol{f}_{0t})^2$, the underlying model is a linear panel data model where the errors have a factor structure (see \cite{bai2009panel} and \cite{moon2015linear}). However, it should be noted that our results do not apply to the cases where the objective functions are not smooth, such as the quantile panel data models considered by \cite{chen2022twostep}.
	\end{rem}
	
	\subsection{The CCE Estimator}
	In this paper, we try to overcome the problems of the fixed-effects estimator mentioned above by taking a CCE approach pioneered by \cite{pesaran2006estimation}. The CCE approach starts by assuming a linear relationship between the regressors and the common factors as follows:
	\begin{equation}\label{cce}
		\boldsymbol{x}_{it} = \boldsymbol{\Gamma}_i \boldsymbol{f}_{0t}+ \boldsymbol{e}_{it} 
	\end{equation}
	for $i=1,\ldots,N, t=1\ldots,T$, where $\boldsymbol{\Gamma}_i$ is a $k\times r$ matrix of non-random constants, and $\boldsymbol{e}_{it} \in \mathbb{R}^k$ is a vector of idiosyncratic components. The key of the CCE estimator is to approximate the common factors by the cross-sectional averages of the regressors: $\bar{ \boldsymbol{x} }_t = N^{-1}\sum_{i=1}^{N} \boldsymbol{x}_{it}.$\footnote{In linear panel data models, \eqref{cce} implies that the dependent variables have a factor structure, thus the cross-sectional averages of the dependent variables are also used to approximate the common factors. However, in nonlinear models, the dependent variables generally do not have a factor structure. Thus, only the cross-sectional averages of the regressors are used.} To ensure that the space of the common factors can be spanned by $\bar{ \boldsymbol{x} }_t$, the following assumption is usually imposed:
	
	\begin{ass} Let $\bar{\boldsymbol{\Gamma}} = N^{-1}\sum_{i=1}^{N}\boldsymbol{\Gamma}_i$, then $\bar{\boldsymbol{\Gamma}}\rightarrow \boldsymbol{\Gamma}_0$ as $N\rightarrow \infty$ with $\rank(\boldsymbol{\Gamma}_0)=r$. 
	\end{ass}
	
	Under Assumption 1 and some other restrictions on the cross-sectional dependence of $\boldsymbol{e}_{it}$, it is easy to show that 
	\[   \bar{\boldsymbol{x}}_t = \boldsymbol{\Gamma}_0 \boldsymbol{f}_{0t} + o_P(1) \text{ for all }t=1,\ldots,T.\]
	When $k=r$, the above equation implies that $\bar{\boldsymbol{x}}_t$ can be used as approximations of $\boldsymbol{f}_{0t}$ since they span the same space asymptotically. However, when $k>r$, using $\bar{\boldsymbol{x}}_t$ as estimators of $\boldsymbol{f}_{0t}$ amounts to overestimating the number of factors. In particular, the fact that $T^{-1}\sum_{t=1}^T  \bar{\boldsymbol{x}}_t\bar{\boldsymbol{x}}_t'$ converges in probability to a singular matrix when $k>r$ implies \textit{asymptotic multicolinearity} of the estimated factors, leading to the problem of \textit{degenerated regressors} as pointed out in \cite{Karabiyik201760}. For linear models, \cite{Karabiyik201760} showed that the standard CCE estimator for $\bm{\beta}_0$ is still consistent but it suffers from extra asymptotic biases due to this problem. 
	
	To overcome the problem of degenerated regressors in the standard CCE method, in this paper we use an alternative approach to estimate the common factors. Let $\hat{\boldsymbol{\Sigma}}_{\bar{\boldsymbol{x}}}=T^{-1}\sum_{t=1}^T  \bar{\boldsymbol{x}}_t\bar{\boldsymbol{x}}_t'$ and $\hat{\boldsymbol{\Psi}}$ be the $k\times r$ matrix of eigenvectors associated with the first $r$ eigenvalues of $\hat{\boldsymbol{\Sigma}}_{\bar{\boldsymbol{x}}}$, then the estimated factors are defined as $\hat{\boldsymbol{f}}_t = \hat{\boldsymbol{\Psi}}' \bar{\boldsymbol{x}}_t$. To establish the properties of $\hat{\boldsymbol{f}}_t$, we need to impose the following assumptions:
	\begin{ass} Let $M>0$ be a generic bounded constant and $\boldsymbol{\Sigma}_{f_0}$ be a $r\times r$ matrix with full rank.\\
		(i) $ \| \boldsymbol{f}_{0t}\| \leq M$ for all $t$.\\
		(ii) $\| \hat{\boldsymbol{\Sigma}}_{f_0} -\boldsymbol{\Sigma}_{f_0} \| =O(T^{-1/2})$, and $\|  \bar{\boldsymbol{\Gamma}} -\boldsymbol{\Gamma}_0  \|= O(N^{-1/2})$;\\ 
		(iii) $\Ex[\boldsymbol{e}_{it}]=0$ for all $i,t$ and $\Ex \| \sqrt{N} \bar{\boldsymbol{e}}_t \|^2 \leq M$ for all $t$, where $\bar{\boldsymbol{e}}_t = N^{-1}\sum_{i=1}^{N}\boldsymbol{e}_{it} $.
	\end{ass}
	\begin{ass}
		The non-zero eigenvalues of $ \boldsymbol{\Gamma}_0 \boldsymbol{\Sigma}_{f_0} \boldsymbol{\Gamma}_0' $ are distinct. 
	\end{ass}
	Moreover, define $\hat{\bm{H}} =\hat{\boldsymbol{\Psi}}'  \bar{\boldsymbol{\Gamma}} $. Let $\bm{D}$ be a $r \times r$ diagonal matrix with the non-zeros eigenvalues of $ \boldsymbol{\Gamma}_0 \boldsymbol{\Sigma}_{f_0} \boldsymbol{\Gamma}_0'$ in decreasing order, and let $\boldsymbol{\Psi}_{0}$ be the matrix of corresponding eigenvectors such that $\boldsymbol{\Gamma}_0 \boldsymbol{\Sigma}_{f_0} \boldsymbol{\Gamma}_0' \boldsymbol{\Psi}_0  = \boldsymbol{\Psi}_0 \bm{D}$. Then it can be shown that:
	
	\begin{prop}  Under Assumptions 1 to 3, as $N,T\rightarrow\infty$, (i) $ \hat{\boldsymbol{f}}_t =\hat{\boldsymbol{H}} \boldsymbol{f}_{0t} +  \hat{\boldsymbol{\Psi}}' \bar{\boldsymbol{e}}_t$ ; (ii) $\hat{\boldsymbol{H}}$ is invertible with probability approaching 1; (iii) $\hat{\boldsymbol{\Psi}} \overset{p}{\rightarrow} \boldsymbol{\Psi}_0 $ and $\hat{\boldsymbol{H}} \overset{p}{\rightarrow}\boldsymbol{H}_0=\boldsymbol{\Psi}_{0}'\boldsymbol{\Gamma}_0$.
	\end{prop}
	
	The above results implies that $\hat{\bm{f}}_t$ is a consistent estimator of $\bm{f}_t$ up to a non-singular normalization matrix.	Given $\hat{\boldsymbol{f}}_t$, the CCE estimator is defined as
	\begin{equation}\label{estimator}
		(\hat{\boldsymbol{\beta}}, \hat{\boldsymbol{\lambda}}_1,\ldots,\hat{\boldsymbol{\lambda}}_N) =\argmax_{\boldsymbol{\beta}\in \mathcal{B},\boldsymbol{\lambda}_i \in \mathcal{A} }\frac{1}{NT}\sum_{i=1}^{N}\sum_{t=1}^{T}    l_{it}( \boldsymbol{\beta}, \boldsymbol{\lambda}_i' \hat{\boldsymbol{f}}_{t}),
	\end{equation}
	where $ l_{it}( \boldsymbol{\beta} , \boldsymbol{\lambda}_i' \hat{\boldsymbol{f}}_{t}) = \log  L(y_{it}, \boldsymbol{\beta}'\boldsymbol{x}_{it} + \boldsymbol{\lambda}_i' \hat{\boldsymbol{f}}_{t})$, and $\mathcal{B}\subset\mathbb{R}^k,  \mathcal{A}\subset\mathbb{R}^r$. Note that in practice, the CCE estimator can be obtained using standard packages in Matlab or R by treating $y_{it }$ as the dependent variable and $( \boldsymbol{x}_{it}, \boldsymbol{1}\{ i=1\} \hat{\boldsymbol{f}}_t, \ldots, \boldsymbol{1}\{ i=N\} \hat{\boldsymbol{f}}_t )$ as the regressors. It should also be mentioned that the computational cost of the CCE estimator is much lower than the fixed-effects estimator, because in the maximization problem \eqref{estimator} there are $k+Nr$ parameters while for the fixed-effects estimator there are $k+(N+T)r$ parameters. More importantly, since the objective function  $\mathcal{L}_{NT}( \boldsymbol{\beta}, \boldsymbol{\Lambda}, \boldsymbol{F} )$ is generally not convex in $( \boldsymbol{\beta}, \boldsymbol{\Lambda}, \boldsymbol{F} ) $, the joint estimation of these estimators normally involves iterative procedures that do not necessary find the global maximum of the objective function. However, it is well known that given $\bm{F}$, the objective function becomes convex in $(\boldsymbol{\beta}, \boldsymbol{\Lambda}) $ in our leading examples (e.g., probit and logit models). Thus, the estimator defined in \eqref{estimator} can be easily obtained using standard optimization methods such as the gradient descent algorithm, without the need of a good initial estimator.

	Finally, the analysis above assumes that $r$ is known. In practice, $r$ needs to be estimated before implementing the CCE method. Observe that if $\hat{\boldsymbol{\Sigma}}_{f_0}=T^{-1}\sum_{t=1}^{T}\boldsymbol{f}_{0t}\boldsymbol{f}_{0t}'$ converges to a positive definite matrix, it is easy to show that $\hat{\boldsymbol{\Sigma}}_{\bar{\boldsymbol{x}}}$ converges in probability to a matrix with rank $r$. Thus, to estimate $r$, we can just estimate the rank of $\hat{\boldsymbol{\Sigma}}_{\bar{\boldsymbol{x}}}$. In particular, let $\hat{\rho}_1,\ldots, \hat{\rho}_k$ be the eigenvalues of $\hat{\boldsymbol{\Sigma}}_{\bar{\boldsymbol{x}}}$ in decreasing order, and let $P_{NT}$ be a sequence of constants converging to 0 as $N,T\rightarrow \infty$, the estimator of $r$ can be simply defined as:
	\[ \hat{r} = \sum_{j=1}^{k} \boldsymbol{1} \{ \hat{\rho}_j \geq P_{NT}\}.\]
	The following result is established in \cite{chen2022twostep}.\footnote{The proofs of Proposition 1 and Proposition 2 are identical to the proofs of Proposition 1 and Lemma 1 in \cite{chen2022twostep}. Thus, they are omitted to save space.}
	\begin{prop} Under Assumptions 1 and 2, we have $P[\hat{r}=r] \rightarrow 1$ as $N,T\rightarrow \infty$ if $P_{NT} \rightarrow 0$ and $P_{NT} \cdot \min\{ \sqrt{N},\sqrt{T}\}\rightarrow \infty$.
	\end{prop}
	
	Given the above result, the true number of factors $r$ can be treated as known in the rest of the paper.\footnote{See footnote 5 of \cite{bai2003inferential}.} 
	
	\begin{rem}An alternative method to estimate the number of factors, inspired by \cite{ahn2013eigenvalue}, is to consider the following estimator based on the ratios of the eigenvalues:
		\[ \tilde{r} = \argmax_{1\leq j\leq k-1 } \hat{\rho}_j /\hat{\rho}_{j+1}.\]
		The advantage of this estimator is that it does not require choosing any tuning parameters. However, the main problem of this estimator is that it replies on the separation of nonzero and zero eigenvalues, so it does not work when $k=r$, because in this case all eigenvalues of $\hat{\boldsymbol{\Sigma}}_{\bar{\boldsymbol{x}}}$ will converge to positive constants. Please see \cite{chen2022twostep} for simulation results on the finite sample performances of $\hat{r}$ and $\tilde{r}$.
	\end{rem}

	\subsection{The Estimator of Average Partial Effects}
	For models with limited dependent variables, the coefficient $\boldsymbol{\beta}_0$ usually cannot capture the partial effects of the regressors, which are the main object of interests for most practitioners. Consider the binary choice models, and let $\boldsymbol{x}^{(0)},\boldsymbol{x}^{(1)}\in\mathbb{R}^k$ be the values of the regressors before and after some policy intervention, the effect of the policy on the \textit{probability of success} is given by
	\[  \delta( \boldsymbol{x}^{(0)},\boldsymbol{x}^{(1)}; \boldsymbol{\beta}_0, \boldsymbol{\lambda}_{0i}' \boldsymbol{f}_{0t}) = G(\boldsymbol{\beta}_0' \boldsymbol{x}^{(1)}+\boldsymbol{\lambda}_{0i}'\boldsymbol{f}_{0t} ) -G(\boldsymbol{\beta}_0' \boldsymbol{x}^{(0)}+\boldsymbol{\lambda}_{0i}'\boldsymbol{f}_{0t} ) . \]
	A typical example is that the first element of $x$ is a binary indicator for some treatment or policy change, then $ \boldsymbol{x}^{(0)} =(0, x_2,\ldots, x_k)$ and $ \boldsymbol{x}^{(1)} =(1, x_2,\ldots, x_k)$. In this case, $ \delta( \boldsymbol{x}^{(0)},\boldsymbol{x}^{(1)}; \boldsymbol{\beta}_0, \boldsymbol{\lambda}_{0i}' \boldsymbol{f}_{0t})$ denotes the partial effect of the treatment for individual $i$ at time $t$, while his/her other characteristics are fixed at $(x_2,\ldots, x_k)$. Similar partial effects can be defined for other nonlinear models, depending on the applications at hand. However, it should be noted that in nonlinear models such partial effects generally depends on $\boldsymbol{x}^{(0)}, \boldsymbol{x}^{(1)}, \boldsymbol{\lambda}_{0i},\boldsymbol{f}_{0t}$.
	
	To summarize the partial effects for all the individuals in the dataset, we consider the following average partial effect (APE):
	\[       \bar{\delta}_{0} ( \boldsymbol{x}^{(0)},\boldsymbol{x}^{(1)} ) = \frac{1}{NT}\sum_{i=1}^{N}\sum_{t=1}^{T}\delta( \boldsymbol{x}^{(0)},\boldsymbol{x}^{(1)}; \boldsymbol{\beta}_0, \boldsymbol{\lambda}_{0i}'\boldsymbol{f}_{0t}), \]
	which is different from the definitions of \cite{chen2020nonlinear} and \cite{boneva2017discrete}, who take expectations of the partial effects with respect to the distribution of the regressors. Note that our definition of APE is closer in spirit to the definition of \cite{hahn2004jackknife}, in the sense that it only averages out the individual and time effects while fixing the values of the regressors. Given $(\hat{\boldsymbol{\beta}}, \hat{\boldsymbol{\Lambda}}, \hat{\boldsymbol{F}} )$, the estimator of APE is simply given by 
	\[  \hat{\delta} ( \boldsymbol{x}^{(0)},\boldsymbol{x}^{(1)} ) = \frac{1}{NT}\sum_{i=1}^{N}\sum_{t=1}^{T}\delta( \boldsymbol{x}^{(0)},\boldsymbol{x}^{(1)}; \hat{\boldsymbol{\beta}}, \hat{\boldsymbol{\lambda}}_{i}' \hat{\boldsymbol{f}}_{t}).   \]

	\section{Asymptotic Results}
	This section presents the main theoretical results of this paper. Sections 3.1 and 3.2 provide the asymptotic distributions of the CCE estimators for $\bm{\beta}_0$ and the APE. Section 3.3 discusses how to correct the asymptotic biases of the estimators. Finally, estimators for the asymptotic variances are proposed in Section 3.4. It should be noted that, as in \cite{chen2020nonlinear} and \cite{chen2022twostep}, the assumptions and the asymptotic results below are all conditional on $(\bm{\Lambda},\bm{F})= (\bm{\Lambda}_0,\bm{F}_0)$.
	
	\subsection{Asymptotic Distribution of the CCE Estimator}
	Let $c_{0,it}= \boldsymbol{\lambda}_{0i}'\boldsymbol{f}_{0t}$. Write $\tilde{\boldsymbol{f}}_{0t} = \boldsymbol{H}_0 \boldsymbol{f}_{0t} $ and $\tilde{\boldsymbol{\lambda}}_{0i} = (\boldsymbol{H}_0^{-1})' \boldsymbol{\lambda}_{0i}$. Note that $ \tilde{\boldsymbol{\lambda}}_{0i}' \tilde{\boldsymbol{f}}_{0t} = \boldsymbol{\lambda}_{0i}'\boldsymbol{f}_{0t}=c_{0,it}$. Let $\mathcal{F}$ be a compact subset of $\mathbb{R}^r$ such that $\tilde{\boldsymbol{f}}_{0t} \in \mathcal{F}$ for all $t$. Let $\mathcal{C}$ be a compact subset of $\mathbb{R}$ such that $\boldsymbol{\lambda}'\boldsymbol{f} \in \mathcal{C}$ for all $\boldsymbol{\lambda} \in \mathcal {A}$ and all $\boldsymbol{f} \in \mathcal{F}$. Moreover, define 
	\[ l_{it}^{(j)}(\boldsymbol{\beta}, \boldsymbol{\lambda}_i'\boldsymbol{f}_t)  = \frac{ \partial^j  \log [ L(y,z)]}{\partial z^j }|_{y=y_{it},z=\boldsymbol{\beta}'\boldsymbol{x}_{it}+\boldsymbol{\lambda}_i'\boldsymbol{f}_t } \text{ for }j=1,\ldots,4, \]
	and
	\[ \underbrace{\boldsymbol{A}_i}_{r\times r}= \frac{1}{T}\sum_{t=1}^{T} \Ex[ l_{it}^{(2)}]\boldsymbol{f}_{0t} \boldsymbol{f}_{0t}' , \quad  \underbrace{\boldsymbol{B}_i}_{k\times r} = \frac{1}{T}\sum_{t=1}^{T} \Ex[ l_{it}^{(2)}\boldsymbol{x}_{it} ] \boldsymbol{f}_{0t}', \quad \dot{\boldsymbol{x}}_{it} =\boldsymbol{x}_{it} - \boldsymbol{B}_i\boldsymbol{A}^{-1}_i \boldsymbol{f}_{0t}  ,\]
	where we suppress the arguments of $l_{it}^{(j)}$ when they are evaluated at $(\boldsymbol{\beta}_0,c_{0,it})$ to simplify the notations. Assume the following conditions hold:
	\begin{ass} Let $p>1$ and $\gamma>0$ be some constants, and let $M(\cdot): \mathbb{R}^k \mapsto \mathbb{R}$ be a function such that $\max_{1\leq i\leq N, 1\leq t\leq T}\Ex [M(\boldsymbol{x}_{it})]^{2p+\gamma} <\infty$ uniformly for all $N,T$. Define $\bm{y}_i^T=(y_{i1},\ldots,y_{iT})$ and $\bm{X}_i^T=(\bm{x}_{i1},\ldots,\bm{x}_{iT})$.\\
		(i) For each $i$, the sequence $\{ (y_{it},\bm{x_{it}}):1\leq t\leq T\}$ is $\alpha$-mixing with mixing coefficient $\alpha_i(j)$, and $\max_{1\leq i\leq N} \alpha_i(j)\leq C\alpha^{j}$ for all $j$ and some $C>0$, $0<\alpha<1$. Moreover, $\{(\bm{y}_i^T,\bm{X}_i^T):1\leq i\leq N\}$ are independent across $i$.\\
		(ii) $\mathcal{B}$ and $\mathcal{A}$ are compact sets. $\boldsymbol{\beta}_0$ is an interior point of $\mathcal{B}$, and $\tilde{\boldsymbol{\lambda}}_{01},\ldots,\tilde{\boldsymbol{\lambda}}_{0N}$ are all interior points of $\mathcal{A}$.\\
		(iii) Define $\bar{l}_{it}(\boldsymbol{\beta},c) = \Ex[ l_{it}(\boldsymbol{\beta}, c) ]$. Then for any $\epsilon>0$, there exists a $\delta(\epsilon)>0$ such that 
		\[     \bar{l}_{it}(\boldsymbol{\beta}_0,c_{0,it}) - \sup_{ \| (\boldsymbol{\beta},\bm{\lambda}) - (\boldsymbol{\beta}_0,\tilde{\bm{\lambda}}_{0i})\|\geq \epsilon } \bar{l}_{it}(\boldsymbol{\beta},\bm{\lambda}'\tilde{\bm{f}}_{0t})  \geq \delta(\epsilon)  \text{ for all }i,t. \]
		(iv) $|l_{it}(\boldsymbol{\beta}, c)  | \leq M(\boldsymbol{x}_{it})$, $|  l_{it}^{(j)}(\boldsymbol{\beta}, c)| \cdot \|\boldsymbol{x}_{it}\|^d \leq M(\boldsymbol{x}_{it})$ for all $\boldsymbol{\beta} \in \mathcal{B}$ and $c \in \mathcal{C}$, for $j=1,2,3, 4$ and $d=0,1,2,3$. Moreover, $\max_{1\leq i\leq N, 1\leq t\leq T}\mathbb{E}\|\bm{x}_{it}\|^{2p+\gamma}<\infty$ uniformly for all $N,T$. \\
		(v) $N/T\rightarrow \kappa^2$ for some $\kappa >0$ as $N,T\rightarrow\infty$.\\
		(vi) There exist a $k\times k$ positive definite matrix $\bm{\Delta} $ such that:
			\[\frac{1}{NT}\sum_{i=1}^{N}\sum_{t=1}^{T} \Ex\left[  l_{it}^{(2)}\dot{\boldsymbol{x}}_{it}\dot{\boldsymbol{x}}_{it}' \right] \rightarrow \boldsymbol{\Delta}    \text{ as }N,T\rightarrow\infty. \]
		(vii) $\boldsymbol{A}_1,\ldots,\boldsymbol{A}_N$ are all invertible for large $T$.
	\end{ass}
	
Most of the conditions in Assumption 4 are standard in the literature, with a few exceptions. First, Assumption 4(i) excludes cross-sectional dependence in the data,\footnote{As stressed in the beginning of this section, these assumptions are made conditional on the factors and factor loadings. Thus, cross-sectional dependence due to the common factors are still allowed.} but it allows for general time-series dependence for each $i$ that is excluded by Assumption 1(i) of \cite{chen2020nonlinear}. Second, unlike Assumption 1(v) of \cite{chen2020nonlinear}, Assumption B of \cite{ando2022bayesian} and Assumption 2.2.C of \cite{gao2023binary}, we don't need $N^{-1}\boldsymbol{\Lambda}_0' \boldsymbol{\Lambda}_0$ to converge to some positive definite matrix. Thus, our assumption allow some columns of $\boldsymbol{\Lambda}_0$ to be 0, meaning that some factors may affect the dependent variables $y_{it}$ only indirectly through the regressors $\bm{x}_{it}$. Third, our moment restrictions on $\bm{x}_{it}$ are generally much weaker than those imposed in existing studies. For example, both \cite{chen2020nonlinear} and \cite{ando2022bayesian} require $\bm{x}_{it}$ to have bounded support, while \cite{gao2023binary} assumes that $\max_{i\leq N,t\leq T}\|\bm{x}_{it}\| = O_p(\log NT)$, which essentially require all the moments of $\bm{x}_{it}$ to exist.
	
	In order to establish the asymptotic distribution of the CCE estimator, the following definition are needed:
	\[\underbrace{\boldsymbol{C}_t }_{k\times r} = \frac{1}{N}\sum_{i=1}^{N} \Ex[ l_{it}^{(2)} \dot{\boldsymbol{x}}_{it} ] \boldsymbol{\lambda}_{0i}',
	\quad \underbrace{\boldsymbol{D}_{t,j}}_{r\times r}=  \frac{1}{N} \sum_{i=1}^{N}\boldsymbol{\lambda}_{0i}\boldsymbol{B}_{i,j} \boldsymbol{A}^{-1}_i   \bar{l}_{it}^{(2)},
	\]
	\[  \underbrace{\boldsymbol{G}_{t,j}}_{r\times r} =\frac{1}{N}\sum_{i=1}^{N} \Ex\left[ l_{it}^{(3)} \dot{\boldsymbol{x}}_{it,j} \right]\boldsymbol{\lambda}_{0i} \boldsymbol{\lambda}_{0i}', \quad  \underbrace{\boldsymbol{Q}_{i}}_{r\times r}= \frac{1}{T}\sum_{t=1}^{T}\sum_{s=1}^{T}  \Ex\left[ l_{it}^{(1)}l_{is}^{(1)}\right]  \boldsymbol{f}_{0t}\boldsymbol{f}_{0s}'  ,  \]
	where $\boldsymbol{B}_{i,j}$ is the $j$th row of $\boldsymbol{B}_{i}$. 
	
	\begin{ass}Let $\boldsymbol{\Upsilon}_0= \boldsymbol{H}_0^{-1}\boldsymbol{\Psi}_0'$ and $\boldsymbol{w}_{it} =l_{it}^{(1)}\dot{\boldsymbol{x}}_{it} +\boldsymbol{C}_t \boldsymbol{\Upsilon}_0 \boldsymbol{e}_{it}$. Then the following limits exist:
		\[\boldsymbol{\Omega} = \lim_{N,T\rightarrow \infty}\frac{1}{NT}\sum_{i=1}^{N}\sum_{t=1}^{T}\sum_{s=1}^{T} \Ex \left[ \boldsymbol{w}_{it}\boldsymbol{w}_{is}'\right],
		\]
		\[ \bm{b}^1=- 0.5 \lim_{N,T\rightarrow\infty}\frac{1}{NT}\sum_{i=1}^{N}\sum_{t=1}^{T}  \Ex[ l_{it}^{(3)} \dot{\boldsymbol{x}}_{it} ] \cdot \boldsymbol{f}_{0t}'  \boldsymbol{A}^{-1}_i \boldsymbol{Q}_{i} \boldsymbol{A}^{-1}_i  \boldsymbol{f}_{0t},\]
		\[\bm{b}^2=\lim_{N,T\rightarrow\infty} \frac{1}{NT}\sum_{i=1}^{N}\sum_{t=1}^{T}\sum_{s=1}^{T} \Ex\left[ l_{it}^{(2)}l_{is}^{(1)} \dot{\boldsymbol{x}}_{it}\right] \cdot \boldsymbol{f}_{0t} ' \boldsymbol{A}^{-1}_i \boldsymbol{f}_{0s},\]
		\[ \bm{d}^{1}=-\lim_{N,T\rightarrow\infty} \frac{1}{NT}\sum_{i=1}^{N}\sum_{t=1}^{T} \Ex\left[ l_{it}^{(2)}\dot{\boldsymbol{x}}_{it}\boldsymbol{e}_{it} ' \right] \boldsymbol{\Upsilon}_0'  \boldsymbol{\lambda}_{0i},\]
		\[ d^2_j=\lim_{N,T\rightarrow\infty}\frac{1}{NT}\sum_{i=1}^{N}\sum_{t=1}^{T} \tr \left(  \Ex[\boldsymbol{e}_{it} \boldsymbol{e}_{it}'] \cdot \boldsymbol{\Upsilon}_0'  ( \boldsymbol{D}_{t,j} -0.5 \boldsymbol{G}_{t,j })\boldsymbol{\Upsilon}_0 \right) \text{ for } j=1,\ldots,k,\]
		and $\bm{d}^2 = (d^2_1,\ldots,d^2_k)'$.	
	\end{ass}

	Then we can show that: 
	\begin{thm} Under Assumptions 1 to 5, 
		\[ \sqrt{NT} (\hat{\boldsymbol{\beta}} -\boldsymbol{\beta}_0) \overset{d}{\rightarrow} \mathcal{N} \left( \kappa \boldsymbol{\Delta}^{-1} \bm{b}+\kappa^{-1} \boldsymbol{\Delta}^{-1}\bm{d},\boldsymbol{\Delta}^{-1}\boldsymbol{\Omega} \boldsymbol{\Delta}^{-1}            \right) \]
		as $N,T\rightarrow\infty$, where $\bm{b}=\bm{b}^1+\bm{b}^2$ and $\bm{d}=\bm{d}^1+\bm{d}^2$.
	\end{thm}
	
The proof of Theorem 1 is based on the following Bahadur representation for $\hat{\bm{\beta}}$:	
\[
\boldsymbol{\Delta} (\hat{\boldsymbol{\beta}} -\boldsymbol{\beta}_0)+o_P(\| \hat{\boldsymbol{\beta}} -\boldsymbol{\beta}_0\|) =- \frac{1}{NT}\sum_{i=1}^{N}\sum_{t=1}^{T} \bm{w}_{it} \\
+\frac{\bm{b}}{T}+	\frac{\bm{d}}{N}+ o_P(T^{-1}),
\]
where the bias term $\bm{b}/T$ is due to the estimation error of $\bm{\hat{\Lambda}}$, and the bias term $\bm{d}/N$ is caused by the estimation error of $\bm{\hat{F}}$. Similar Bahadur representations were established by \cite{fernandez2016individual} and \cite{chen2020nonlinear} for nonlinear panel data models, and by \cite{chen2022twostep} for quantile panel data models. This representation provides the theoretical basis for the analytical and split-panel jackknife bias corrections that will be discussed in Section 3.3 below.

In the next two remarks, we compare Theorem 1 above with the main results of \cite{hahn2004jackknife} and \cite{chen2020nonlinear}. Following these two papers, we assume that there is no time series dependence to facilitate the comparison. Note that in this case, the expressions of $\bm{Q}_i$, $\bm{\Omega}$ and $\bm{b}^2$ reduce to
\[\boldsymbol{Q}_{i}= \frac{1}{T}\sum_{t=1}^{T} \Ex\left[ l_{it}^{(1)}\right]^2  \boldsymbol{f}_{0t}\boldsymbol{f}_{0t}' ,\quad \boldsymbol{\Omega} = \lim_{N,T\rightarrow \infty}\frac{1}{NT}\sum_{i=1}^{N}\sum_{t=1}^{T} \Ex \left[ \boldsymbol{w}_{it}\boldsymbol{w}_{it}'\right],\]
\[\bm{b}^2=\lim_{N,T\rightarrow\infty} \frac{1}{NT}\sum_{i=1}^{N}\sum_{t=1}^{T}\Ex\left[ l_{it}^{(2)}l_{it}^{(1)} \dot{\boldsymbol{x}}_{it}\right] \cdot \boldsymbol{f}_{0t} ' \boldsymbol{A}^{-1}_i \boldsymbol{f}_{0t}. \]
	
	\begin{rem} As mentioned in Remark 1, Theorem 1 also holds for extremum estimators where $l_{it}(\boldsymbol{\beta},c)$ is some smooth object function such that $ (\boldsymbol{\beta}_0,c_{0,it})$ uniquely maximizes $\Ex[l_{it}(\boldsymbol{\beta},c)]$. When $l_{it}$ is the log likelihood function as we have assumed, the expressions for the biases can be further simplified. Note that Bartlett identity gives $\Ex[ l_{it}^{(1)}]^2=-\Ex[ l_{it}^{(2)}]$, therefore $\boldsymbol{Q}_{i} = - \boldsymbol{A}_i$ and 
		\[ \bm{b}=  -\lim_{N,T\rightarrow\infty} \frac{1}{NT}\sum_{i=1}^{N}\sum_{t=1}^{T} \Ex\left[ \left(l_{it}^{(2)}l_{it}^{(1)}+0.5l_{it}^{(3)}\right)  \dot{\boldsymbol{x}}_{it}\right] \cdot \boldsymbol{f}_{0t} ' \boldsymbol{Q}^{-1}_{i} \boldsymbol{f}_{0t}.\]
		For the binary choice models, it can be shown that 
		\[ \Ex\left[ l_{it}^{(2)}l_{it}^{(1)}+0.5l_{it}^{(3)} |\boldsymbol{x}_{it} \right]  = -0.5 \frac{g(z_{it})g^{(1)}(z_{it}) }{G(z_{it}) (1-G(z_{it}))},   \]
		where $g(z) =\partial G(z)/\partial z$ and $g^{(j)}(z) =\partial^j g(z)/\partial z^j $. In particular, for logit models, it can be shown that $g(z)= G(z)(1-G(z))$, thus 
		\[ \bm{b}=  0.5 \lim_{N,T\rightarrow\infty} \frac{1}{NT}\sum_{i=1}^{N}\sum_{t=1}^{T} \Ex\left[ g^{(1)}(z_{it})  \dot{\boldsymbol{x}}_{it}\right] \cdot \boldsymbol{f}_{0t} ' \boldsymbol{Q}^{-1}_{i} \boldsymbol{f}_{0t}.\]
		For probit models, we have $g^{(1)}(z) =-z g(z)$, thus 
		\[  \Ex\left[ l_{it}^{(2)}l_{it}^{(1)}+0.5l_{it}^{(3)} |\boldsymbol{x}_{it} \right]  = 0.5z_{it}  \frac{( g(z_{it}))^2 }{G(z_{it}) (1-G(z_{it}))} = -0.5 z_{it}\cdot \Ex\left[ l_{it}^{(2)} |\boldsymbol{x}_{it} \right],\]
		and
		\[  \bm{b} =  0.5 \lim_{N,T\rightarrow\infty} \frac{1}{NT}\sum_{i=1}^{N}\sum_{t=1}^{T} \Ex\left[ \Ex\left[ l_{it}^{(2)} |\boldsymbol{x}_{it} \right] z_{it} \dot{\boldsymbol{x}}_{it}\right] \cdot \boldsymbol{f}_{0t} ' \boldsymbol{Q}^{-1}_{i} \boldsymbol{f}_{0t}.\]
		For poisson models, it is easy to show that $\Ex[l_{it}^{(3)} |\boldsymbol{x}_{it} ] = 2/z_{it}^2$ and that $ \Ex[ l_{it}^{(2)}l_{it}^{(1)} |\boldsymbol{x}_{it} ] = -1/z_{it}^2 $, it then follows that $\bm{b}^{1} +\bm{b}^{2}=0$. However, in general we have $\bm{d}^{1} +\bm{d}^{2}\neq 0$, thus for poisson models there is still a need to correct the asymptotic bias for the estimated coefficients --- this is different from the fixed effects estimators of \cite{chen2020nonlinear}. 
	\end{rem}
	
	\begin{rem} The bias terms $\bm{d}^{1}, \bm{d}^{2}$ and the term $\boldsymbol{C}_t \boldsymbol{\Upsilon}_0 \boldsymbol{e}_{it}$ in the definition of $\boldsymbol{w}_{it}$ come from the estimation errors of $\hat{\boldsymbol{f}}_t$. Thus, if $\boldsymbol{f}_{0t}$ are observed, these terms will disappear. In this case, we have 
		\[ \boldsymbol{\Omega}= \lim_{N,T\rightarrow \infty}\frac{1}{NT}\sum_{i=1}^{N}\sum_{t=1}^{T} \Ex \left[  (l_{it}^{(1)})^2 \cdot \dot{\boldsymbol{x}}_{it}\dot{\boldsymbol{x}}_{it}'\right] = -\boldsymbol{\Delta} ,\]
		and thus  
		\[ \sqrt{NT} (\hat{\boldsymbol{\beta}} -\boldsymbol{\beta}_0) \overset{d}{\rightarrow} \mathcal{N} \left( \kappa \boldsymbol{\Delta}^{-1} \bm{b}, - \boldsymbol{\Delta}^{-1}        \right).
		\]
		Moreover, if $r=1$ and $\boldsymbol{f}_{0t}=1$ for all $t$, the model reduces to the standard nonlinear panel models with only individual effects, and we have
		\[ \boldsymbol{A}_i= \frac{1}{T}\sum_{t=1}^{T}  \Ex[ l_{it}^{(2)}\boldsymbol{x}_{it} ] , \quad  \boldsymbol{B}_i = \frac{1}{T}\sum_{t=1}^{T} \Ex[ l_{it}^{(2)}\boldsymbol{x}_{it} ], \quad \boldsymbol{Q}_{i}=-\boldsymbol{A}_i, \quad  \dot{\boldsymbol{x}}_{it} =\boldsymbol{x}_{it} -\boldsymbol{B}_i/\boldsymbol{A}_i , \]
		and
		\[  \bm{b} = \lim_{N,T\rightarrow\infty}  \frac{1}{NT}\sum_{i=1}^{N}\sum_{t=1}^{T} \Ex\left[ \left(l_{it}^{(2)}l_{it}^{(1)}+0.5l_{it}^{(3)}\right)  \dot{\boldsymbol{x}}_{it}\right]/\boldsymbol{A}_i.\]
		Under stationarity assumptions, the above expression coincides with the asymptotic biases of the fixed effects estimator derived in \cite{hahn2004jackknife}.
	\end{rem}
	
	\subsection{Asymptotic Distribution of the APE Estimator}
	To derive the asymptotic distribution of the APE estimator, we first define the partial derivatives of $\delta$: $ \boldsymbol{\delta}^{\boldsymbol{\beta}}(\boldsymbol{\beta},c ) = \partial  \delta (\boldsymbol{x}^{(0)},\boldsymbol{x}^{(1)}; \boldsymbol{\beta},c )  /  \partial \boldsymbol{\beta},   \delta^{c}( \boldsymbol{\beta},c ) = \partial  \delta ( \boldsymbol{x}^{(0)},\boldsymbol{x}^{(1)};\boldsymbol{\beta},c )  /  \partial c$. Moreover, $\boldsymbol{\delta}^{\boldsymbol{\beta} c}( \boldsymbol{\beta},c )$, $\boldsymbol{\delta}^{\boldsymbol{\beta} \boldsymbol{\beta}}( \boldsymbol{\beta},c )$, $\delta^{c c}(\boldsymbol{\beta},c )$ and $\delta^{c cc}( \boldsymbol{\beta},c ) $ can be defined in a similar fasion.
	For simplicity, write $\boldsymbol{\delta}^{\boldsymbol{\beta}}_{0,it} = \boldsymbol{\delta}^{\boldsymbol{\beta}}(\boldsymbol{\beta}_0, c_{0,it} ) $, $\delta^{c}_{0,it} = \delta^{c}( \boldsymbol{\beta}_0, c_{0,it} ) $ and $\delta^{cc}_{0,it} = \delta^{cc}( \boldsymbol{\beta}_0, c_{0,it} ). $
	In addition, define 
	\[\boldsymbol{\gamma} =  \lim_{N,T\rightarrow\infty} \left( \frac{1}{NT}\sum_{i=1}^{N}\sum_{t=1}^{T} \boldsymbol{\delta}^{\boldsymbol{\beta}}_{0,it}  -\frac{1}{N}\sum_{i=1}^{N} \boldsymbol{B}_i \boldsymbol{A}_i^{-1}\boldsymbol{\gamma}_i \right),\quad \boldsymbol{\gamma}_i =  \frac{1}{T}\sum_{t=1}^{T} \delta^{c}_{0,it} \boldsymbol{f}_{0t} , \quad \boldsymbol{\gamma}_t= \frac{1}{N} \sum_{i=1}^{N}  \delta^{c}_{0,it}\boldsymbol{\lambda}_{0i},\]
	\[ \boldsymbol{R}_t =\frac{1}{N}\sum_{i=1}^{N} \boldsymbol{\lambda}_{0i} \boldsymbol{\gamma}_i' \boldsymbol{A}_i^{-1} \bar{l}_{it}^{(2)},   \quad  \boldsymbol{W}_t = \frac{1}{N}\sum_{i=1}^{N}\left( \delta^{cc}_{0,it} -0.5  \bar{l}_{it}^{(3)} \cdot \boldsymbol{\gamma}_i' \boldsymbol{A}^{-1}_i \boldsymbol{f}_{0t} \right) \boldsymbol{\lambda}_{0i}\boldsymbol{\lambda}_{0i}' , \]
	and assume that:
	\begin{ass}There exists a $M<\infty$ such that $ \|\boldsymbol{\delta}^{\boldsymbol{\beta} c}( \boldsymbol{\beta},c )\| \leq M$, $\|\boldsymbol{\delta}^{\boldsymbol{\beta} \boldsymbol{\beta} }( \boldsymbol{\beta},c )\| \leq M$, $\|\boldsymbol{\delta}^{cc \boldsymbol{\beta}}( \boldsymbol{\beta},c )\| \leq M$ and $\|\delta^{cc c}( \boldsymbol{\beta},c )\| \leq M$ for all $\boldsymbol{\beta} \in \mathcal{B}$ and $c\in\mathcal{C}$.
	\end{ass}
	
\begin{ass}
			There following limits exist:
			\[b^3 = \lim_{N,T\rightarrow\infty}\frac{1}{NT}\sum_{i=1}^{N}\sum_{t=1}^{T}  \left( \delta^{cc}_{0,it} -0.5  \bar{l}_{it}^{(3)} \cdot \boldsymbol{\gamma}_i' \boldsymbol{A}^{-1}_i \boldsymbol{f}_{0t} \right) \cdot \boldsymbol{f}_{0t}'  \boldsymbol{A}^{-1}_i \boldsymbol{Q}_{i} \boldsymbol{A}^{-1}_i  \boldsymbol{f}_{0t},\]
			\[b^4 = \lim_{N,T\rightarrow\infty}\frac{1}{NT}\sum_{i=1}^{N}\sum_{t=1}^{T}\sum_{s=1}^{T}\Ex \left[ l_{it}^{(2)}l_{is}^{(1)} \right]\cdot  \boldsymbol{\gamma}_i' \boldsymbol{A}^{-1}_i \boldsymbol{f}_{0t} \cdot  \boldsymbol{f}_{0t} ' \boldsymbol{A}^{-1}_i \boldsymbol{f}_{0s},\]
			\[d^3 = - \lim_{N,T\rightarrow\infty} \frac{1}{NT}\sum_{i=1}^{N}\sum_{t=1}^{T}\boldsymbol{\gamma}_{i}' \boldsymbol{A}^{-1}_i\boldsymbol{f}_{0t} \cdot \boldsymbol{\lambda}_{0i} '\boldsymbol{\Upsilon}_0 \cdot \Ex\left[  l_{it}^{(2)}\boldsymbol{e}_{it} \right],\]
			\[d^4 = \lim_{N,T\rightarrow\infty} \frac{1}{NT}\sum_{i=1}^{N}\sum_{t=1}^{T}  \tr\left[ \boldsymbol{\Upsilon}_0'   (\boldsymbol{W}_t-\boldsymbol{R}_t) \boldsymbol{\Upsilon}_0 \cdot \Ex \left[ \boldsymbol{e}_{it}\boldsymbol{e}_{it}'\right] \right].\]
	\end{ass}
	Then, it can be shown that:
	\begin{thm}Under Assumptions 1 to 7, 
		\[ \sqrt{NT} \left[ \hat{\delta} ( \boldsymbol{x}^{(0)},\boldsymbol{x}^{(1)} ) -  \bar{\delta}_0 ( \boldsymbol{x}^{(0)},\boldsymbol{x}^{(1)} ) \right] \overset{d}{\rightarrow}  \mathcal{N}( \kappa \cdot  b_{APE}+ \kappa^{-1} \cdot d_{APE}, \sigma^2) \]
		 as $N,T\rightarrow\infty$, where 
		\[\sigma^2= \lim_{N,T\rightarrow\infty}\frac{1}{NT} \sum_{i=1}^{N} \sum_{t=1}^{T}\sum_{s=1}^{T} \Ex[v_{it}v_{is}], \quad v_{it} = \boldsymbol{\gamma}' \boldsymbol{\Delta}^{-1}\boldsymbol{w}_{it} + (\boldsymbol{R}_t \boldsymbol{f}_{0t} -\boldsymbol{\gamma}_t)' \boldsymbol{\Upsilon}_0 \boldsymbol{e}_{it} + l_{it}^{(1)}\boldsymbol{\gamma}_i' \boldsymbol{A}_i^{-1}\boldsymbol{f}_{0t}, \] 
		\[b_{APE} =  \boldsymbol{\gamma}' \boldsymbol{\Delta}^{-1} ( \bm{b}^1 +\bm{b}^2)  + b^3 +b^4,   \quad d_{APE}=  \boldsymbol{\gamma}' \boldsymbol{\Delta}^{-1} ( \bm{d}^1 +\bm{d}^2)  + d^3 +d^4 .\]
	\end{thm}

	\subsection{Bias Correction}
	
	In order to make valid inference, we need to eliminate the asymptotic biases of the CCE estimator and the APE estimator. This can be done by either analytical bias correction or by split-panel jackknife method. 
	\subsubsection{Analytical Bias Correction}
	For analytical bias correction, we need to construct consistent estimators of $\boldsymbol{\Delta}, \bm{b}^1,\bm{b}^2,\bm{d}^1,\bm{d}^2$ and $\boldsymbol{\gamma},b^3,b^4,d^3,d^4$. First, consider the bias correction of $\bm{\hat{\beta}}$ and define:
	\[\hat{l}_{it}^{(j)}=l_{it}^{(j)}(\hat{\boldsymbol{\beta}},\hat{c}_{it}) , \quad \hat{\boldsymbol{A}}_i= \frac{1}{T}\sum_{t=1}^{T} \hat{l}_{it}^{(2)} \hat{\boldsymbol{f}}_{t} \hat{\boldsymbol{f}}_{t}' , \quad  \hat{\boldsymbol{B}}_i = \frac{1}{T}\sum_{t=1}^{T}  \hat{l}_{it}^{(2)}\boldsymbol{x}_{it}  \hat{\boldsymbol{f}}_{t}', \quad \hat{\dot{\boldsymbol{x}}}_{it} =\boldsymbol{x}_{it} - \hat{\boldsymbol{B}}_i \hat{\boldsymbol{A}}^{-1}_i \hat{\boldsymbol{f}}_{t}  ,\]
	\[ \hat{\boldsymbol{\Delta}} = \frac{1}{NT}\sum_{i=1}^{N}\sum_{t=1}^{T}  \hat{l}_{it}^{(2)} \hat{\dot{\boldsymbol{x}}}_{it} \hat{\dot{\boldsymbol{x}}}_{it}', \quad  \hat{\boldsymbol{D}}_{t,j} =  \frac{1}{N} \sum_{i=1}^{N}  \hat{\boldsymbol{\lambda}}_{i} \hat{\boldsymbol{B}}_{i,j} \hat{\boldsymbol{A}}^{-1}_i   \hat{l}_{it}^{(2)}, \quad \hat{\boldsymbol{G}}_{t,j} =\frac{1}{N}\sum_{i=1}^{N} \hat{l}_{it}^{(3)} \hat{\dot{\boldsymbol{x}}}_{it,j} \hat{\boldsymbol{\lambda}}_{i} \hat{\boldsymbol{\lambda}}_{i}',\quad  \hat{\boldsymbol{\Upsilon}}= \hat{\boldsymbol{\Psi}}' , \]
	\[\hat{\boldsymbol{Q}}_i=\frac{1}{T}\sum_{t=1}^{T}\sum_{s=1}^{T}\hat{l}_{it}^{(1)}\hat{l}_{is}^{(1)}\hat{\boldsymbol{f}}_{t}\hat{\boldsymbol{f}}_{s}'k\left(\frac{t-s}{L}\right), \quad \hat{\boldsymbol{e}}_{it} = \boldsymbol{x}_{it} - \hat{\boldsymbol{\Gamma}}_i \hat{\boldsymbol{f}}_t, \quad \hat{\boldsymbol{\Gamma}}_i'  =  \left( \sum_{t=1}^{T} \hat{\boldsymbol{f}}_t \hat{\boldsymbol{f}}_t '\right)^{-1}\left( \sum_{t=1}^{T} \hat{\boldsymbol{f}}_t \boldsymbol{x}_{it} '\right) ,\]
where $L\rightarrow\infty$ as $N,T\rightarrow\infty$ and $k(x)=(1-|x|)\boldsymbol{1}\{|x|\leq1\}$ is the Bartlett kernel function, corresponding to the HAC estimators of  \cite{newey1987}. Then the estimators of the biases of $\bm{\hat{\beta}}$ can be constructed as follows:
	\[ \hat{\bm{b}}^{1}= -0.5\frac{1}{NT}\sum_{i=1}^{N}\sum_{t=1}^{T}  \hat{l}_{it}^{(3)} \hat{\dot{\boldsymbol{x}}}_{it}   \hat{\boldsymbol{f}}_{t}'  \hat{\boldsymbol{A}}^{-1}_i \hat{\boldsymbol{Q}}_i \hat{\boldsymbol{A}}^{-1}_i \hat{\boldsymbol{f}}_{t} ,\quad \hat{\bm{b}}^{2} = \frac{1}{NT}\sum_{i=1}^{N}\sum_{t=1}^{T}\sum_{s=1}^{T}  \hat{l}_{it}^{(2)} \hat{l}_{is}^{(1)} \hat{\dot{\boldsymbol{x}}}_{it} \hat{\boldsymbol{f}}_{t}'  \hat{\boldsymbol{A}}^{-1}_i  \hat{\boldsymbol{f}}_{s}k\left(\frac{t-s}{L}\right) ,\]
	\[ \hat{\bm{d}}^{1}= - \frac{1}{NT}\sum_{i=1}^{N}\sum_{t=1}^{T}  \hat{l}_{it}^{(2)} \hat{\dot{\boldsymbol{x}}}_{it} \hat{\boldsymbol{e}}_{it} '  \hat{\boldsymbol{\Upsilon}}'  \hat{\boldsymbol{\lambda}}_{i}, \quad \hat{d}^{2}_j = \frac{1}{NT}\sum_{i=1}^{N}\sum_{t=1}^{T} \tr \left(  \hat{\boldsymbol{e}}_{it} \hat{\boldsymbol{e}}_{it}' \cdot  \hat{\boldsymbol{\Upsilon}}'  ( \hat{\boldsymbol{D}} _{t,j} -0.5 \hat{\boldsymbol{G}}_{t,j }) \hat{\boldsymbol{\Upsilon}} \right), \]
and $\hat{\bm{d}}^2=(d_1^2,\dots,d_k^2)'$, $\hat{\bm{b}}=\hat{\bm{b}}^1+\hat{\bm{b}}^2$, $\hat{\bm{d}}=\hat{\bm{d}}^1+\hat{\bm{d}}^2$. The bias-corrected CCE estimator is then defined as
		\[  \hat{ \boldsymbol{\beta}}_{ABC} = \hat{\boldsymbol{\beta}} - \hat{\boldsymbol{\Delta}}^{-1}\left(\frac{\hat{\bm{b}}}{T}    +  \frac{\hat{\bm{d}}}{N}\right) .\]
		
Next, consider the bias correction of the APE estimator and define $\hat{\delta}^{c}_{it} = \delta^c(\hat{\boldsymbol{\beta}},\hat{c}_{it})$, $\hat{\boldsymbol{\delta}}^{\boldsymbol{\beta}}_{it} = \boldsymbol{\delta}^{\boldsymbol{\beta}}(\hat{\boldsymbol{\beta}},\hat{c}_{it})$, $\hat{\delta}^{cc}_{it} = \delta^{cc}(\hat{\boldsymbol{\beta}},\hat{c}_{it})$,
	\[ \hat{\boldsymbol{\gamma}}_i =  \frac{1}{T}\sum_{t=1}^{T} \hat{\delta}^{c}_{it} \hat{\boldsymbol{f}}_{t} , \quad \hat{\boldsymbol{\gamma}}_t= \frac{1}{N} \sum_{i=1}^{N}  \hat{\delta}^{c}_{it}\hat{\boldsymbol{\lambda}}_{i}, \quad \hat{\boldsymbol{\gamma}} =   \frac{1}{NT}\sum_{i=1}^{N}\sum_{t=1}^{T} \hat{\boldsymbol{\delta}}^{\boldsymbol{\beta}}_{it}  -\frac{1}{N}\sum_{i=1}^{N} \hat{\boldsymbol{B}}_i \hat{\boldsymbol{A}}_i^{-1} \hat{\boldsymbol{\gamma}}_i, \]
	\[ \hat{\boldsymbol{R}}_t =\frac{1}{N}\sum_{i=1}^{N} \hat{\boldsymbol{\lambda}}_{i} \hat{\boldsymbol{\gamma}}_i' \hat{\boldsymbol{A}}_i^{-1} \hat{l}_{it}^{(2)},   \quad  \hat{\boldsymbol{W}}_t = \frac{1}{N}\sum_{i=1}^{N}\left( \hat{\delta}^{cc}_{it} -0.5  \hat{l}_{it}^{(3)} \cdot \hat{\boldsymbol{\gamma}}_i' \hat{\boldsymbol{A}}^{-1}_i \hat{\boldsymbol{f}}_{t} \right) \hat{\boldsymbol{\lambda}}_{i} \hat{\boldsymbol{\lambda}}_{i}' , \]
				\[\hat{b}^3 = \frac{1}{NT}\sum_{i=1}^{N}\sum_{t=1}^{T}  \left( \hat{\delta}^{cc}_{it} -0.5  \hat{l}_{it}^{(3)} \cdot \hat{\boldsymbol{\gamma}}_i' \hat{\boldsymbol{A}}^{-1}_i \hat{\boldsymbol{f}}_{t} \right) \cdot \hat{\boldsymbol{f}}_{t}'  \hat{\boldsymbol{A}}^{-1}_i \hat{\boldsymbol{Q}}_{i} \hat{\boldsymbol{A}}^{-1}_i  \hat{\boldsymbol{f}}_{t},\]
			\[\hat{b}^4 = \frac{1}{NT}\sum_{i=1}^{N}\sum_{t=1}^{T}\sum_{s=1}^{T} \hat{ l}_{it}^{(2)} \hat{l}_{is}^{(1)} \cdot  \hat{\boldsymbol{\gamma}}_i'  \hat{\boldsymbol{A}}^{-1}_i \hat{\boldsymbol{f}}_{t} \cdot  \hat{ \boldsymbol{f}}_{t} '  \hat{\boldsymbol{A}}^{-1}_i  \hat{\boldsymbol{f}}_{s}k\left(\frac{t-s}{L}\right),\]
			\[\hat{d}^3 = - \frac{1}{NT}\sum_{i=1}^{N}\sum_{t=1}^{T} \hat{\boldsymbol{\gamma}}_{i}' \hat{\boldsymbol{A}}^{-1}_i \hat{\boldsymbol{f}}_{t} \cdot \hat{\boldsymbol{\lambda}}_{i} ' \hat{\boldsymbol{\Upsilon}} \cdot   \hat{l}_{it}^{(2)} \hat{\boldsymbol{e}}_{it} , \quad \hat{d}^4 = \frac{1}{NT}\sum_{i=1}^{N}\sum_{t=1}^{T}  \tr\left[ \hat{\boldsymbol{\Upsilon}}'   ( \hat{\boldsymbol{W}}_t - \hat{\boldsymbol{R}}_t) \hat{\boldsymbol{\Upsilon}} \cdot  \hat{\boldsymbol{e}}_{it} \hat{\boldsymbol{e}}_{it}' \right].\]
The bias-corrected APE estimator is then defined as	
\[
\hat{\delta}_{ABC} ( \boldsymbol{x}^{(0)},\boldsymbol{x}^{(1)} ) =\hat{\delta} ( \boldsymbol{x}^{(0)},\boldsymbol{x}^{(1)} ) - \frac{ \hat{\boldsymbol{\gamma}}' \hat{\boldsymbol{\Delta}}^{-1} \hat{ \bm{b}} + \hat{b}^3 +\hat{b}^4}{T}
-\frac{ \hat{\boldsymbol{\gamma}}' \hat{\boldsymbol{\Delta}}^{-1}  \hat{\bm{d}} + \hat{d}^3 +\hat{d}^4}{N}.
\]
		It can be shown that the above bias-corrected estimators are free of asymptotic biases.  
		\begin{thm}
			If $L\rightarrow\infty$ and $LT^{1/(2p)-1/2}\rightarrow0$ as $N,T\rightarrow\infty$, then under Assumptions 1 to 5 it holds that $\hat{\boldsymbol{\Delta}} = \boldsymbol{\Delta} + o_P(1)$, $\hat{\bm{b}} = \bm{b} + o_P(1)$, $\hat{\bm{d}} = \bm{d} + o_P(1)$ and therefore
			\[   \sqrt{NT} (\hat{\boldsymbol{\beta}}_{ABC} -\boldsymbol{\beta}_0) \overset{d}{\rightarrow} \mathcal{N} \left( 0,\boldsymbol{\Delta}^{-1}\boldsymbol{\Omega} \boldsymbol{\Delta}^{-1}            \right) \text{ as }N,T\rightarrow\infty.  \]
Moreover, under Assumptions 1 to 7, it holds that $\hat{\bm{\gamma}} = \bm{\gamma}+o_P(1)$, $\hat{b}^3=b^3+o_P(1)$, $\hat{b}^4=b^4+o_P(1)$, $\hat{d}^3=d^3+o_P(1)$, $\hat{d}^4=d^4+o_P(1)$	and therefore
\[ 		 \sqrt{NT} \left[ \hat{\delta}_{ABC} ( \boldsymbol{x}^{(0)},\boldsymbol{x}^{(1)} ) -  \bar{\delta}_0 ( \boldsymbol{x}^{(0)},\boldsymbol{x}^{(1)} ) \right] \overset{d}{\rightarrow}  \mathcal{N}(0, \sigma^2)  \text{ as }N,T\rightarrow\infty .\]		
	\end{thm}
	
	\subsubsection{Split-Panel Jackknife Bias Correction}
	Following \cite{dhaene2015split}, \cite{fernandez2016individual} and \cite{chen2020nonlinear}, bias correction can also achieved by a simple procedure called split-panel jackknife (SPJ, hereafter). Let $\hat{\boldsymbol{\beta}}_{N/2,T}^1$ and $\hat{\boldsymbol{\beta}}_{N/2,T}^2$ be the CCE estimators using subsamples $\{ (i,t):i=1,\ldots, N/2; t=1,\ldots,T\}$ and  $\{ (i,t):i=N/2+1,\ldots, N; t=1,\ldots,T\}$ respectively. Similarly, let $\hat{\boldsymbol{\beta}}_{N,T/2}^1$ and $\hat{\boldsymbol{\beta}}_{N,T/2}^2$ be the CCE estimators using subsamples $\{ (i,t):i=1,\ldots, N; t=1,\ldots,T/2\}$ and  $\{ (i,t):i=1,\ldots, N; t=T/2+1,\ldots,T\}$ respectively. Define 
	\[   \hat{ \boldsymbol{\beta}}_{SPJ} = 3\hat{\boldsymbol{\beta}} -\frac{1}{2}\left(\hat{\boldsymbol{\beta}}_{N/2,T}^1+\hat{\boldsymbol{\beta}}_{N/2,T}^2 \right) -\frac{1}{2}\left(\hat{\boldsymbol{\beta}}_{N,T/2}^1+\hat{\boldsymbol{\beta}}_{N,T/2}^2\right) . \]
	Then under some type of homogeneity conditions to guarantee that the asymptotic biases of $ \hat{\boldsymbol{\beta}}_{N/2,T}^1,\hat{\boldsymbol{\beta}}_{N/2,T}^2, \hat{\boldsymbol{\beta}}_{N,T/2}^1,\hat{\boldsymbol{\beta}}_{N,T/2}^2$ and $\hat{\boldsymbol{\beta}}$ all converge to the same limit, it can be shown that 
	\[   \sqrt{NT} (\hat{\boldsymbol{\beta}}_{SPJ} -\boldsymbol{\beta}_0) \overset{d}{\rightarrow} \mathcal{N} \left( 0,\boldsymbol{\Delta}^{-1}\boldsymbol{\Omega} \boldsymbol{\Delta}^{-1}            \right) \text{ as }N,T\rightarrow\infty.  \]
A bias-corrected estimator using the SPJ for the APE can be defined in a similar fashion. 	
	
	\subsection{Estimating the Variances}
	For the CCE estimator, define 
	\[ \hat{\boldsymbol{C}}_t  = \frac{1}{N}\sum_{i=1}^{N}  \hat{l}_{it}^{(2)} \hat{\dot{\boldsymbol{x}}}_{it}  \hat{\boldsymbol{\lambda}}_{i}', \quad \hat{\boldsymbol{w}} _{it} = \hat{l}_{it}^{(1)} \hat{\dot{\boldsymbol{x}}}_{it} + \hat{\boldsymbol{C}}_t  \hat{\boldsymbol{\Upsilon}} \hat{\boldsymbol{e}}_{it}, \quad \hat{\boldsymbol{\Omega}} =\frac{1}{NT}\sum_{i=1}^{N}\sum_{t=1}^{T}\sum_{s=1}^{T} \hat{\boldsymbol{w}} _{it}\hat{\boldsymbol{w}} _{is}'k\left(\frac{t-s}{L}\right) .\]
	For the APE estimator, define 
	\[\hat{v}_{it} = \hat{\boldsymbol{\gamma}}' \hat{\boldsymbol{\Delta}}^{-1}\hat{\boldsymbol{w}}_{it} + (\hat{\boldsymbol{R}}_t \hat{\boldsymbol{f}}_{t} -\hat{\boldsymbol{\gamma}}_t)' \hat{\boldsymbol{\Upsilon}} \hat{\boldsymbol{e}}_{it} + \hat{l}_{it}^{(1)}\hat{\boldsymbol{\gamma}}_i' \hat{\boldsymbol{A}}_i^{-1}\hat{\boldsymbol{f}}_{t}, \quad
	\hat{\sigma}^2 = \frac{1}{NT}\sum_{i=1}^{N}\sum_{t=1}^{T}\sum_{s=1}^{T} \hat{v}_{it}\hat{v}_{is}k\left(\frac{t-s}{L}\right).
	\]
	It can be show that:
	\begin{thm} Under Assumptions 1 to 7, $\|\hat{\boldsymbol{\Delta}}^{-1}\hat{\boldsymbol{\Omega}} \hat{\boldsymbol{\Delta}}^{-1} - \boldsymbol{\Delta}^{-1}\boldsymbol{\Omega} \boldsymbol{\Delta}^{-1}   \| =o_P(1)$ and $\hat{\sigma}^2 -\sigma^2=o_P(1)$ if $L\rightarrow\infty$ and $LT^{1/(2p)-1/2}\rightarrow0$ as $N,T\rightarrow\infty$.
	\end{thm}
	The above result ensures that we can make asymptotically valid inference based on the estimated variances and bias-corrected estimators.  
	
	\begin{rem}
	As pointed out in \cite{chen2020nonlinear}, the optimal choice of the bandwidth parameter $L$ in nonlinear panel data models remains a challenging open question, and there is no consensus in the literature regarding this choice in practice. For example, \cite{hahn2011bias} and \cite{galvao2016smoothed} recommended $L=1$, whereas \cite{fernandez2016individual} suggested a sensitivity analysis starting from $L=0$. In the next section, the choice of $L$ is examined by means of Monte Carlo simulations. We find that setting $L=1$ works really well in sample samples when the time-series dependence is moderate, and therefore recommend this choice for empirical applications. 
	\end{rem}

\section{Simulations}
In this section, the finite sample performance of the proposed estimation method is evaluated using Monte Carlo simulations. The data generating process (DGP) of $y_{it}$ is given by:
	\[   y_{it} = \boldsymbol{1} \{ x_{it,1} +x_{it,2}+x_{it,3}+x_{it,4}+\lambda_{i,1}f_{t,1}  + \lambda_{i,2}f_{t,2} - \epsilon_{it} \geq 0 \} ,  \]  
where $\epsilon_{it}$ are i.i.d with the standard logistic distributions, $f_{t,1} = 0.3+0.7 f_{t-1,1} + u_{1t} , f_{t,2} = 0.6+0.4 f_{t-1,2} + u_{2t},$ $u_{1t},u_{2t}\sim \text{i.i.d } \mathcal{N}(0,1)$ and  $\lambda_{i,1},\lambda_{i,2}\sim \text{i.i.d } \mathcal{N}(1,1)$. In addition, the covariates are generated by
	\[  x_{it,1} = \theta_{1i}f_{t,1}+f_{t,2}+e_{it,1} , \quad x_{it,2} = \theta_{2i} f_{t,2} +e_{it,2}, \quad  x_{it,3} = 1.5e_{it,3}, \quad x_{it,4} = e_{it,4},\]
where $\theta_{1i},\theta_{2i}\sim \text{i.i.d } \mathcal{N}(1,1)$. As for $e_{it,j}$, $j=1,2,3,4$, two cases are considered : (i) $e_{it,j}\sim \text{i.i.d } \mathcal{N}(0,1)$; (ii) $e_{it,j} = 0.6e_{i,t-1,j}+h_{it,j}$ where $h_{it,j}\sim \text{i.i.d } \mathcal{N}(0,1)$. For the first case, there is no serial dependence in $(y_{it},\bm{x}_{it})$ conditional on $\{\bm{\lambda}_i\}$ and $\{\bm{f}_t\}$. For the second case, we need to take into account the time-series dependence when constructing the bias-corrected estimators and estimating the asymptotic variances.  

Our focus is whether the bias correction methods proposed in Section 3.3 can effectively reduce the bias of the CCE estimator, and therefore improve the empirical coverage rate of the confidence interval. To this end, we compare three estimators: the CCE estimator without bias correction $\hat{\bm{\beta}}$, the CCE estimator with analytical bias correction $\hat{\bm{\beta}}_{ABC}$, and the CCE estimator with SPJ bias correction $\hat{\bm{\beta}}_{SPJ}$. Table 2 below reports the biases and standard errors of these three estimators for the above model, along with the empirical coverage rates of their confidence intervals from 500 replications for $N,T\in\{50,100,200\}$. For the DGP without serial dependence, we set $L=0$ and the results are reported in the upper panel. For the DGP with serial dependence, the results with $L=1,2,3$ are reported in the lower three panels. Moreover, to save space, we only show results for the coefficient of $x_{it,1}$, and the results for the other three coefficients (which are very similar) are available upon request. 

Based on the results presented in Table 2, several key observations can be made. Firstly, both bias correction methods are found to be effective in significantly reducing the biases associated with CCE estimators. The analytical bias correction method is observed to perform better in models without serial dependence, whilst the SPJ method is more effective in models with serial dependence. Secondly, the standard errors of the CCE estimators are found to not be inflated in most cases, and in some instances the standard errors of the bias-corrected estimators are even lower. This suggests that bias correction does not come at the expense of increased uncertainty. Thirdly, the smaller biases of the bias-corrected estimators result in empirical coverage rates of their confidence intervals that are closer to their nominal rates of 95$\%$. Lastly, it is found that increasing $L$ from 1 to 3 does not result in improved performance of the bias-corrected estimators in models with serial dependence, thus supporting the recommendation of  \cite{hahn2011bias} and \cite{galvao2016smoothed} that using $L=1$ in practice is a reasonably good choice.

	\section{Application}
In a recent study, \cite{ma2019nonfinancial} documented that a sizable fraction of financial activities comes from firms that simultaneously issue in one financial market and repurchase in another.\footnote{Most studies discussed the capital market-driven firm financing activities in a single market, see \cite{Baker2000}, \cite{BAKER2003261}, \cite{HONG2008119} and \cite{DHT2012}, among others.} For example, using the U.S. data from 1985 to 2015, the author found that about 45$\%$ of equity repurchases in value come from firms that concurrently net issue debt, and about 35$\%$ of net debt issuance comes from the firms that net repurchase equity. The central question of that paper is how such cross-market arbitrage of the nonfinancial firms is affected by relative valuations across debt and equity markets. In this section, we revisit this important question by employing the proposed estimation method in this paper. Our main objective is to compare the estimation results of panel logistic models with only individual effects, as assumed in \cite{ma2019nonfinancial}, with models featuring interactive effects while also demonstrating the efficacy of the bias correction methods in both type of models.

	Following \cite{ma2019nonfinancial}, let the dependent variable $y_{it}$ be an indicator that identifies instances in which a firm $i$ both issues debt and repurchases equity at time period $t$:  $y_{it} = \mathbf{1}\{ s_{it}>0, d_{it}>0 \} $, where $s_{it}$ is the net equity repurchases, defined as the net purchase of Common and Preferred Stock (i.e., PRSTKC-SSTK) of firm $i$ in quarter $t$, and $d_{it}$ is the net debt issuance, defined as long-term debt issuance (DLTIS) minus long-term debt reduction (DLTR).
	
The explanatory variables in this study consist of three measures of valuations in both the debt and equity markets. To measure the debt market valuation, firm-level spreads are constructed since most firms have more than one outstanding bond. A bond's \textit{credit spread} is defined as the yield difference between its yield and the contemporaneous yield on the nearest-maturity Treasury, while the \textit{term spread} is defined as the yield difference between the nearest-maturity Treasury and the three-month Treasury bill. The firm-level credit and term spreads are then calculated as the equal-weighted averages of its outstanding bonds' spreads. Meanwhile, the \textit{book-to-market ratio} (B/P) is used as a measure of a firm's valuation condition in the equity market.\footnote{As discussed in \cite{ma2019nonfinancial} and \cite{DHT2012}, the value-to-price ratio is more appropriate if we want to study the effect of equity market valuation on $y_{it}$. We use the book-to-market ratio here mainly because we don't have access to some of the datasets needed to construct the value-to-price ratio.}  Other firm-level variables that may impact financing decisions, such as net income, cash holding, financing flows driven by investment plans (CAPX), deviations from target capital structure (measured by ex ante distance to target leverage), asset growth (which captures a firm's expansion tendency), and firm size, will also be controlled for in the analysis.\footnote{See \cite{ma2019nonfinancial} for detailed construction of these variables.}

Our analysis draws on three primary data sources: Compustat for firm-level balance sheet and cash flow variables, CRSP for equity market valuation data, and the Trade Reporting and Compliance Engine (TRACE) database for bond pricing data. As per established literature, all flow variables, such as net issuance, in a given quarter are normalized by lagged assets at the end of the previous quarter, while all stock variables, such as cash holdings, are normalized by assets in the same quarter. Our final sample consists of quarterly data for 145 nonfinancial firms over the period 2013 to 2022, corresponding to a balanced panel with $T=40$ and $5800$ total observations.

	We consider the following models and estimators.
	
	\textbf{Model A:} (only individual effects)   $\mathbb{E}[ y_{it} | \boldsymbol{x}_{it},\alpha_i] =  1/(1+e^{-\bm{\beta}_0'\bm{x}_{it}-\alpha_i }    )$ \\
	(A1) Conditional MLE (same as \cite{ma2019nonfinancial});\\
	(A2) fixed-effects estimator, analytical bias correction with $L = 1$ (\cite{hahn2004jackknife});\\
	(A3) fixed-effects estimator, SPJ bias correction (\cite{dhaene2015split}).
	
	\textbf{Model B:} (interactive effects)   $\mathbb{E}[ y_{it} | \boldsymbol{x}_{it},\boldsymbol{\lambda}_{i},\boldsymbol{f}_{t}] =  1/(1+e^{-\bm{\beta}_0'\bm{x}_{it}-\boldsymbol{\lambda}_{i}'\boldsymbol{f}_{t} }    )$ \\
	(B1) CCE estimator, no bias correction;\\
	(B2) CCE estimator, analytical bias correction with $L= 1$;\\
	(B3) CCE estimator, SPJ bias correction.
	
Model A is the traditional panel logistic model with only individual effects, used as the benchmark model for comparison. For this model, we consider three estimation methods: (A1) the classical conditional maximum likelihood estimator of \cite{andersen1970asymptotic}, which is also the estimator used by \cite{ma2019nonfinancial}; (A2) the fixed-effects estimator proposed by \cite{hahn2004jackknife}, where the asymptotic bias is corrected using analytical bias correction; (A3) the same fixed-effects estimator with SPJ bias correction proposed by \cite{dhaene2015split}. 

Compared with Model A, Model B introduces interactive fixed effects and the CCE framework. As in the simulations, three different estimators proposed in this paper are considered: (B1) the CCE estimator without bias correction; (B2) the CCE estimator 
with analytical bias correction where the bandwidth parameter is chosen as $L=1$; (B3) the CCE estimator with SPJ bias correction. Using the method proposed in Section 2.2 with $P_{NT}= \min\{N,T\}^{-1/3}$, the estimated number of factors is equal to 4.

Table 3 below displays our estimation results. Overall, the results are consistent with those of \cite{ma2019nonfinancial}, despite our use of a different sample period. Specifically, coefficients for the credit spread and term spread are negative, while the coefficient for B/P ratio is positive. These findings suggest that concurrently issuing debt and repurchasing equity is more likely when the cost of debt is low and the cost of equity is high.

However, in Model B, the coefficients of the credit spread and B/P ratio are significantly larger in absolute value, while the coefficient for the term spread is less robust across different estimation methods. Notably, in most coefficients of Model B, significant differences are evident between the CCE estimators and their bias-corrected counterparts. Overall, our empirical application underscores the importance of interactive fixed effects in nonlinear panel data models and highlights the efficacy of the proposed bias correction methods.

	\section{Conclusion}
In this paper, we introduced a novel CCE estimator for nonlinear panel data models with interactive fixed effects and homogeneous coefficients, filling an important gap in the literature. We established the theoretical properties of the estimator and demonstrated the effectiveness of both analytical and split-panel jackknife methods for eliminating its asymptotic bias. Monte Carlo simulations confirmed that the proposed estimators provide reliable results in finite samples. We also presented an empirical application that shows the applicability of our method for analyzing the cross-market arbitrage behavior of nonfinancial firms. Although our proposed method shows promising results, further research is needed to investigate the optimal choice of bandwidth parameter for the HAC estimators in the presence of serial dependence. Overall, our findings suggest that our proposed estimator is a useful tool for empirical researchers interested in analyzing nonlinear panel data models with interactive fixed effects.
\newpage

{\footnotesize
		\begin{center}
			\begin{threeparttable}[!]
				\caption{Bias Corrections of the CCE Estimators}
				\begin{tabular}{ccccccccccc}
					\hline
					&  & \multicolumn{3}{c}{Bias} & \multicolumn{3}{c}{Std Error} & \multicolumn{3}{c}{ Coverage Rate $(95\%)$ } \\
					\cmidrule(r){3-5} \cmidrule(r){6-8} \cmidrule(r){9-11}
					& $(N,T)$ & $\hat{\boldsymbol{\beta}}$  & $\hat{\boldsymbol{\beta}}_{ABC}$ & $\hat{\boldsymbol{\beta}}_{SPJ}$ & $\hat{\boldsymbol{\beta}}$  & $\hat{\boldsymbol{\beta}}_{ABC}$ & $\hat{\boldsymbol{\beta}}_{SPJ}$   & $\hat{\boldsymbol{\beta}}$  & $\hat{\boldsymbol{\beta}}_{ABC}$ & $\hat{\boldsymbol{\beta}}_{SPJ}$ \\
					\hline
					i.i.d & $(50,50)$  &0.132 &	0.018 &	-0.055 & 0.134 &	0.216 &	0.151 &	0.842 &	0.910 &	0.870 
					\\
					& $(50,100)$  & 0.055 &	0.005 &	-0.022 & 0.081 &	0.089 &	0.080 &	0.886 &	0.934 &	0.922 
					 \\
					& $(50,200)$  & 0.011 &	0.003 &	-0.005 & 0.054 &	0.055 &	0.057 &	0.946 &	0.940 &	0.924 
					 \\
					& $(100,50)$  &  0.148 & -0.027 & -0.036 &	0.092 & 	0.156 &	0.096 &	0.646 &	0.906 &	0.904 
					 \\
					& $(100,100)$  &  0.062 & 0.000 &	-0.020 &	0.056 &	0.053 &	0.056 &	0.788 &	0.974 &	0.932 
					 \\
					& $(100,200)$  & 0.026 &	0.004 &	-0.002 &	0.039 &	0.038 &	0.039 &	0.906 &	0.944 &	0.932 
					 \\
					& $(200,50)$  & 0.164 &	-0.027 &	-0.030 &	0.074 &	0.071 &	0.075 &	0.326 &	0.918 &	0.894 
					\\
					& $(200,100)$  & 0.067 &	0.001 &	-0.017 &	0.040 &	0.037 &	0.039 &	0.580 &	0.966 &	0.912 
					 \\
					& $(200,200)$  & 0.026 &	-0.001 &	-0.006 &	0.026 &	0.025 &	0.026 &	0.848 &	0.958 &	0.946 
					 \\
					\hline
					$L=1$ & $(50,50)$  & 0.136 &	0.071 &	-0.059 &	0.123 &	0.454 &	0.147 &	0.792 &	0.806 &	0.826 
					 \\
					& $(50,100)$  & 0.049 &	0.017 &	-0.024 &	0.068 &	0.068 &	0.072 &	0.892 &	0.942 &	0.920 
					 \\
					& $(50,200)$  &  0.009 &	0.008 &	-0.003 &	0.046 &	0.048 &	0.050 &	0.934 &	0.926 &	0.916 
					 \\
					& $(100,50)$  & 0.153 &	0.065 &	-0.037 &	0.085 &	0.124 &	0.092 &	0.530 &	0.884 &	0.870 
					 \\
					& $(100,100)$  & 0.065 &	0.016 &	-0.017 &	0.048 &	0.047 &	0.048 &	0.722 &	0.938 &	0.940 
					 \\
					& $(100,200)$  & 0.020 &	0.003 &	-0.007 &	0.034 &	0.034 &	0.034 &	0.900 &	0.930 &	0.932 
					 \\
					& $(200,50)$  & 0.167 &	0.054 &	-0.034 &	0.061 &	0.063 &	0.066 &	0.176 &	0.860 &	0.868 
					 \\
					& $(200,100)$  & 0.067 &	0.014 &	-0.017 &	0.036 &	0.034 &	0.035 &	0.496 &	0.936 &	0.912 
					 \\
					& $(200,200)$  & 0.028 &	0.004 &	-0.004 &	0.024 &	0.023 &	0.023 &	0.768 &	0.952 &	0.944 
					\\
					\hline
					$L=2$ & $(50,50)$   &0.116 &	0.057 &	-0.073 &	0.117 &	0.290 &	0.138 &	0.824 &	0.852 &	0.814 
					\\
					& $(50,100)$  & 0.056 &	0.025 &	-0.016 &	0.071 &	0.072 &	0.075 &	0.848 &	0.940 &	0.918 
					\\
					& $(50,200)$  & 0.006 &	0.007 &	-0.006 &	0.048 &	0.049 &	0.052 &	0.934 &	0.928 &	0.912 
					\\
					& $(100,50)$  &  0.157 &	0.057 &	-0.037 &	0.090 &	0.205 &	0.096 &	0.512 &	0.856 &	0.848 
					 \\
					& $(100,100)$  & 0.064 &	0.016 &	-0.019 &	0.051 &	0.049 &	0.053 &	0.736 &	0.942 &	0.886 
					\\
					& $(100,200)$  & 0.020 &	0.003 &	-0.007 &	0.034 &	0.034 &	0.034 &	0.898 &	0.930 &	0.934 
					 \\
					& $(200,50)$  & 0.166 &	0.057 &	-0.033 &	0.064 &	0.070 &	0.062 &	0.180 &	0.828 &	0.894 
					 \\
					& $(200,100)$  & 0.068 &	0.016 &	-0.015 &	0.035 &	0.034 &	0.035 &	0.488 &	0.926 &	0.926 
					 \\
					& $(200,200)$  & 0.028 &	0.004 &	-0.004 &	0.024 &	0.023 &	0.023 &	0.770 &	0.952 &	0.944 
					 \\
					\hline 
					$L=3$ & $(50,50)$& 0.134 &	0.085 &	-0.062 &	0.121 &	0.494 &	0.147 &	0.792 &	0.802 &	0.824 
					 \\
					& $(50,100)$  & 0.055 &	0.023 &	-0.020 &	0.073 &	0.078 &	0.078 &	0.858 &	0.922 &	0.904 
					 \\
					& $(50,200)$  & 0.006 &	0.007 &	-0.008 &	0.047 &	0.049 &	0.052 &	0.942 &	0.932 &	0.912 
					\\
					& $(100,50)$  & 0.152 &	0.058 &	-0.046 &	0.082 &	0.100 &	0.092 &	0.512 &	0.882 &	0.860 
					\\
					& $(100,100)$  & 0.065 &	0.019 &	-0.014 &	0.048 &	0.046 &	0.048 &	0.746 &	0.946 &	0.928 
					 \\
					& $(100,200)$  & 0.019 &	0.003 &	-0.007 &	0.034 &	0.034 &	0.034 &	0.902 &	0.928 &	0.936 
					 \\
					& $(200,50)$  & 0.161 &	0.057 &	-0.036 &	0.065 &	0.068 &	0.071 &	0.236 &	0.826 &	0.836 
					 \\
					& $(200,100)$  & 0.065 &	0.014 &	-0.018 &	0.036 &	0.034 &	0.035 &	0.514 &	0.928 &	0.910 
					\\
					& $(200,200)$  & 0.025 &	0.002 &	-0.006 &	0.024 &	0.023 &	0.023 &	0.808 &	0.950 &	0.938 
					 \\
					\hline
				\end{tabular}
				\begin{tablenotes}
\item Note: The DGP is given by: $y_{it} = \boldsymbol{1} \{ x_{it,1} +x_{it,2}+x_{it,3}+x_{it,4}+\lambda_{i,1}f_{t,1}  + \lambda_{i,2}f_{t,2} - \epsilon_{it} \geq 0 \} $, where $\epsilon_{it}$ are i.i.d with the standard logistic distributions, $f_{t,1} = 0.3+0.7 f_{t-1,1} + u_{1t} , f_{t,2} = 0.6+0.4 f_{t-1,2} + u_{2t},$ $u_{1t},u_{2t}\sim \text{i.i.d } \mathcal{N}(0,1)$ and  $\lambda_{i,1},\lambda_{i,2}\sim \text{i.i.d } \mathcal{N}(1,1)$ .The covariates are generated by $x_{it,1} = \theta_{1i}f_{t,1}+f_{t,2}+e_{it,1} , x_{it,2} = \theta_{2i} f_{t,2} +e_{it,2},  x_{it,3} = 1.5e_{it,3},  x_{it,4} = e_{it,4}$, where $\theta_{1i},\theta_{2i}\sim \text{i.i.d } \mathcal{N}(1,1)$. As for $e_{it,j}$, $j=1,2,3,4$, two cases are considered : (i) $e_{it,j}\sim \text{i.i.d } \mathcal{N}(0,1)$; (ii) $e_{it,j} = 0.6e_{i,t-1,j}+h_{it,j}$ where $h_{it,j}\sim \text{i.i.d } \mathcal{N}(0,1)$. The above table reports the biases and standard errors of three estimators, along with the empirical coverage rates of their confidence intervals from 500 replications.
              \end{tablenotes}
			\end{threeparttable}
		\end{center}
	}

			\begin{center}
\begin{threeparttable}
	\caption{Estimation Results of the Empirical Application  }
		\begin{tabular}{ccccccc}
			\toprule
			 
			  & \multicolumn{3}{c}{Model A: Individual Effects}        & \multicolumn{3}{c}{Model B: Interactive Effects}     \\ \cmidrule(r){1-1} \cmidrule(r){2-4}  \cmidrule(r){5-7}
			\textbf{Methods} & (A1)     & (A2)     & (A3)     & (B1)    & (B2)    & (B3)    \\ \cmidrule(r){1-1} \cmidrule(r){2-4}  \cmidrule(r){5-7}
			L.Credit spread & -0.3316 & -0.3281 & -0.4235 & -0.4259 & -0.5056 & -0.5741 \\
			& [-4.32]    & [-4.22]    & [-5.44]    & [-4.56]   & [-5.41]   & [-6.15]   \\
			L.Term spread   & -0.0998 & -0.0971 & -0.1586 & 0.0963  & 0.2296  & -0.0216 \\
			& [-1.79]    & [-1.72]    & [-2.81]    & [0.80]    & [1.90]    & [-0.18]   \\
			L.B/P           & 0.3893 & 0.3918 & 1.2477 & 0.2412  & 0.6255  & 0.5330  \\
			& [1.81]     & [1.80]      & [5.73]     & [0.76]    & [1.98]    & [1.69]    \\
			Net income      & -0.0612 & -0.0608 & -0.0457 & -0.0689 & -0.0647 & -0.0321 \\
			& [-2.74]    & [-2.69]    & [-2.02]    & [-2.36]   & [-2.21]   & [-1.10]   \\
			L.Cash holding  & -0.0250 & -0.0245 & 0.0002 & -0.0362 & -0.0393 & 0.0514  \\
			& [-3.44]    & [-3.32]    & [0.03]     & [-2.98]   & [-3.24]   & [4.23]    \\
			CAPX            & 0.2780 & 0.2764 & 0.2914 & 0.3284  & 0.0589  & -0.0408 \\
			& [4.03]     & [3.91]     & [4.12]     & [3.61]    & [0.65]    & [-0.45]   \\
			L.Leverage dev  & -0.0257 & -0.0244 & -0.0059 & -0.0396 & -0.0377 & 0.0182  \\
			& [-4.05]    & [-3.79]    & [-0.92]    & [-4.03]   & [-3.85]   & [1.85]    \\
			L.Size          & -0.5672 & -0.5638 & -0.8717 & 0.0142  & 0.2995  & 1.2508  \\
			& [-3.59]    & [-3.52]    & [-5.45]    & [0.05]    & [1.00]    & [4.16]    \\
			L.Asset growth  & -0.0032 & -0.0031 & 0.0000   & -0.0083 & -0.0117 & -0.0135 \\
			& [-1.11]    & [-1.03]    & [0.00]     & [-1.95]   & [-2.74]   & [-3.15]   \\ \bottomrule
		\end{tabular}
		{\small
		\begin{tablenotes}
\item Note: This table presents the estimation results of the empirical application ($t$ statistics are shown in brackets). The dependent variable $y_{it}$ is an indicator that identifies instances in which a firm $i$ both issues debt and repurchases equity at time period $t$. For any right-hand-side variable X, L.X means the one-period-lag of X. Model A is the benchmark model with only individual effects, and three estimators for this model are considered: (A1) the classical conditional maximum likelihood estimator; (A2) the fixed-effects estimator where the asymptotic bias is corrected using analytical bias correction; (A3) the fixed-effects estimator with SPJ bias correction. Model B introduces interactive fixed effects and the CCE framework. Three different estimators proposed in this paper are considered: (B1) the CCE estimator without bias correction; (B2) the CCE estimator 
with analytical bias correction where the bandwidth parameter is chosen as $L=1$; (B3) the CCE estimator with SPJ bias correction. The estimated number of factors is equal to 4.
              \end{tablenotes}}
\end{threeparttable}	
\end{center}

	\appendix
	\numberwithin{equation}{section}
	\newpage

	\section{Proofs of the Main Results}

		\subsection{Proof of Theorem 1}

		
		\begin{lemma} Let $\{X_i\}$, $i=1,2,\ldots,$ be a sequence of random variables such that $\Ex[X_i] =0$ for all $i$. Suppose one of the following conditions holds:
			(a) $\{X_i\}$ is independent and $ \sup_{1 \leq i \leq n}\Ex |X_i|^{2p}< \infty$ for some $p\geq 1$ and all $n$;
	     (b) $\{X_i\}$ is $\alpha$-mixing with coefficients $\alpha(k)$ satisfying $\alpha(k)\leq C\cdot\alpha^k$ for all $k$, some $C>0$ and $0<\alpha<1$, and $ \sup_{1\leq i \leq n}\Ex |X_i|^{2p+\gamma}< \infty$ for some $p\geq 1,\gamma>0$ and all $n$. Then as $n\rightarrow\infty$ we have			
				\[(i) \quad \Ex  \left| \sum_{i=1}^{n}X_i   \right|^{2p} =O(n^{p}) \quad \text{and} \quad (ii) \quad P  \left[  \left| \frac{1}{n}\sum_{i=1}^{n}X_i   \right| >C \right] =O(n^{-p}).\]
		\end{lemma}
		\begin{proof}

				First note that (ii) follows from Markov's inequality once (i) holds. Next, when condition (a) holds, (i) is directly implied by Rosenthal's inequality (see \cite{rosenthal1970subspaces}). 
				
				Now suppose condition (b) is satisfied. For $p=1$, Corollary 1.1 and (1.25a) of \cite{rio2017asymptotic} implies that
				\[\Ex\left(\sum_{i=1}^{n}X_i\right)^2\leq\sum_{i=1}^{n}\sum_{j=1}^{n}\left|\Ex(X_iX_j)\right|\leq a_\gamma \left(\sum_{i=1}^{n}\|X_i\|_{2+\gamma}^2\right)\cdot\left(\sum_{k=0}^{+\infty}(k+1)^{2/\gamma}\alpha(k)\right)^{\gamma/(2+\gamma)},\]
				where $a_\gamma$ is a positive constant only depends on $\gamma$, and $\|X\|_{q}=\left(\Ex|X|^q\right)^{1/q}$ is the $L_q$-norm of a random variable $X$. Then (i) follows from condition (b). 
				
				When $p>1$, by Theorem 6.3, equation (6.4), Corollary 1.1, equation (1.25a) and (C.9) of \cite{rio2017asymptotic}, for any $0<\epsilon\leq2p+\gamma-2$ we have
				\begin{multline*}
					\mathbb{E}\left|\sum_{i=1}^{n}X_i\right|^{2p}\leq a_{p,\epsilon}\left(\sum_{i=1}^{n}\|X_i\|_{2+\epsilon}^2\right)^p\left(\sum_{k=0}^{+\infty}(k+1)^{2/\epsilon}\alpha(k)\right)^{\frac{\epsilon p}{2+\epsilon}}
					+ nb_p\sum_{k=0}^{+\infty}(k+1)^{2p-2}\int_{0}^{\alpha(k)}Q_{(n)}^{2p}(u)du,
				\end{multline*}
				where $a_{p,\epsilon},b_{p}$ are some positive constants only depend on $p$ and $\epsilon$, $Q_X(u)=\inf\{v:\mathbb{P}(|X|>v)\leq u\}$ and $Q_{(n)}(u)=\sup_{1\leq i \leq n}Q_{X_i}(u)$. For the second term on the right-hand side of the above equation, we have $Q_{X_i}(u)\leq \|X_i\|_{2p+\gamma}\cdot u^{-1/(2p+\gamma)}$ since 
				\[\mathbb{P}\left(|X_i|>v\right)\leq \frac{\mathbb{E}|X_i|^{2p+\gamma}}{v^{2p+\gamma}}\quad \Rightarrow \quad \mathbb{P}\left(|X_i|>\|X_i\|_{2p+\gamma}\cdot u^{-1/(2p+\gamma)}\right)\leq u.\]
				Thus, it holds that $Q_{(n)}(u)\leq \sup_{1\leq i \leq n}\|X_i\|_{2p+\gamma}\cdot u^{-1/(2p+\gamma)}$ and
				\[\int_{0}^{\alpha(k)}Q_{(n)}^{2p}(u)du \leq \frac{2p+\gamma}{\gamma}\cdot\left(\sup_{1\leq i \leq n}\|X_i\|_{2p+\gamma}\right)^{2p}\cdot\alpha(k)^{\gamma/(2p+\gamma)}.\]
				That is, for the case $p>1$,
				\begin{multline*}
					\mathbb{E}\left|\sum_{i=1}^{n}X_i\right|^{2p}\leq a_{p,\epsilon}\left(\sum_{i=1}^{n}\|X_i\|_{2+\epsilon}^2\right)^p\left(\sum_{k=0}^{+\infty}(k+1)^{2/\epsilon}\alpha(k)\right)^{\frac{\epsilon p}{2+\epsilon}} \\
					+ n\cdot\frac{(2p+\gamma)b_p}{\gamma}\cdot\left(\sup_{1\leq i \leq n}\|X_i\|_{2p+\gamma}\right)^{2p}\cdot\left(\sum_{k=0}^{+\infty}(k+1)^{2p-2}\alpha(k)^{\gamma/(2p+\gamma)}\right),
				\end{multline*}
				which together with condition (b) lead to $\mathbb{E}\left|\sum_{i=1}^{n}X_i\right|^{2p}= O_P(n^{p})+o_P(n^{p})$.
		\end{proof}

		
		\begin{lemma}If $(X_i,Y_i)$, $i=1,2,\ldots,$ is a sequence of random vectors in $\mathbb{R}^2$ such that $\Ex[ |X_i|^{2}\log (1+|X_i|)]< \infty$ and $\Ex[ |Y_i|^{2}\log (1+|Y_i|)]< \infty$ for all $i$. Suppose $\{(X_i,Y_i)\}$ is $\alpha$-mixing with coefficients $\alpha(j)$ satisfying $\alpha(j)\leq C\cdot \alpha^j$ for all $j$, some $C>0$ and $0<\alpha<1$, then:
				\[\sum_{i=1}^{n}\sum_{j=1}^{n} \left|Cov(X_i,Y_j)\right|=O(n).\]
		\end{lemma}
		\begin{proof}
		By arguments similar to the proof of Corollary 1.1 in \cite{rio2017asymptotic}, it can be shown that
				\[\sum_{i=1}^{n}\sum_{j=1}^{n} \left|Cov(X_i,Y_j)\right|\leq 2 \sum_{i=1}^{n}\int_{0}^{1}[\alpha^{-1}(u)]Q_{X_{i}}^{2}(u)du+2\sum_{j=1}^{n}\int_{0}^{1}[\alpha^{-1}(u)]Q_{Y_{j}}(u)du,\]
				where for some positive integer $q$ and random variable $Z$, integral $\int_{0}^{1}[\alpha^{-1}(u)]^{q-1}Q_{Z}^{q}(u)du$ can be viewed as some weighted moment of $|Z|$ as in \cite{rio2017asymptotic}. Then the assumptions of this lemma and (C.17) of \cite{rio2017asymptotic} imply the boundedness of all the integrals involved, which leads to the desired result.	\end{proof}

		
\begin{lemma}
				Let $X_i$, $i=1,2,\ldots,$ be a sequence of random variables such that $\Ex[X_i] =0$ and $ \Ex[ |X_i|^{2p}(\log (1+|X_i|))^{2p-1}]< \infty$ for some positive integer $p$ and all $i$. Suppose $\{X_i\}$ is $\alpha$-mixing with coefficients $\alpha(j)$ satisfying $\alpha(j)\leq C\cdot\alpha^j$ for all $j$, some $C>0$ and $0<\alpha<1$, then it holds that
				\[\sum_{1\leq i_1 \leq \cdots \leq i_{2p} \leq n}\left|\Ex\left[X_{i_1}\dots X_{i_{2p}}\right]\right|=O(n^{p})\]
			\end{lemma}
			\begin{proof}
				By (2.15), (2.20) and $\mathcal{H}(q)$ in the proof of Theorem 2.2 in \cite{rio2017asymptotic}, we have
				\[\sum_{1\leq i_1 \leq \cdots \leq i_{2p} \leq n}\left|\Ex\left[X_{i_1}\dots X_{i_{2p}}\right]\right|\leq a_p \left(\sum_{k=1}^{n}\int_{0}^{1}[\alpha^{-1}(u)]Q_{X_{i_k}}^{2}(u)du\right)^p+b_p\sum_{k=1}^{n}\int_{0}^{1}[\alpha^{-1}(u)]^{2p-1}Q_{X_{i_k}}^{2p}(u)du.\]
				Then, similar to the proof of Lemma 2, the assumptions $ \Ex[ |X_i|^{2p}(\log (1+|X_i|))^{2p-1}]< \infty$, $\alpha(j)\leq C\cdot\alpha^j$  and (C.17) of \cite{rio2017asymptotic} imply the boundedness of all the integrals involved, which leads to the desired result.
		\end{proof}
		
		
		\begin{lemma}[Consistency] Under Assumptions 1 to 4, we have 
			\[ \| \hat{\boldsymbol{\beta}} -\boldsymbol{\beta}_0 \| = o_P(1) \quad \text{ and } \quad \max_{1\leq i\leq N} \| \hat{\boldsymbol{\lambda}}_i - \tilde{\boldsymbol{\lambda}}_{0i}\| =o_P(1).\]
		\end{lemma}
		\begin{proof}
			\textbf{Step 1: consistency of $\hat{\bm{\beta}}$}

			First, by definition, $\Lx_{NT}(\boldsymbol{\beta}_0,  \tilde{\boldsymbol{\Lambda}}_0, \hat{\boldsymbol{F}}) - \Lx_{NT}( \hat{\boldsymbol{\beta}}, \hat{\boldsymbol{\Lambda}}, \hat{\boldsymbol{F}}) \leq 0 $. Adding and subtracting terms, we have
			\begin{equation*}
				\frac{1}{NT}\sum_{i=1}^{N} \sum_{t=1}^{T}  \left[ \bar{l}_{it}(\boldsymbol{\beta}_0,c_{0,it})  - \bar{l}_{it}(\hat{\boldsymbol{\beta}},\hat{\boldsymbol{\lambda}}_i'  \tilde{\boldsymbol{f}}_{0t}  ) \right] \leq I+II+III+IV
			\end{equation*}
			where 
			\[ I =- \frac{1}{NT}\sum_{i=1}^{N} \sum_{t=1}^{T}  \left[ l_{it}(\boldsymbol{\beta}_0,c_{0,it})  - \bar{l}_{it}(\boldsymbol{\beta}_0,c_{0,it} ) \right] , \quad II = \frac{1}{NT}\sum_{i=1}^{N} \sum_{t=1}^{T}  \left[ l_{it}( \hat{\boldsymbol{\beta}},\hat{\boldsymbol{\lambda}}_i' \tilde{\boldsymbol{f}}_{0t})  - \bar{l}_{it}(\hat{\boldsymbol{\beta}},\hat{\boldsymbol{\lambda}}_i' \tilde{\boldsymbol{f}}_{0t}  ) \right] ,\]
			\[ III =\Lx_{NT}(\boldsymbol{\beta}_0, \tilde{\boldsymbol{\Lambda}}_0, \tilde{\boldsymbol{F}}_0) -\Lx_{NT}(\boldsymbol{\beta}_0,  \tilde{\boldsymbol{\Lambda}}_0, \hat{\boldsymbol{F}}) , \quad IV = \Lx_{NT}(\hat{\boldsymbol{\beta}}, \hat{\boldsymbol{\Lambda}}, \hat{\boldsymbol{F}}) -\Lx_{NT}(\hat{\boldsymbol{\beta}}, \hat{\boldsymbol{\Lambda}}, \tilde{\boldsymbol{F}}_0).  \]
			By Assumption 4(iii), for any $\epsilon>0$, $\| \hat{\boldsymbol{\beta}} - \boldsymbol{\beta}_0   \| > \epsilon$ implies that there exists a $\delta>0$ such that 
			\[\bar{l}_{it}(\boldsymbol{\beta}_0,c_{0,it})  - \bar{l}_{it}(\hat{\boldsymbol{\beta}},\hat{\boldsymbol{\lambda}}_i' \tilde{\boldsymbol{f}}_{0t}  ) \geq \delta \text{ for all }i,t.   \]
			It then follows that 
			\begin{equation*}
				P\left[ \| \hat{\boldsymbol{\beta}} - \boldsymbol{\beta}_0   \| > \epsilon \right] \leq P\left[ |I| > \delta/4 \right]+P\left[ |II| > \delta/4 \right]+P\left[ |III| > \delta/4 \right]+P\left[ |IV| > \delta/4 \right].
			\end{equation*}
			Note that by Assumption 4(i) and Lemma 1, 
			\begin{multline*}
				\mathbb{E}\left[\left(\frac{1}{\sqrt{NT}}\sum_{i=1}^{N}\sum_{t=1}^{T}[l_{it}(\boldsymbol{\beta}_0,c_{0,it})-\bar{l}_{it}(\boldsymbol{\beta}_0,c_{0,it})]\right)^2\right] 
				\\	= \frac{1}{NT}\sum_{i=1}^{N}\mathbb{E}\left(\sum_{t=1}^{T}[l_{it}(\beta_0,c_{0,it})-\bar{l}_{it}(\beta_0,c_{0,it})]\right)^2=O(1).
			\end{multline*}
			It then follows that $I = O_P(1/\sqrt{NT})$ and $P\left[ |I| > \delta/4 \right] \rightarrow 0$. 
			
			Second, by Assumption 4 we have
			\begin{multline*} |III| =\left|\frac{1}{NT}\sum_{i=1}^{N} \sum_{t=1}^{T} l_{it}(\boldsymbol{\beta}_0, \tilde{\boldsymbol{\lambda}}_{0i}' \hat{\boldsymbol{f}}_t) -  l_{it}(\boldsymbol{\beta}_0, \tilde{\boldsymbol{\lambda}}_{0i}'  \tilde{\boldsymbol{f}}_{0t}) \right|  
				\leq \frac{1}{NT}\sum_{i=1}^{N} \sum_{t=1}^{T} M(\boldsymbol{x}_{it}) \cdot \| \tilde{\boldsymbol{\lambda}}_{0i}\| \cdot \|\hat{\boldsymbol{f}}_t - \tilde{\boldsymbol{f}}_{0t} \| \\
				\lesssim \max_{1\leq t\leq T} \| \hat{\boldsymbol{f}}_t - \tilde{\boldsymbol{f}}_{0t}\| \cdot \frac{1}{NT}\sum_{i=1}^{N} \sum_{t=1}^{T} M(\bm{x}_{it}).
			\end{multline*}
			Since $\hat{\boldsymbol{f}}_t - \tilde{\boldsymbol{f}}_{0t} = \hat{\boldsymbol{\Psi}}' \bar{\boldsymbol{e}}_t+(\hat{\bm{H}}-\bm{H}_0)\bm{f}_{0t} $, we have
			\[ \max_{1\leq t\leq T} \| \hat{\boldsymbol{f}}_t - \tilde{\boldsymbol{f}}_{0t}\|  \leq   N^{-1/2 } \cdot \|\hat{\boldsymbol{\Psi}}\| \cdot  \max_{1\leq t\leq T} \| \sqrt{N} \bar{\boldsymbol{e}}_t \|  +o_P(1) = O_P(T^{1/2p}N^{-1/2})+o_P(1)=o_P(1) , \]
			because $\max_{1\leq t\leq T} \| \sqrt{N} \bar{\boldsymbol{e}}_t \| =O_P(T^{1/2p}) $ by Assumption 4(iv) and Lemma 1. It then follows that $III=o_P(1) $ and thus $P\left[ |III| > \delta/4 \right] \rightarrow 0$. Similarly, it can be shown that 
			$P\left[ |IV| > \delta/4 \right] \rightarrow 0$. 
			
			Third,
			\[ P\left[ |II| > \delta/4 \right]
			\leq \sum_{i=1}^{N}  P\left[  \sup_{\boldsymbol{\beta}\in\mathcal{B},\boldsymbol{\lambda}\in\mathcal{A}} \left|   \frac{1}{T}\sum_{t=1}^{T}  \left( l_{it}( \boldsymbol{\beta},\boldsymbol{\lambda}' \tilde{\boldsymbol{f}}_{0t})  - \bar{l}_{it}(\boldsymbol{\beta},\boldsymbol{\lambda}' \tilde{\boldsymbol{f}}_{0t}  ) \right) \right|> \delta/4 \right].\]
			
			Write $\boldsymbol{\theta}=(\boldsymbol{\beta},\boldsymbol{\lambda})$ and $\Theta = \mathcal{B} \times \mathcal{A}$. Let 
			\[ 0<\omega<\delta/\left(24\cdot \max_{i,t} \Ex[M(\bm{x}_{it})]\cdot C_{\mathcal{F}}\right),\] 
			where $C_{\mathcal{F}}=1+\max_{f\in\mathcal{F}}\|f\|$, and let $\Theta_1,\ldots, \Theta_J$ be a partition of $\Theta$ such that $\|\boldsymbol{\theta}_k -\boldsymbol{\theta}_l\|\leq \omega$ for any $\boldsymbol{\theta}_k,\boldsymbol{\theta}_l \in \Theta_j$ and any $1\leq j\leq J$. For any $\boldsymbol{\theta}=(\boldsymbol{\beta},\boldsymbol{\lambda})\in\Theta $, there exists $1\leq j\leq J$ and $\boldsymbol{\theta}^{\ast}=(\boldsymbol{\beta}^{\ast},\boldsymbol{\lambda}^{\ast})\in \Theta_j$ such that $\|\boldsymbol{\theta} -\boldsymbol{\theta}^{\ast}\|\leq \omega$, implying that 
			\[
			\left|l_{it}(\bm{\beta},\bm{\lambda}'\tilde{\bm{f}}_{0t})-l_{it}(\bm{\beta}^{\ast},(\bm{\lambda}^{\ast})'\tilde{\bm{f}}_{0t})\right|\leq \left|l_{it}^{(1)}(\dot{\bm{\beta}},\dot{\bm{\lambda}}'\tilde{\bm{f}}_{0t})\right|\cdot ( \|\bm{x}_{it}\|+ \|\tilde{\bm{f}}_{0t}\|)\cdot \|\bm{\theta}-\bm{\theta}^{\ast}\| \leq \omega \cdot M(\bm{x}_{it})C_{\mathcal{F}}
			\]
			where $(\dot{\bm{\beta}},\dot{\bm{\lambda}})$ lies between $\bm{\theta}$ and $\bm{\theta}^{\ast}$. It follows that
			\begin{multline*}
				\left|   \frac{1}{T}\sum_{t=1}^{T}  \left( l_{it}( \boldsymbol{\beta},\boldsymbol{\lambda}' \tilde{\boldsymbol{f}}_{0t})  - \bar{l}_{it}(\boldsymbol{\beta},\boldsymbol{\lambda}' \tilde{\boldsymbol{f}}_{0t}  ) \right) \right|
				\leq 
				\left|   \frac{1}{T}\sum_{t=1}^{T}  \left( l_{it}( \boldsymbol{\beta}^{\ast},(\boldsymbol{\lambda}^{\ast})' \tilde{\boldsymbol{f}}_{0t})  - \bar{l}_{it}(\boldsymbol{\beta}^{\ast},(\boldsymbol{\lambda}^{\ast})'\tilde{\boldsymbol{f}}_{0t}  ) \right) \right| \\
				+\omega \cdot C_{\mathcal{F}}\left|   \frac{1}{T}\sum_{t=1}^{T}( M(\bm{x}_{it})-\Ex M(\bm{x}_{it})) \right|  +2\omega \cdot C_{\mathcal{F}}\frac{1}{T}\sum_{t=1}^{T}\Ex M(\bm{x}_{it}).
			\end{multline*}
			Thus,
			\begin{multline}\label{A1}
				P\left[  \sup_{\boldsymbol{\beta}\in\mathcal{B},\boldsymbol{\lambda}\in\mathcal{A}} \left|   \frac{1}{T}\sum_{t=1}^{T}  \left( l_{it}( \boldsymbol{\beta},\boldsymbol{\lambda}' \tilde{\boldsymbol{f}}_{0t})  - \bar{l}_{it}(\boldsymbol{\beta},\boldsymbol{\lambda}' \tilde{\boldsymbol{f}}_{0t}  ) \right) \right|> \delta/4 \right] \\
				\leq \sum_{j=1}^{J} P\left[   \left|   \frac{1}{T}\sum_{t=1}^{T}  \left( l_{it}( \boldsymbol{\beta}_j^{\ast},(\boldsymbol{\lambda}_j^{\ast})' \tilde{\boldsymbol{f}}_{0t})  - \bar{l}_{it}( \boldsymbol{\beta}_j^{\ast},(\boldsymbol{\lambda}_j^{\ast})' \tilde{\boldsymbol{f}}_{0t} ) \right) \right|> \delta/12 \right]\\
				+ P\left[   \left | \frac{1}{T}\sum_{t=1}^{T} M(\bm{x}_{it})-\Ex M(\bm{x}_{it})  \right|> \delta/(12\omega\cdot C_{\mathcal{F}}) \right] 
				+P\left[  \frac{1}{T}\sum_{t=1}^{T}\Ex M(\bm{x}_{it}) > \delta/(24\omega\cdot C_{\mathcal{F}}) \right], 
			\end{multline}	
			where $(\boldsymbol{\beta}_j^{\ast}, \boldsymbol{\lambda}_j^{\ast}) \in \Theta_j$. The last term on the right-hand side of \eqref{A1} is 0 by the definition of $\omega$. It follows Assumption 4(iv) and Lemma 1 that the first two terms on the right-hand side of \eqref{A1} are $O(T^{-p})$. Thus, we have $P\left[ |II| > \delta/4 \right] = O(N/T^p)=o(1)$. Therefore, it can be concluded that $P[ \| \hat{\boldsymbol{\beta}} - \boldsymbol{\beta}_0   \| > \epsilon ] \rightarrow 0$.
			
			\vspace{0.5cm}
			
			\textbf{Step 2: uniform consistency of $\bm{\hat{\lambda}_i}$}	
			
			Define $ \Lx_{i,T}(\boldsymbol{\beta}, \boldsymbol{\lambda}, \boldsymbol{F}) = T^{-1}\sum_{t=1}^{T} l_{it}(\boldsymbol{\beta}, \boldsymbol{\lambda}'\boldsymbol{f}_t )$, then we have $\Lx_{i,T}(\hat{\boldsymbol{\beta}}, \hat{\boldsymbol{\lambda}}_i, \hat{\boldsymbol{F}}) \geq \Lx_{i,T}(\hat{\boldsymbol{\beta}}, \tilde{\boldsymbol{\lambda}}_{0i}, \hat{\boldsymbol{F}}) $ for all $i$. Adding and subtracting terms gives:
			\begin{multline*}
				\frac{1}{T}\sum_{t=1}^{T }  \left[\bar{ l}_{it}(\boldsymbol{\beta}_0, \tilde{ \boldsymbol{\lambda}}_{0i}' \tilde{\boldsymbol{f}}_{0t}) - \bar{ l}_{it}(\hat{\boldsymbol{\beta}}, \hat{\boldsymbol{\lambda}}_i' \tilde{\boldsymbol{f}}_{0t})  \right]     \leq  
				\frac{1}{T}\sum_{t=1}^{T }\left[ l_{it}(\hat{\boldsymbol{\beta}}, \hat{\boldsymbol{\lambda}}_i' \tilde{\boldsymbol{f}}_{0t})- \bar{ l}_{it}(\hat{\boldsymbol{\beta}}, \hat{\boldsymbol{\lambda}}_i' \tilde{\boldsymbol{f}}_{0t}) \right] \\  
				- \frac{1}{T}\sum_{t=1}^{T }\left[ l_{it}(\hat{\boldsymbol{\beta}}, c_{0,it})  - \bar{ l}_{it}(\hat{\boldsymbol{\beta}}, c_{0,it})  \right] -\frac{1}{T}\sum_{t=1}^{T }  \left[\bar{ l}_{it}(\hat{\boldsymbol{\beta}}, c_{0,it}) - \bar{ l}_{it}(\boldsymbol{\beta}_0, c_{0,it})  \right]  \\
				+ \left[ \Lx_{i,T}(\hat{\boldsymbol{\beta}}, \hat{\boldsymbol{\lambda}}_i, \hat{\boldsymbol{F}}) - \Lx_{i,T}(\hat{\boldsymbol{\beta}}, \hat{\boldsymbol{\lambda}}_i, \tilde{\boldsymbol{F}}_0)\right]    
				- \left[  \Lx_{i,T}(\hat{\boldsymbol{\beta}},\tilde{ \boldsymbol{\lambda}}_{0i}, \hat{\boldsymbol{F}}) -  \Lx_{i,T}(\hat{\boldsymbol{\beta}}, \tilde{ \boldsymbol{\lambda}}_{0i}, \tilde{\boldsymbol{F}}_0)\right].
			\end{multline*}
			By Assumption 4(iii), for any $\epsilon>0$, $\max_{1\leq i \leq N} \| \hat{\boldsymbol{\lambda}}_i  - \tilde{ \boldsymbol{\lambda}}_{0i}\| \geq \epsilon$ implies that there exists a $\delta>0$ such that 
			\[ \frac{1}{T}\sum_{t=1}^{T }  \left[\bar{ l}_{it}({\boldsymbol{\beta}}_0, \tilde{ \boldsymbol{\lambda}}_{0i}' \tilde{\boldsymbol{f}}_{0t}) - \bar{ l}_{it}( \hat{{\boldsymbol{\beta}}}, \hat{\boldsymbol{\lambda}}_i' \tilde{\boldsymbol{f}}_{0t})  \right]  \geq \delta  \text{ for some } i \leq N.\]
			Thus, 		
			
			\begin{multline}\label{A2}
				P\left[\max_{1\leq i \leq N} \| \hat{\boldsymbol{\lambda}}_i  - \tilde{\boldsymbol{\lambda}}_{0i}\| \geq \epsilon \right] \leq 
				P\left[ \max_{1\leq i\leq N}  \left|\frac{1}{T}\sum_{t=1}^{T }\left[ l_{it}(\hat{\boldsymbol{\beta}}, \hat{\boldsymbol{\lambda}}_i' \tilde{\boldsymbol{f}}_{0t})- \bar{ l}_{it}(\hat{\boldsymbol{\beta}}, \hat{\boldsymbol{\lambda}}_i' \tilde{\boldsymbol{f}}_{0t}) \right] \right|\geq \delta/5 \right]\\
				+P\left[\max_{1\leq i\leq N}  \left| \frac{1}{T}\sum_{t=1}^{T }\left[ l_{it}(\hat{\boldsymbol{\beta}}, c_{0,it})  - \bar{ l}_{it}(\hat{\boldsymbol{\beta}}, c_{0,it})  \right]  \right|\geq \delta/5 \right]\\
				+P\left[\max_{1\leq i\leq N}  \left| \frac{1}{T}\sum_{t=1}^{T }\left[ \bar{l}_{it}(\boldsymbol{\beta}_0, c_{0,it})  - \bar{ l}_{it}(\hat{\boldsymbol{\beta}}, c_{0,it})  \right]  \right|\geq \delta/5 \right]\\
				+P\left[\max_{1\leq i\leq N}  \left| \Lx_{i,T}(\hat{\boldsymbol{\beta}}, \hat{\boldsymbol{\lambda}}_i, \hat{\boldsymbol{F}}) - \Lx_{i,T}(\hat{\boldsymbol{\beta}}, \hat{\boldsymbol{\lambda}}_i, \tilde{\boldsymbol{F}}_0) \right|\geq \delta/5 \right] \\ 
				+P\left[\max_{1\leq i\leq N}  \left| \Lx_{i,T}(\hat{\boldsymbol{\beta}}, \tilde{\boldsymbol{\lambda}}_{0i}, \hat{\boldsymbol{F}}) -  \Lx_{i,T}(\hat{\boldsymbol{\beta}}, \tilde{\boldsymbol{\lambda}}_{0i}, \tilde{\boldsymbol{F}}_0) \right|\geq \delta/5 \right] .
			\end{multline}
			Similar to the proof of step 1, we can show that the first two terms on the right-hand side of \eqref{A2} are both $O(N/T^{p}) =o(1)$. Note that 
			\[\max_{1\leq i\leq N} \left| \Lx_{i,T}(\hat{\boldsymbol{\beta}}, \hat{\boldsymbol{\lambda}}_i, \hat{\boldsymbol{F}}) - \Lx_{i,T}(\hat{\boldsymbol{\beta}}, \hat{\boldsymbol{\lambda}}_i, \tilde{\boldsymbol{F}}_0) \right| \lesssim \max_{1\leq t\leq T} \| \hat{\boldsymbol{f}}_t - \tilde{\boldsymbol{f}}_{0t}\| \cdot  \max_{1\leq i\leq N}  \frac{1}{T} \sum_{t=1}^{T} M(\bm{x}_{it}) .\]
			We have shown that $\max_{1\leq t\leq T} \| \hat{\boldsymbol{f}}_t - \tilde{\boldsymbol{f}}_{0t}\| =o_P(1)$. Moreover, 
			\[ \max_{1\leq i\leq N}  \frac{1}{T} \sum_{t=1}^{T} M(\bm{x}_{it}) \leq\max_{1\leq i\leq N}  \left| \frac{1}{T} \sum_{t=1}^{T} \left[ M(\bm{x}_{it}) -\Ex M(\bm{x}_{it}) \right]  \right|  + \max_{i,t} \Ex M(\bm{x}_{it}) = O_P(N^{1/2p}T^{-1/2}) +O(1). \]
			Thus, the fourth term on the right-hand side of \eqref{A2} is $o(1)$. It can be shown in a similar way that the last term on the right-hand side of \eqref{A2} is also $o(1)$. Finally, by the consistency of $\hat{\bm{\beta}}$,
			\[\max_{1\leq i\leq N}  \left| \frac{1}{T}\sum_{t=1}^{T }\left[ \bar{l}_{it}(\boldsymbol{\beta}_0, c_{0,it})  - \bar{ l}_{it}(\hat{\boldsymbol{\beta}}, c_{0,it}) \right] \right| \leq  \|\hat{\bm{\beta}}-\bm{\beta}_0 \| \cdot \max_{i,t}\Ex M(\bm{x}_{it})=o_P(1),
			\]
			it follows that the third term on the right-hand side of \eqref{A2} is $o(1)$. Then the desired result follows. \end{proof}

		Now define $\check{\boldsymbol{f}}_{0t} = \hat{\boldsymbol{H}} \boldsymbol{f}_{0t} $ and $\check{\boldsymbol{\lambda}}_{0i} = (\hat{\bm{H}}^{-1})' \boldsymbol{\lambda}_{0i}$. Note that $ \check{\boldsymbol{\lambda}}_{0i}' \check{\boldsymbol{f}}_{0t} = \boldsymbol{\lambda}_{0i}'\boldsymbol{f}_{0t}=c_{0,it}$. Write $\hat{c}_{it} = \hat{\bm{\lambda}}_i' \hat{\bm{f}}_t$. Note that Lemma 4 also implies that 
		\[ 
		\max_{1\leq i\leq N} \| \hat{\boldsymbol{\lambda}}_i - \check{\boldsymbol{\lambda}}_{0i}\| =o_P(1).
		\]
		
		\vspace{0.5cm}	
		
		\begin{lemma}Under Assumptions 1 to 4, we have 
			\[  \| \hat{\boldsymbol{\beta}} -\boldsymbol{\beta}_0 \| = O_P((NT)^{-1/2})+O_P(N^{-1})+O_P(T^{-1}) +o_P\left( \frac{1}{N}\sum_{i=1}^{N} \| \hat{\boldsymbol{\lambda}}_i  -\check{\boldsymbol{\lambda}}_{0i}\|      \right). \]
		\end{lemma}
		
		\begin{proof}
			Expanding the first order conditions around $(\boldsymbol{\beta}_0, c_{0,it})$ gives:
			\begin{multline}\label{A3}
				0= \frac{1}{NT}\sum_{i=1}^{N}\sum_{t=1}^{T} \boldsymbol{x}_{it}\cdot l_{it}^{(1)}(\hat{\boldsymbol{\beta}}, \hat{c}_{it} ) = \frac{1}{NT}\sum_{i=1}^{N}\sum_{t=1}^{T}  l_{it}^{(1)}\boldsymbol{x}_{it}
				+ \left( \frac{1}{NT}\sum_{i=1}^{N}\sum_{t=1}^{T}  l_{it}^{(2)}\boldsymbol{x}_{it}\boldsymbol{x}_{it}' \right) (\hat{\boldsymbol{\beta}} -\boldsymbol{\beta}_0) \\
				+ 0.5(\hat{\boldsymbol{\beta}} -\boldsymbol{\beta}_0) '\left( \frac{1}{NT}\sum_{i=1}^{N}\sum_{t=1}^{T}  l_{it}^{(3)}(\ast) \cdot \boldsymbol{x}_{it}\boldsymbol{x}_{it}\boldsymbol{x}_{it}' \right) (\hat{\boldsymbol{\beta}} -\boldsymbol{\beta}_0)
				+  \left( \frac{1}{NT}\sum_{i=1}^{N}\sum_{t=1}^{T}  l_{it}^{(3)}(\ast) \cdot \boldsymbol{x}_{it}\boldsymbol{x}_{it}' (\hat{c}_{it} -c_{0,it})  \right) (\hat{\boldsymbol{\beta}} -\boldsymbol{\beta}_0) \\
				+\frac{1}{NT}\sum_{i=1}^{N}\sum_{t=1}^{T}  l_{it}^{(2)} \boldsymbol{x}_{it} \cdot(\hat{c}_{it} -c_{0,it}) + 0.5 \frac{1}{NT}\sum_{i=1}^{N}\sum_{t=1}^{T}  l_{it}^{(3)}(\ast) \cdot \boldsymbol{x}_{it} \cdot(\hat{c}_{it} -c_{0,it})^2,
			\end{multline}
			\begin{multline}\label{A4}
				0= \frac{1}{T}\sum_{t=1}^{T} \hat{\boldsymbol{f}}_t \cdot l_{it}^{(1)}(\hat{\boldsymbol{\beta}}, \hat{c}_{it} ) =  \frac{1}{T}\sum_{t=1}^{T}  l_{it}^{(1)}\hat{\boldsymbol{f}}_t + \left( \frac{1}{T}\sum_{t=1}^{T}  l_{it}^{(2)}\hat{\boldsymbol{f}}_t \boldsymbol{x}_{it}' \right) (\hat{\boldsymbol{\beta}} -\boldsymbol{\beta}_0)\\
				+0.5(\hat{\boldsymbol{\beta}} -\boldsymbol{\beta}_0) '\left( \frac{1}{T}\sum_{t=1}^{T}  l_{it}^{(3)}(\ast) \cdot \boldsymbol{x}_{it}\hat{\boldsymbol{f}}_t\boldsymbol{x}_{it}' \right) (\hat{\boldsymbol{\beta}} -\boldsymbol{\beta}_0)+\left( \frac{1}{T}\sum_{t=1}^{T}  l_{it}^{(3)}(\ast) \cdot \hat{\boldsymbol{f}}_t\boldsymbol{x}_{it}' (\hat{c}_{it} -c_{0,it})  \right) (\hat{\boldsymbol{\beta}} -\boldsymbol{\beta}_0) \\
				+\frac{1}{T}\sum_{t=1}^{T}  l_{it}^{(2)} \hat{\boldsymbol{f}}_{t} \cdot(\hat{c}_{it} -c_{0,it})+0.5 \frac{1}{T}\sum_{t=1}^{T}  l_{it}^{(3)}(\ast) \cdot \hat{\boldsymbol{f}}_{t} \cdot(\hat{c}_{it} -c_{0,it})^2,
			\end{multline}
			where $l_{it}^{(3)}(\ast)  =l_{it}^{(3)}(\boldsymbol{\beta}^{\ast},c_{it}^{\ast})$, and $(\boldsymbol{\beta}^{\ast},c_{it}^{\ast})$ is between $(\boldsymbol{\beta}_{0},c_{0,it})$ and $(\hat{\boldsymbol{\beta}},\hat{c}_{it})$. Given Assumption 4 and Lemma 4, it is easy to show that \footnote{ For a sequence of random variables $z_1,\ldots, z_N$, $ \max_{1\leq i\leq N} \|z_i\|= O_P(1)$ is written as $z_i =\bar{O}_P(1)$. The notation $\bar{o}_P(1)$ is defined similarly.} 
			\[  (\hat{\boldsymbol{\beta}} -\boldsymbol{\beta}_0) '\frac{1}{NT}\sum_{i=1}^{N}\sum_{t=1}^{T}  l_{it}^{(3)}(\ast) \cdot \boldsymbol{x}_{it}\boldsymbol{x}_{it}\boldsymbol{x}_{it}'  = o_P(1), \quad  \frac{1}{NT}\sum_{i=1}^{N}\sum_{t=1}^{T}  l_{it}^{(3)}(\ast) \cdot \boldsymbol{x}_{it}\boldsymbol{x}_{it}' (\hat{c}_{it} -c_{0,it}) =o_P(1), \]
			\[ (\hat{\boldsymbol{\beta}} -\boldsymbol{\beta}_0) '\frac{1}{T}\sum_{t=1}^{T}  l_{it}^{(3)}(\ast) \cdot \boldsymbol{x}_{it}\hat{\boldsymbol{f}}_t\boldsymbol{x}_{it}'  =\bar{o}_P(1), \quad  \frac{1}{T}\sum_{t=1}^{T}  l_{it}^{(3)}(\ast) \cdot \hat{\boldsymbol{f}}_t\boldsymbol{x}_{it}' (\hat{c}_{it} -c_{0,it}) = \bar{o}_P(1). \]
			Moreover, since $\hat{c}_{it} -c_{0,it} = (\hat{\boldsymbol{\lambda}}_i - \check{\boldsymbol{\lambda}}_{0i})'\hat{\boldsymbol{f}}_t + \check{\boldsymbol{\lambda}}_{0i}'(\hat{\boldsymbol{f}}_t - \check{\boldsymbol{f}}_{0t})$, equations \eqref{A3} and \eqref{A4} can be written as 
			\begin{multline}\label{A5}
				\left( \frac{1}{NT}\sum_{i=1}^{N}\sum_{t=1}^{T}  l_{it}^{(2)}\boldsymbol{x}_{it}\boldsymbol{x}_{it}' \right) (\hat{\boldsymbol{\beta}} -\boldsymbol{\beta}_0) =o_P(\| \hat{\boldsymbol{\beta}} -\boldsymbol{\beta}_0\|) - \frac{1}{NT}\sum_{i=1}^{N}\sum_{t=1}^{T}  l_{it}^{(1)}\boldsymbol{x}_{it}
				-  \frac{1}{N}\sum_{i=1}^{N}  \left( \frac{1}{T }\sum_{t=1}^{T}  l_{it}^{(2)} \boldsymbol{x}_{it}\hat{\boldsymbol{f}}_t' \right)  (\hat{\boldsymbol{\lambda}}_i - \check{\boldsymbol{\lambda}}_{0i}) \\ 
				-\frac{1}{T}\sum_{t=1}^{T} \left(\frac{1}{N}\sum_{i=1}^{N} l_{it}^{(2)} \boldsymbol{x}_{it}\check{\boldsymbol{\lambda}}_{0i}'\right)(\hat{\boldsymbol{f}}_t - \check{\boldsymbol{f}}_{0t}) 
				- 0.5 \frac{1}{NT}\sum_{i=1}^{N}\sum_{t=1}^{T}  l_{it}^{(3)}(\ast) \cdot \boldsymbol{x}_{it} \cdot   \hat{\boldsymbol{f}}_t' (\hat{\boldsymbol{\lambda}}_i - \check{\boldsymbol{\lambda}}_{0i}) (\hat{\boldsymbol{\lambda}}_i - \check{\boldsymbol{\lambda}}_{0i})'\hat{\boldsymbol{f}}_t  \\
				-  \frac{1}{NT}\sum_{i=1}^{N}\sum_{t=1}^{T}  l_{it}^{(3)}(\ast) \cdot \boldsymbol{x}_{it} \cdot    \check{\boldsymbol{\lambda}}_{0i}'(\hat{\boldsymbol{f}}_t - \check{\boldsymbol{f}}_{0t}) (\hat{\boldsymbol{\lambda}}_i - \check{\boldsymbol{\lambda}}_{0i})'\hat{\boldsymbol{f}}_t -
				0.5 \frac{1}{NT}\sum_{i=1}^{N}\sum_{t=1}^{T}  l_{it}^{(3)}(\ast) \cdot \boldsymbol{x}_{it} \cdot  \check{\boldsymbol{\lambda}}_{0i}'(\hat{\boldsymbol{f}}_t - \check{\boldsymbol{f}}_{0t}) (\hat{\boldsymbol{f}}_t - \check{\boldsymbol{f}}_{0t})'\check{\boldsymbol{\lambda}}_{0i},
			\end{multline}
			\begin{multline}\label{A6}
				\left( \frac{1}{T}\sum_{t=1}^{T}  l_{it}^{(2)} \hat{\boldsymbol{f}}_{t} \hat{\boldsymbol{f}}_t '  \right) (\hat{\boldsymbol{\lambda}}_i - \check{\boldsymbol{\lambda}}_{0i}) = \bar{o}_P(\| \hat{\boldsymbol{\beta}} -\boldsymbol{\beta}_0\|)- \frac{1}{T}\sum_{t=1}^{T}  l_{it}^{(1)}\hat{\boldsymbol{f}}_t - \left( \frac{1}{T}\sum_{t=1}^{T}  l_{it}^{(2)}\hat{\boldsymbol{f}}_t \boldsymbol{x}_{it}' \right) (\hat{\boldsymbol{\beta}} -\boldsymbol{\beta}_0)  \\
				-\left( \frac{1}{T}\sum_{t=1}^{T}  l_{it}^{(2)} \hat{\boldsymbol{f}}_{t} (\hat{\boldsymbol{f}}_t - \check{\boldsymbol{f}}_{0t})' \right) \check{\boldsymbol{\lambda}}_{0i}
				- 0.5 \frac{1}{T}\sum_{t=1}^{T}  l_{it}^{(3)}(\ast) \cdot \hat{\boldsymbol{f}}_t \cdot   \hat{\boldsymbol{f}}_t '(\hat{\boldsymbol{\lambda}}_i - \check{\boldsymbol{\lambda}}_{0i}) (\hat{\boldsymbol{\lambda}}_i - \check{\boldsymbol{\lambda}}_{0i})'\hat{\boldsymbol{f}}_t  \\
				-  \frac{1}{T}\sum_{t=1}^{T}  l_{it}^{(3)}(\ast) \hat{\boldsymbol{f}}_t \cdot   \check{\boldsymbol{\lambda}}_{0i}'(\hat{\boldsymbol{f}}_t - \check{\boldsymbol{f}}_{0t}) (\hat{\boldsymbol{\lambda}}_i - \check{\boldsymbol{\lambda}}_{0i})'\hat{\boldsymbol{f}}_t -
				0.5 \frac{1}{T}\sum_{t=1}^{T}  l_{it}^{(3)}(\ast) \hat{\boldsymbol{f}}_t \cdot  \check{\boldsymbol{\lambda}}_{0i}'(\hat{\boldsymbol{f}}_t - \check{\boldsymbol{f}}_{0t}) (\hat{\boldsymbol{f}}_t - \check{\boldsymbol{f}}_{0t})'\check{\boldsymbol{\lambda}}_{0i}.
			\end{multline}
			
			Define 
			\[  \check{\boldsymbol{A}}_i= \hat{\boldsymbol{H}}\boldsymbol{A}_i\hat{\boldsymbol{H}}'.\]
			It is easy to show that $T^{-1}\sum_{t=1}^{T}  l_{it}^{(2)} \hat{\boldsymbol{f}}_{t} \hat{\boldsymbol{f}}_t '  -  \check{\boldsymbol{A}}_i=\bar{o}_P(1)$ and $T^{-1}\sum_{t=1}^{T}  l_{it}^{(2)}\hat{\boldsymbol{f}}_t \boldsymbol{x}_{it}'  - \hat{\boldsymbol{H}} \boldsymbol{B}_i' =\bar{o}_P(1)$. 
			
			Thus, from \eqref{A6} we can show that 
			\begin{multline}\label{A7}
				\hat{\boldsymbol{\lambda}}_i - \check{\boldsymbol{\lambda}}_{0i} = \bar{o}_P(\| \hat{\boldsymbol{\beta}} -\boldsymbol{\beta}_0\|) 
				-  \check{\boldsymbol{A}}_i^{-1}\left( \frac{1}{T}\sum_{t=1}^{T}  l_{it}^{(2)} \hat{\boldsymbol{f}}_{t} \hat{\boldsymbol{f}}_t ' -  \check{\boldsymbol{A}}_i\right) (\hat{\boldsymbol{\lambda}}_i - \check{\boldsymbol{\lambda}}_{0i} ) 
				-  (\hat{\boldsymbol{H}}')^{-1}  \boldsymbol{A}^{-1}_i\cdot \frac{1}{T}\sum_{t=1}^{T}  l_{it}^{(1)}\boldsymbol{f}_{0t}\\ 
				-  \check{\boldsymbol{A}}_i^{-1}\cdot \frac{1}{T}\sum_{t=1}^{T}  l_{it}^{(1)}( \hat{\boldsymbol{f}}_t - \check{\boldsymbol{f}}_{0t}) -  (\hat{\boldsymbol{H}}')^{-1}  \boldsymbol{A}^{-1}_i \boldsymbol{B}_i'  (\hat{\boldsymbol{\beta}} -\boldsymbol{\beta}_0) -  \check{\boldsymbol{A}}_i^{-1}\cdot\left( \frac{1}{T}\sum_{t=1}^{T}  l_{it}^{(2)} \hat{\boldsymbol{f}}_t  (\hat{\boldsymbol{f}}_t - \check{\boldsymbol{f}}_{0t})' \right) \check{\boldsymbol{\lambda}}_{0i} \\
				- 0.5 \frac{1}{T}\sum_{t=1}^{T}  l_{it}^{(3)}(\ast) \cdot  \check{\boldsymbol{A}}_i^{-1}\hat{\boldsymbol{f}}_t \cdot   \hat{\boldsymbol{f}}_t '(\hat{\boldsymbol{\lambda}}_i - \check{\boldsymbol{\lambda}}_{0i}) (\hat{\boldsymbol{\lambda}}_i - \check{\boldsymbol{\lambda}}_{0i})'\hat{\boldsymbol{f}}_t  
				-  \frac{1}{T}\sum_{t=1}^{T}  l_{it}^{(3)}(\ast)  \check{\boldsymbol{A}}_i^{-1}\hat{\boldsymbol{f}}_t \cdot   \check{\boldsymbol{\lambda}}_{0i}'(\hat{\boldsymbol{f}}_t - \check{\boldsymbol{f}}_{0t}) (\hat{\boldsymbol{\lambda}}_i - \check{\boldsymbol{\lambda}}_{0i})'\hat{\boldsymbol{f}}_t  \\
				-0.5 \frac{1}{T}\sum_{t=1}^{T}  l_{it}^{(3)}(\ast)  \check{\boldsymbol{A}}_i^{-1}\hat{\boldsymbol{f}}_t \cdot  \check{\boldsymbol{\lambda}}_{0i}'(\hat{\boldsymbol{f}}_t - \check{\boldsymbol{f}}_{0t}) (\hat{\boldsymbol{f}}_t - \check{\boldsymbol{f}}_{0t})'\check{\boldsymbol{\lambda}}_{0i}.
			\end{multline}
			Plugging \eqref{A7} into \eqref{A5} gives 
			\begin{multline}\label{A8}
				\left( \frac{1}{NT}\sum_{i=1}^{N}\sum_{t=1}^{T}  l_{it}^{(2)}\boldsymbol{x}_{it}\boldsymbol{x}_{it}'-\frac{1}{N}\sum_{i=1}^{N} \boldsymbol{B}_i\boldsymbol{A}^{-1}_i\boldsymbol{B}_i'  \right) (\hat{\boldsymbol{\beta}} -\boldsymbol{\beta}_0) +o_P(\| \hat{\boldsymbol{\beta}} -\boldsymbol{\beta}_0\|) =- \frac{1}{NT}\sum_{i=1}^{N}\sum_{t=1}^{T}  l_{it}^{(1)}\dot{\boldsymbol{x}}_{it}\\ 
				+  \frac{1}{T}\sum_{t=1}^{T}  \left(      \frac{1}{N}\sum_{i=1}^{N}l_{it}^{(1)} \boldsymbol{B}_i\boldsymbol{A}^{-1}_i \right)\hat{\boldsymbol{H}}^{-1}(\hat{\boldsymbol{f}}_t -\check{\boldsymbol{f}}_{0t})  \\
				+\frac{1}{NT}\sum_{i=1}^{N}\sum_{t=1}^{T}  (\hat{\boldsymbol{f}}_t - \check{\boldsymbol{f}}_{0t})'  \check{\boldsymbol{\lambda}}_{0i}   \boldsymbol{B}_i \boldsymbol{A}^{-1}_i \hat{\boldsymbol{H}}^{-1}  l_{it}^{(2)}  \hat{\boldsymbol{f}}_t   
				-\frac{1}{T}\sum_{t=1}^{T} \left(\frac{1}{N}\sum_{i=1}^{N} l_{it}^{(2)} \boldsymbol{x}_{it}\check{\boldsymbol{\lambda}}_{0i}'\right)(\hat{\boldsymbol{f}}_t - \check{\boldsymbol{f}}_{0t}) \\
				-0.5 \frac{1}{NT}\sum_{i=1}^{N}\sum_{t=1}^{T}  l_{it}^{(3)}(\ast) \cdot(\boldsymbol{x}_{it} -\boldsymbol{B}_i \boldsymbol{A}^{-1}_i \hat{\boldsymbol{H}}^{-1}\hat{\boldsymbol{f}}_t) \cdot  \check{\boldsymbol{\lambda}}_{0i}'(\hat{\boldsymbol{f}}_t - \check{\boldsymbol{f}}_{0t}) (\hat{\boldsymbol{f}}_t - \check{\boldsymbol{f}}_{0t})'\check{\boldsymbol{\lambda}}_{0i}.\\
				- 0.5 \frac{1}{NT}\sum_{i=1}^{N}\sum_{t=1}^{T}  l_{it}^{(3)}(\ast) \cdot (\boldsymbol{x}_{it} -\boldsymbol{B}_i \boldsymbol{A}^{-1}_i \hat{\boldsymbol{H}}^{-1}\hat{\boldsymbol{f}}_t) \cdot   \hat{\boldsymbol{f}}_t' (\hat{\boldsymbol{\lambda}}_i - \check{\boldsymbol{\lambda}}_{0i}) (\hat{\boldsymbol{\lambda}}_i - \check{\boldsymbol{\lambda}}_{0i})'\hat{\boldsymbol{f}}_t  \\
				-  \frac{1}{NT}\sum_{i=1}^{N}\sum_{t=1}^{T}  l_{it}^{(3)}(\ast) \cdot (\boldsymbol{x}_{it} -\boldsymbol{B}_i \boldsymbol{A}^{-1}_i \hat{\boldsymbol{H}}^{-1}\hat{\boldsymbol{f}}_t) \cdot    \check{\boldsymbol{\lambda}}_{0i}'(\hat{\boldsymbol{f}}_t - \check{\boldsymbol{f}}_{0t}) (\hat{\boldsymbol{\lambda}}_i - \check{\boldsymbol{\lambda}}_{0i})'\hat{\boldsymbol{f}}_t \\
				+\frac{1}{N}\sum_{i=1}^{N} \boldsymbol{B}_i \boldsymbol{A}^{-1}_i\hat{\boldsymbol{H}}^{-1}\left( \frac{1}{T}\sum_{t=1}^{T}  l_{it}^{(2)} \hat{\boldsymbol{f}}_{t} \hat{\boldsymbol{f}}_t ' -  \check{\boldsymbol{A}}_i\right) (\hat{\boldsymbol{\lambda}}_i - \check{\boldsymbol{\lambda}}_{0i} ) \\
				-  \frac{1}{N}\sum_{i=1}^{N}  \left( \frac{1}{T }\sum_{t=1}^{T}  l_{it}^{(2)} \boldsymbol{x}_{it}\hat{\boldsymbol{f}}_t'  - \boldsymbol{B}_i \hat{\boldsymbol{H}}'\right)  (\hat{\boldsymbol{\lambda}}_i - \check{\boldsymbol{\lambda}}_{0i}).
			\end{multline}
			
			\noindent \textbf{Step 1:}
			It can be shown that under Assumption 4, 
			\begin{multline}\label{A9}
				\frac{1}{NT}\sum_{i=1}^{N}\sum_{t=1}^{T}  l_{it}^{(2)}\boldsymbol{x}_{it}\boldsymbol{x}_{it}'-\frac{1}{N}\sum_{i=1}^{N} \boldsymbol{B}_i\boldsymbol{A}^{-1}_i\boldsymbol{B}_i'  \overset{p}{\rightarrow}
				\frac{1}{NT}\sum_{i=1}^{N}\sum_{t=1}^{T}  \Ex\left[   l_{it}^{(2)}\boldsymbol{x}_{it}\boldsymbol{x}_{it}' -\boldsymbol{B}_i\boldsymbol{A}^{-1}_i\boldsymbol{B}_i'   \right] \\ 
				= \frac{1}{NT}\sum_{i=1}^{N}\sum_{t=1}^{T} \Ex\left[  l_{it}^{(2)}\dot{\boldsymbol{x}}_{it}\dot{\boldsymbol{x}}_{it}' \right] \rightarrow \boldsymbol{\Delta} . 
			\end{multline}

			\noindent \textbf{Step 2:} By Assumption 4 and Lemma 1 it is easy to show that
			\begin{equation}\label{A10} \frac{1}{NT}\sum_{i=1}^{N}\sum_{t=1}^{T}  l_{it}^{(1)}\dot{\boldsymbol{x}}_{it} =O_P\left( \frac{1}{\sqrt{NT}}\right).\end{equation}

			\noindent \textbf{Step 3:} The $j$th element of the second term on the right-hand side of \eqref{A8} is 
			\[  \tr \left[   \frac{1}{T}\sum_{t=1}^{T} (\hat{\boldsymbol{f}}_t -\check{\boldsymbol{f}}_{0t}) \left(       \frac{1}{N}\sum_{i=1}^{N}l_{it}^{(1)} \boldsymbol{B}_{i,j}\boldsymbol{A}^{-1}_i \right) \cdot \hat{\boldsymbol{H}}^{-1}      \right]\]
			where $\boldsymbol{B}_{i,j}$ is the $j$th row of $\boldsymbol{B}_i$. Since $\hat{\boldsymbol{f}}_t -\check{\boldsymbol{f}}_{0t} =\hat{\boldsymbol{\Psi}}' N^{-1} \sum_{i=1}^{N} \boldsymbol{e}_{it} $, it follows that
			\[  \frac{1}{T}\sum_{t=1}^{T} (\hat{\boldsymbol{f}}_t -\check{\boldsymbol{f}}_{0t}) \left(       \frac{1}{N}\sum_{i=1}^{N}l_{it}^{(1)} \boldsymbol{B}_{i,j}\boldsymbol{A}^{-1}_i \right)
			= \hat{\boldsymbol{\Psi}}' \cdot\frac{1}{N} \cdot  \frac{1}{T}\sum_{t=1}^{T} \left( \frac{1}{\sqrt{N}}\sum_{i=1}^{N} \boldsymbol{e}_{it}\right) \left(       \frac{1}{\sqrt{N}}\sum_{i=1}^{N}l_{it}^{(1)} \boldsymbol{B}_{i,j}\boldsymbol{A}^{-1}_i \right).
			\]
			Note that
			\[ \Ex\left[  \left( \frac{1}{\sqrt{N}}\sum_{i=1}^{N} \boldsymbol{e}_{it}\right) \left(       \frac{1}{\sqrt{N}}\sum_{i=1}^{N}l_{it}^{(1)} \boldsymbol{B}_{i,j}\boldsymbol{A}^{-1}_i \right)    \right]
			= \frac{1}{N}\sum_{i=1}^{N} \Ex\left[l_{it}^{(1)}\boldsymbol{e}_{it}\right] \boldsymbol{B}_{i,j}\boldsymbol{A}^{-1}_i  =0, 
			\]
			and for $m$th element of $\boldsymbol{e}_{it}$ (denoted by $e_{it,m}$) and $p$th column of $\boldsymbol{A}^{-1}_i$ (denoted by $\boldsymbol{A}^{-1}_{i,p}$), we have 
			\begin{align*}
				& \text{Var} \left[\frac{1}{T}\sum_{t=1}^{T} \left( \frac{1}{\sqrt{N}}\sum_{i=1}^{N} e_{it,m}\right) \left(       \frac{1}{\sqrt{N}}\sum_{i=1}^{N}l_{it}^{(1)} \boldsymbol{B}_{i,j}\boldsymbol{A}_{i,p}^{-1} \right) \right] \\ =&\frac{1}{N^2T^2}\Ex\left[\left(\sum_{t=1}^{T}  \sum_{i=1}^{N} \sum_{q=1}^{N}l_{it}^{(1)}e_{qt,m} \boldsymbol{B}_{i,j}\boldsymbol{A}_{i,p}^{-1}\right)^2\right]  \\
				=&\frac{1}{N^2T^2}\sum_{t_1=1}^T\sum_{t_2=1}^T\sum_{i_1=1}^N\sum_{i_2=1}^N\sum_{q_1=1}^N\sum_{q_2=1}^N\Ex\left[l_{i_{1}t_{1}}^{(1)}l_{i_{2}t_{2}}^{(1)}e_{q_{1}t_{1},m}e_{q_{2}t_{2},m}\right] \boldsymbol{B}_{i_{1},j}\boldsymbol{A}_{i_{1},p}^{-1}\boldsymbol{B}_{i_{2},j}\boldsymbol{A}_{i_{2},p}^{-1} \\
				=& \frac{1}{N^2T^2}\sum_{i=1}^{N}\sum_{q=1}^{N} \Ex\left[\left(\sum_{t=1}^Tl_{it}^{(1)}e_{qt,m}\right)^2\right]\left(\boldsymbol{B}_{i,j}\boldsymbol{A}_{i,p}^{-1}\right)^2 \\
				&+ \frac{1}{N^2T^2}\sum_{i=1}^{N}\sum_{q\neq i}^{N}\sum_{t_1=1}^T\sum_{t_2=1}^T \text{Cov}\left(l_{it_1}^{(1)}e_{qt_1,m},l_{qt_2}^{(1)}e_{it_2,m}\right)\boldsymbol{B}_{i,j}\boldsymbol{A}_{i,p}^{-1}\boldsymbol{B}_{q,j}\boldsymbol{A}_{q,p}^{-1}\\
				=&O(T^{-1})=o(1)
			\end{align*}
			by Lemma 1 and Lemma 2. Thus, it can be concluded that 
			\begin{equation}\label{A11} \frac{1}{T}\sum_{t=1}^{T}  \left(      \frac{1}{N}\sum_{i=1}^{N}l_{it}^{(1)} \boldsymbol{B}_i\boldsymbol{A}^{-1}_i \right)\hat{\boldsymbol{H}}^{-1}(\hat{\boldsymbol{f}}_t -\check{\boldsymbol{f}}_{0t}) = o_P(N^{-1}).
			\end{equation}

			\noindent \textbf{Step 4:} Define 
			\[\boldsymbol{C}_t^{\ast} = \frac{1}{N}\sum_{i=1}^{N} \Ex[ l_{it}^{(2)}\boldsymbol{x}_{it} ] \boldsymbol{\lambda}_{0i}',   \]
			then the fourth term on the right-hand side of \eqref{A8} can be written as
			\[ -\frac{1}{T}\sum_{t=1}^{T}\boldsymbol{C}_t^{\ast} \hat{\boldsymbol{H}}^{-1} (\hat{\boldsymbol{f}}_t - \check{\boldsymbol{f}}_{0t}) -\frac{1}{T}\sum_{t=1}^{T} \left(\frac{1}{N}\sum_{i=1}^{N} l_{it}^{(2)} \boldsymbol{x}_{it}\boldsymbol{\lambda}_{0i}' -\boldsymbol{C}_t^{\ast} \right)\hat{\boldsymbol{H}}^{-1}(\hat{\boldsymbol{f}}_t - \check{\boldsymbol{f}}_{0t}). \]
			First, 
			\[\frac{1}{T}\sum_{t=1}^{T}\boldsymbol{C}_t^{\ast} \hat{\boldsymbol{H}}^{-1} (\hat{\boldsymbol{f}}_t - \check{\boldsymbol{f}}_{0t}) =\frac{1}{NT}\sum_{i=1}^{N}\sum_{t=1}^{T}\boldsymbol{C}_t^{\ast} \hat{\boldsymbol{H}}^{-1} \hat{\boldsymbol{\Psi}}' \boldsymbol{e}_{it} = \frac{1}{NT}\sum_{i=1}^{N}\sum_{t=1}^{T}\boldsymbol{C}_t^{\ast} \boldsymbol{H}_0^{-1} \boldsymbol{\Psi}_0' \boldsymbol{e}_{it} +o_P(1/\sqrt{NT}). \]
			Second, similar to the proof of step 3, it can be shown that 
			\begin{align*} 
				&\frac{1}{T}\sum_{t=1}^{T} \left(\frac{1}{N}\sum_{i=1}^{N} l_{it}^{(2)} \boldsymbol{x}_{it}\boldsymbol{\lambda}_{0i}' -\boldsymbol{C}_t^{\ast} \right)\hat{\boldsymbol{H}}^{-1}(\hat{\boldsymbol{f}}_t - \check{\boldsymbol{f}}_{0t})\\
				=& \frac{1}{N}\cdot \frac{1}{T}\sum_{t=1}^{T}\left(\frac{1}{\sqrt{N}}\sum_{i=1}^{N} \left[  l_{it}^{(2)} \boldsymbol{x}_{it} - \Ex[l_{it}^{(2)} \boldsymbol{x}_{it} ] \right]\boldsymbol{\lambda}_{0i}'\right)\hat{\boldsymbol{H}}^{-1} \hat{\boldsymbol{\Psi}}' \left(\frac{1}{\sqrt{N}}\sum_{i=1}^{N} \boldsymbol{e}_{it}\right) \\
				=& \frac{1}{N}\cdot \frac{1}{NT}\sum_{i=1}^{N}\sum_{t=1}^{T} \Ex\left[ l_{it}^{(2)}\boldsymbol{x}_{it}\boldsymbol{e}_{it} ' \right] \boldsymbol{\Psi}_0(\boldsymbol{H}_0^{-1})'  \boldsymbol{\lambda}_{0i}   +o_P(N^{-1}).
			\end{align*}
			
			Thus,
			\begin{multline}
				-\frac{1}{T}\sum_{t=1}^{T} \left(\frac{1}{N}\sum_{i=1}^{N} l_{it}^{(2)} \boldsymbol{x}_{it}\check{\boldsymbol{\lambda}}_{0i}'\right)(\hat{\boldsymbol{f}}_t - \check{\boldsymbol{f}}_{0t})  = -\frac{1}{NT}\sum_{i=1}^{N}\sum_{t=1}^{T}\boldsymbol{C}_t^{\ast} \boldsymbol{H}_0^{-1} \boldsymbol{\Psi}_0' \boldsymbol{e}_{it} \\
				-\frac{1}{N}\cdot \frac{1}{NT}\sum_{i=1}^{N}\sum_{t=1}^{T} \Ex\left[ l_{it}^{(2)}\boldsymbol{x}_{it}\boldsymbol{e}_{it} ' \right] \boldsymbol{\Psi}_0(\boldsymbol{H}_0^{-1})'  \boldsymbol{\lambda}_{0i}  +o_P(T^{-1})=O_P((NT)^{-1/2})+O_P(N^{-1}).
			\end{multline}
			
			\noindent \textbf{Step 5:} For the third term on the right-hand side of \eqref{A8}, its $j$th element can be written as
			\[ \tr\left[ \hat{\boldsymbol{H}}^{-1}  \cdot \frac{1}{T} \sum_{t=1}^{T} \hat{\boldsymbol{f}}_t   (\hat{\boldsymbol{f}}_t - \check{\boldsymbol{f}}_{0t})' (\hat{\boldsymbol{H}}^{-1})'  \left(  \frac{1}{N} \sum_{i=1}^{N}\boldsymbol{\lambda}_{0i}\boldsymbol{B}_{i,j} \boldsymbol{A}^{-1}_i   l_{it}^{(2)} \right)  \right]\]
			Note that 
			\begin{align*}
				& \hat{\boldsymbol{H}}^{-1}  \cdot \frac{1}{T} \sum_{t=1}^{T} \hat{\boldsymbol{f}}_t   (\hat{\boldsymbol{f}}_t - \check{\boldsymbol{f}}_{0t})' (\hat{\boldsymbol{H}}^{-1})'  \left(  \frac{1}{N} \sum_{i=1}^{N}\boldsymbol{\lambda}_{0i}\boldsymbol{B}_{i,j} \boldsymbol{A}^{-1}_i   l_{it}^{(2)} \right) \\
				= &  \frac{1}{T} \sum_{t=1}^{T} \boldsymbol{f}_{0t}   (\hat{\boldsymbol{f}}_t - \check{\boldsymbol{f}}_{0t})' (\hat{\boldsymbol{H}}^{-1})'  \left(  \frac{1}{N} \sum_{i=1}^{N}\boldsymbol{\lambda}_{0i}\boldsymbol{B}_{i,j} \boldsymbol{A}^{-1}_i   l_{it}^{(2)} \right) \\
				&+\hat{\boldsymbol{H}}^{-1}  \cdot \frac{1}{T} \sum_{t=1}^{T}(\hat{\boldsymbol{f}}_t - \check{\boldsymbol{f}}_{0t})  (\hat{\boldsymbol{f}}_t - \check{\boldsymbol{f}}_{0t})' (\hat{\boldsymbol{H}}^{-1})'  \left(  \frac{1}{N} \sum_{i=1}^{N}\boldsymbol{\lambda}_{0i}\boldsymbol{B}_{i,j} \boldsymbol{A}^{-1}_i   l_{it}^{(2)} \right) .
			\end{align*}
			
			First, it can be shown that 
			\begin{align*}
				&\frac{1}{T} \sum_{t=1}^{T} \boldsymbol{f}_{0t}   (\hat{\boldsymbol{f}}_t - \check{\boldsymbol{f}}_{0t})' (\hat{\boldsymbol{H}}^{-1})'  \left(  \frac{1}{N} \sum_{i=1}^{N}\boldsymbol{\lambda}_{0i}\boldsymbol{B}_{i,j} \boldsymbol{A}^{-1}_i   l_{it}^{(2)} \right)  \\
				= &  \frac{1}{NT} \sum_{i=1}^{N}\sum_{t=1}^{T} \boldsymbol{f}_{0t}  \boldsymbol{e}_{it}'   (\hat{\boldsymbol{H}}^{-1}\hat{\boldsymbol{\Psi}}')'  \boldsymbol{D}_{t,j} + \frac{1}{N} \cdot \frac{1}{T} \sum_{t=1}^{T} \boldsymbol{f}_{0t}   \left(\frac{1}{\sqrt{N}}\sum_{i=1}^{N}\boldsymbol{e}_{it}' \right) (\hat{\boldsymbol{H}}^{-1}\hat{\boldsymbol{\Psi}}')'  \left(  \frac{1}{\sqrt{N}} \sum_{i=1}^{N}\boldsymbol{\lambda}_{0i}\boldsymbol{B}_{i,j} \boldsymbol{A}^{-1}_i   \left( l_{it}^{(2)} - \bar{l}_{it}^{(2)} \right) \right) \\
				=  &\frac{1}{NT} \sum_{i=1}^{N}\sum_{t=1}^{T} \boldsymbol{f}_{0t}  \boldsymbol{e}_{it}'   (\boldsymbol{H}_0^{-1}\boldsymbol{\Psi}_0')'  \boldsymbol{D}_{t,j}   + \frac{1}{N}  \cdot \frac{1}{NT}\sum_{i=1}^{N}\sum_{t=1}^{T} \boldsymbol{f}_{0t}\cdot \Ex\left[  l_{it}^{(2)}\boldsymbol{e}_{it}' \right](\boldsymbol{H}_0^{-1}\boldsymbol{\Psi}_0')' \boldsymbol{\lambda}_{0i}\boldsymbol{B}_{i,j} \boldsymbol{A}^{-1}_i   +o_P(N^{-1}).
			\end{align*}

			Second,
			\begin{align*}
				&\hat{\boldsymbol{H}}^{-1}  \cdot \frac{1}{T} \sum_{t=1}^{T}(\hat{\boldsymbol{f}}_t - \check{\boldsymbol{f}}_{0t})  (\hat{\boldsymbol{f}}_t - \check{\boldsymbol{f}}_{0t})' (\hat{\boldsymbol{H}}^{-1})'  \left(  \frac{1}{N} \sum_{i=1}^{N}\boldsymbol{\lambda}_{0i}\boldsymbol{B}_{i,j} \boldsymbol{A}^{-1}_i   l_{it}^{(2)} \right) \\ 
				= &  (\hat{\boldsymbol{H}}^{-1}\hat{\boldsymbol{\Psi}}') \cdot \frac{1}{N}  \cdot \frac{1}{T}\sum_{t=1}^{T} \left(\frac{1}{\sqrt{N}}\sum_{i=1}^{N}\boldsymbol{e}_{it} \right)\left(\frac{1}{\sqrt{N}}\sum_{i=1}^{N}\boldsymbol{e}_{it}' \right)(\hat{\boldsymbol{H}}^{-1}\hat{\boldsymbol{\Psi}}')' \boldsymbol{D}_{t,j}\\
				& +O_P(T^{1/2p} N^{-1/2})\cdot O_P\left( \frac{1}{T}\sum_{t=1}^{T}\| \hat{\boldsymbol{f}}_t - \check{\boldsymbol{f}}_{0t}\|^2 \right) \\
				=&(\boldsymbol{H}_0^{-1}\boldsymbol{\Psi}_0') \cdot \frac{1}{N}  \cdot   \frac{1}{NT}\sum_{i=1}^{N}\sum_{t=1}^{T}  \Ex[\boldsymbol{e}_{it} \boldsymbol{e}_{it}'](\boldsymbol{H}_0^{-1}\boldsymbol{\Psi}_0')' \boldsymbol{D}_{t,j} + o_P(N^{-1})
			\end{align*}
			since
				\[\frac{1}{T}\sum_{t=1}^{T}\| \hat{\boldsymbol{f}}_t - \check{\boldsymbol{f}}_{0t}\|^2 = O_P(1)\cdot\frac{1}{N}  \cdot \frac{1}{T}\sum_{t=1}^{T} \left\|\frac{1}{\sqrt{N}}\sum_{i=1}^{N}\boldsymbol{e}_{it} \right\|^2 = O_P(N^{-1}).\]

			Thus, the $j$th element of 
			\[    \frac{1}{NT}\sum_{i=1}^{N}\sum_{t=1}^{T}  (\hat{\boldsymbol{f}}_t - \check{\boldsymbol{f}}_{0t})'  \check{\boldsymbol{\lambda}}_{0i}   \boldsymbol{B}_i \boldsymbol{A}^{-1}_i \hat{\boldsymbol{H}}^{-1}  l_{it}^{(2)}  \hat{\boldsymbol{f}}_t  \] 
			is given by 
			\begin{multline}
				\frac{1}{NT} \sum_{i=1}^{N}\sum_{t=1}^{T} \boldsymbol{f}_{0t}'\boldsymbol{D}_{t,j}' (\boldsymbol{H}_0^{-1}\boldsymbol{\Psi}_0')  \boldsymbol{e}_{it}     + \frac{1}{N}  \cdot \frac{1}{NT}\sum_{i=1}^{N}\sum_{t=1}^{T}\boldsymbol{B}_{i,j} \boldsymbol{A}^{-1}_i\boldsymbol{f}_{0t} \cdot \boldsymbol{\lambda}_{0i} '(\boldsymbol{H}_0^{-1}\boldsymbol{\Psi}_0') \cdot \Ex\left[  l_{it}^{(2)}\boldsymbol{e}_{it} \right] \\
				+ \tr\left[ \frac{1}{N}  \cdot   \frac{1}{NT}\sum_{i=1}^{N}\sum_{t=1}^{T}  \Ex[\boldsymbol{e}_{it} \boldsymbol{e}_{it}'] \cdot (\boldsymbol{H}_0^{-1}\boldsymbol{\Psi}_0')' \cdot \boldsymbol{D}_{t,j} \cdot (\boldsymbol{H}_0^{-1}\boldsymbol{\Psi}_0') \right] +o_P(N^{-1}).
			\end{multline}
			
			Therefore, we have
			\[
			\frac{1}{NT}\sum_{i=1}^{N}\sum_{t=1}^{T}  (\hat{\boldsymbol{f}}_t - \check{\boldsymbol{f}}_{0t})'  \check{\boldsymbol{\lambda}}_{0i}   \boldsymbol{B}_i \boldsymbol{A}^{-1}_i \hat{\boldsymbol{H}}^{-1}  l_{it}^{(2)}  \hat{\boldsymbol{f}}_t 
			=O_P((NT)^{-1/2})+O_P(N^{-1}).
			\]		
			
			\noindent \textbf{Step 6:} The fifth term on the right-hand side of \eqref{A8} is equal to 
			\begin{multline*}
				-0.5 \frac{1}{NT}\sum_{i=1}^{N}\sum_{t=1}^{T}  l_{it}^{(3)} \dot{\boldsymbol{x}}_{it} \cdot  \check{\boldsymbol{\lambda}}_{0i}'(\hat{\boldsymbol{f}}_t - \check{\boldsymbol{f}}_{0t}) (\hat{\boldsymbol{f}}_t - \check{\boldsymbol{f}}_{0t})'\check{\boldsymbol{\lambda}}_{0i} + o_P(\|\hat{\boldsymbol{\beta}} -\boldsymbol{\beta}_0\|) \\
				+ O_P(1)\cdot\left(\max_{1\leq i\leq N} \| \hat{\boldsymbol{\lambda}}_i -\check{\boldsymbol{\lambda}}_{0i}\| + \max_{1\leq t\leq T} \| \hat{\boldsymbol{f}}_t -\check{\boldsymbol{f}}_{0t}\|\right)\cdot \left( \frac{1}{NT}\sum_{i=1}^{N}\sum_{t=1}^{T}M(\boldsymbol{x}_{it})\| \hat{\boldsymbol{f}}_t -\check{\boldsymbol{f}}_{0t}\|^2\right)
				.\end{multline*}
			
			First, similar to the proof of step 3, it can be shown that:
			\begin{align*}
				&\frac{1}{NT}\sum_{i=1}^{N}\sum_{t=1}^{T}  l_{it}^{(3)} \dot{\boldsymbol{x}}_{it,j} \cdot  \check{\boldsymbol{\lambda}}_{0i}'(\hat{\boldsymbol{f}}_t - \check{\boldsymbol{f}}_{0t}) (\hat{\boldsymbol{f}}_t - \check{\boldsymbol{f}}_{0t})'\check{\boldsymbol{\lambda}}_{0i}\\ 
				= &   \frac{1}{N}\cdot \frac{1}{T}\sum_{t=1}^{T}  \left( \frac{1}{\sqrt{N} }\sum_{i=1}^{N}\boldsymbol{e}_{it}  \right)' (\hat{\boldsymbol{H}}^{-1} \hat{\boldsymbol{\Psi}}')' \left( \frac{1}{N}\sum_{i=1}^{N} l_{it}^{(3)} \dot{\boldsymbol{x}}_{it,j} \boldsymbol{\lambda}_{0i} \boldsymbol{\lambda}_{0i}' \right)(\hat{\boldsymbol{H}}^{-1} \hat{\boldsymbol{\Psi}}')\left( \frac{1}{\sqrt{N} }\sum_{i=1}^{N}\boldsymbol{e}_{it}  \right) \\
				=& \frac{1}{N}\cdot \frac{1}{T}\sum_{t=1}^{T}  \left( \frac{1}{\sqrt{N} }\sum_{i=1}^{N}\boldsymbol{e}_{it}  \right)' (\boldsymbol{H}_0^{-1} \boldsymbol{\Psi}_0')' \boldsymbol{G}_{t,j} (\boldsymbol{H}_0^{-1} \boldsymbol{\Psi}_0')\left( \frac{1}{\sqrt{N} }\sum_{i=1}^{N}\boldsymbol{e}_{it}  \right)  +o_P(N^{-1}) \\
				=&\frac{1}{N}\cdot \frac{1}{NT}\sum_{i=1}^{N}\sum_{t=1}^{T}  \tr\left[ (\boldsymbol{H}_0^{-1} \boldsymbol{\Psi}_0')' \cdot  \boldsymbol{G}_{t,j} \cdot (\boldsymbol{H}_0^{-1} \boldsymbol{\Psi}_0') \cdot \Ex \left[ \boldsymbol{e}_{it}\boldsymbol{e}_{it}'\right] \right]+o_P(N^{-1}).
			\end{align*}

			Second, we have
				\[\frac{1}{NT}\sum_{i=1}^{N}\sum_{t=1}^{T}M(\boldsymbol{x}_{it})\| \hat{\boldsymbol{f}}_t -\check{\boldsymbol{f}}_{0t}\|^2\leq \left( \max_{1\leq t \leq T}\frac{1}{N}\sum_{i=1}^{N}M(\boldsymbol{x}_{it})  \right)\cdot \left( \frac{1}{T}\sum_{t=1}^{T}\| \hat{\boldsymbol{f}}_t -\check{\boldsymbol{f}}_{0t}\|^2  \right) = O_P(N^{-1}),\]
				where the equality follows from $\max_{1\leq t \leq T}N^{-1}\sum_{i=1}^{N}M(\boldsymbol{x}_{it}) = O_P(1)$ (similar to the proof of Lemma 4) and $T^{-1}\sum_{t=1}^{T}\| \hat{\boldsymbol{f}}_t -\check{\boldsymbol{f}}_{0t}\|^2 = O_P(N^{-1})$ (see the proof of step 5). That is,
				\[\left(\max_{1\leq i\leq N} \| \hat{\boldsymbol{\lambda}}_i -\check{\boldsymbol{\lambda}}_{0i}\| + \max_{1\leq t\leq T} \| \hat{\boldsymbol{f}}_t -\check{\boldsymbol{f}}_{0t}\|\right)\cdot \left( \frac{1}{NT}\sum_{i=1}^{N}\sum_{t=1}^{T}M(\boldsymbol{x}_{it})\| \hat{\boldsymbol{f}}_t -\check{\boldsymbol{f}}_{0t}\|^2\right)=o_P(N^{-1}).\]
			
			Therefore, the $j$th element of the fifth term on the right-hand side of \eqref{A8} is
			\begin{multline}\label{A14}
				-0.5\frac{1}{N}\cdot \frac{1}{NT}\sum_{i=1}^{N}\sum_{t=1}^{T}  \tr\left[ (\boldsymbol{H}_0^{-1} \boldsymbol{\Psi}_0')' \cdot  \boldsymbol{G}_{t,j} \cdot (\boldsymbol{H}_0^{-1} \boldsymbol{\Psi}_0') \cdot \Ex \left[ \boldsymbol{e}_{it}\boldsymbol{e}_{it}'\right] \right]+o_P(N^{-1})+o_P(\|\hat{\boldsymbol{\beta}} -\boldsymbol{\beta}_0\|).
			\end{multline}
			
			\noindent \textbf{Step 7:} It is easy to show that the last four terms on the right-hand side of \eqref{A8} are all $o_P\left( N^{-1} \sum_{i=1}^{N}\| \hat{\boldsymbol{\lambda}}_i -\check{\boldsymbol{\lambda}}_{0i}\| \right)$. Combining all the above results, we have 
			\begin{equation*}
				\boldsymbol{\Delta} (\hat{\boldsymbol{\beta}} -\boldsymbol{\beta}_0) +o_P(\| \hat{\boldsymbol{\beta}} -\boldsymbol{\beta}_0\|) =O_P((NT)^{-1/2})+  O_P(T^{-1}) +O_P(N^{-1})+ o_P\left( \frac{1}{N} \sum_{i=1}^{N}\| \hat{\boldsymbol{\lambda}}_i -\check{\boldsymbol{\lambda}}_{0i}\| \right),
			\end{equation*}
			which gives the desired result since $\boldsymbol{\Delta} >0$. \end{proof}

		\vspace{0.5cm}
		
		\begin{lemma}Under Assumptions 1 to 4, we have 
				\[    \frac{1}{N}\sum_{i=1}^{N} \| \hat{\boldsymbol{\lambda}}_i  -\check{\boldsymbol{\lambda}}_{0i}\|  = O_P( T^{ -\frac{1}{2} }).\]
		\end{lemma}
		\begin{proof}Plugging the result of Lemma 5 into \eqref{A7} we have
			\begin{multline}\label{A15}
					\frac{1}{N}\sum_{i=1}^{N} \| \hat{\boldsymbol{\lambda}}_i  -\check{\boldsymbol{\lambda}}_{0i}\| \leq O_P(T^{-1}) 
					+ o_P\left( \frac{1}{N} \sum_{i=1}^{N}\| \hat{\boldsymbol{\lambda}}_i -\check{\boldsymbol{\lambda}}_{0i}\| \right) + 
					O_P(1)\cdot \frac{1}{N}\sum_{i=1}^{N}\left\|\frac{1}{T}\sum_{t=1}^{T}  l_{it}^{(1)}\boldsymbol{f}_{0t}\right\| \\
					+O_P(1)\cdot \frac{1}{N}\sum_{i=1}^{N} \left\| \frac{1}{T}\sum_{t=1}^{T}  l_{it}^{(1)}( \hat{\boldsymbol{f}}_t - \check{\boldsymbol{f}}_{0t})\right\|  +  O_P(1)\cdot \frac{1}{N}\sum_{i=1}^{N} \left\| \frac{1}{T}\sum_{t=1}^{T}  l_{it}^{(2)} \hat{\boldsymbol{f}}_t  (\hat{\boldsymbol{f}}_t - \check{\boldsymbol{f}}_{0t})' \right\| \\
					+O_P(1)\cdot \frac{1}{N}\sum_{i=1}^{N} \left\|\frac{1}{T}\sum_{t=1}^{T}  l_{it}^{(3)}(\ast)  \check{\boldsymbol{A}}_i^{-1}\hat{\boldsymbol{f}}_t \cdot  \check{\boldsymbol{\lambda}}_{0i}'(\hat{\boldsymbol{f}}_t - \check{\boldsymbol{f}}_{0t}) (\hat{\boldsymbol{f}}_t - \check{\boldsymbol{f}}_{0t})'\check{\boldsymbol{\lambda}}_{0i}\right\|.
			\end{multline}
			
			First, by Lemma 1 
			\[ \Ex  \left\|\frac{1}{T}\sum_{t=1}^{T}  l_{it}^{(1)}\boldsymbol{f}_{0t} \right\|^{2p}  = O(T^{-p}), \]
		for all $i$, thus it holds that 
				\[\frac{1}{N} \sum_{i=1}^{N} \Ex\left\|\frac{1}{T}\sum_{t=1}^{T}  l_{it}^{(1)}\boldsymbol{f}_{0t} \right\| \leq \frac{1}{N} \sum_{i=1}^{N} \left(\Ex\left\|\frac{1}{T}\sum_{t=1}^{T}  l_{it}^{(1)}\boldsymbol{f}_{0t} \right\|^{2p}\right)^{1/2p} = O(T^{-1/2}),\]
				and therefore
				\[ \frac{1}{N} \sum_{i=1}^{N} \left\|\frac{1}{T}\sum_{t=1}^{T}  l_{it}^{(1)}\boldsymbol{f}_{0t} \right\|  = O_P(T^{-1/2}).\]
				
			Second, note that
				\[\frac{1}{T}\sum_{t=1}^{T}  l_{it}^{(1)}( \hat{\boldsymbol{f}}_t - \check{\boldsymbol{f}}_{0t}) = \hat{\boldsymbol{\Psi}}\cdot \frac{1}{NT}\sum_{t=1}^{T}\sum_{j=1}^{N}  l_{it}^{(1)} \boldsymbol{e}_{jt}\] 
				and
				\[\left\|\frac{1}{NT}\sum_{t=1}^{T}\sum_{j=1}^{N}  l_{it}^{(1)} \boldsymbol{e}_{jt} \right\|^{2p}\leq 2^{2p-1}\cdot\left(\left\|\frac{1}{T}\sum_{t=1}^{T}  l_{it}^{(1)} \frac{1}{N}\sum_{j\neq i}^{N} \boldsymbol{e}_{jt} \right\|^{2p} + \left\|\frac{1}{NT}\sum_{t=1}^{T}  l_{it}^{(1)} \boldsymbol{e}_{it} \right\|^{2p}\right)\]
				since $(a+b)^{k}\leq 2^{k-1} (a^k+b^k)$ for any $k\geq 1$ and $a,b\geq0$. By the uniform boundedness of $\Ex\|\boldsymbol{e}_{jt}\|^{2p+\gamma}$ and Rosenthal inequality it can be shown that $\max_{1\leq t\leq T}\Ex\left\| N^{-1/2}\sum_{j\neq i}^{N}\boldsymbol{e}_{jt}\right\|^{2p+\gamma}<\infty$, which further implies that 
				\[\max_{1\leq t\leq T}\Ex\left\|  l_{it}^{(1)} \frac{1}{\sqrt{N}}\sum_{j\neq i}^{N} \boldsymbol{e}_{jt} \right\|^{2p+\gamma}\leq \max_{1\leq t\leq T}\Ex\left|l_{it}^{(1)}\right|^{2p+\gamma}\cdot \max_{1\leq t\leq T}\Ex\left\|\frac{1}{\sqrt{N}}\sum_{j\neq i}^{N} \boldsymbol{e}_{jt}\right\|^{2p+\gamma}<\infty.\]
				Let $\alpha_i^*(j)$ be the strong mixing coefficients of $\{l_{it}^{(1)} \cdot N^{-1/2}\sum_{j\neq i}^{N} \boldsymbol{e}_{jt}\}$, then Theorem 5.2 of \cite{bradley2005basic} and Assumption 4 imply that $\alpha_i^*(j)\leq N\cdot C\alpha^j$. Similar to the proof of Lemma 1, by Rosenthal type inequality for dependent sequence we have
				\begin{multline*}
					\Ex\left\|\sum_{t=1}^{T}  l_{it}^{(1)} \frac{1}{\sqrt{N}}\sum_{j\neq i}^{N} \boldsymbol{e}_{jt} \right\|^{2p} \\
					\lesssim T^pN^{p-\frac{2p}{2p+\gamma}}\cdot\left(\frac{1}{T}\sum_{t=1}^{T}\left\|l_{it}^{(1)} \frac{1}{\sqrt{N}}\sum_{j\neq i}^{N} \boldsymbol{e}_{jt}\right\|_{2p+\gamma}^2\right)^p\left(\sum_{k=0}^{+\infty}(k+1)^{\frac{2}{2p+\gamma-2}}\alpha^k\right)^{p-\frac{2p}{2p+\gamma}}\\
					+ TN^{\frac{\gamma}{2p+\gamma}}\cdot\left(\max_{1\leq t \leq T}\left\|l_{it}^{(1)} \frac{1}{\sqrt{N}}\sum_{j\neq i}^{N} \boldsymbol{e}_{jt}\right\|_{2p+\gamma}\right)^{2p}\cdot\left(\sum_{k=0}^{+\infty}(k+1)^{2p-2}\alpha^{k\cdot\frac{\gamma}{2p+\gamma}}\right) \\
					= O\left(T^pN^{p-\frac{2p}{2p+\gamma}}\right)+ O\left(TN^{\frac{\gamma}{2p+\gamma}}\right) = O\left(T^pN^{p-\frac{2p}{2p+\gamma}}\right)
			\end{multline*}

		Besides, it is easy to show that $\Ex\left\|\sum_{t=1}^{T}  l_{it}^{(1)} \boldsymbol{e}_{it} \right\|^{2p}=O(T^p)$. Thus it follows that
				\[\Ex  \left\|\frac{1}{NT}\sum_{j=1}^{N}\sum_{t=1}^{T}  l_{it}^{(1)} \boldsymbol{e}_{jt} \right\|^{2p}  = O(T^{-p}N^{-2p/(2p+\gamma)})\]
				and
				\[\frac{1}{N}\sum_{i=1}^{N}  \left\| \frac{1}{T}\sum_{t=1}^{T}  l_{it}^{(1)}( \hat{\boldsymbol{f}}_t - \check{\boldsymbol{f}}_{0t}) \right\| = O_P(T^{-1/2}N^{-1/(2p+\gamma)})=o_P(T^{-1/2}).\]

			Third,
			\begin{multline}
				\frac{1}{T}\sum_{t=1}^{T}  l_{it}^{(2)} \hat{\boldsymbol{f}}_t  (\hat{\boldsymbol{f}}_t - \check{\boldsymbol{f}}_{0t})' = \hat{\boldsymbol{H}}\cdot  \frac{1}{T}\sum_{t=1}^{T}  l_{it}^{(2)} \boldsymbol{f}_{0t}  (\hat{\boldsymbol{f}}_t - \check{\boldsymbol{f}}_{0t})' +\frac{1}{T}\sum_{t=1}^{T}  l_{it}^{(2)} ( \hat{\boldsymbol{f}}_t - \check{\boldsymbol{f}}_{0t}) (\hat{\boldsymbol{f}}_t - \check{\boldsymbol{f}}_{0t})' \\
				= \hat{\boldsymbol{H}}\cdot  \frac{1}{NT}\sum_{t=1}^{T}\sum_{j=1}^{N}  l_{it}^{(2)} \boldsymbol{f}_{0t}\boldsymbol{e}_{jt}' \cdot \hat{\boldsymbol{\Psi}} + \frac{1}{T}\sum_{t=1}^{T}  l_{it}^{(2)} ( \hat{\boldsymbol{f}}_t - \check{\boldsymbol{f}}_{0t}) (\hat{\boldsymbol{f}}_t - \check{\boldsymbol{f}}_{0t})'.
			\end{multline}
			Similar to the previous step, we can show that 
			\[\Ex  \left\|\frac{1}{NT}\sum_{j=1}^{N}\sum_{t=1}^{T}  l_{it}^{(2)} \boldsymbol{f}_{0t}\boldsymbol{e}_{jt}' \right\|^{2p}  = O(T^{-p}N^{-2p/(2p+\gamma)})\]
				and
				\[\frac{1}{N}\sum_{i=1}^{N}\left\|\frac{1}{NT}\sum_{j=1}^{N}\sum_{t=1}^{T}  l_{it}^{(2)} \boldsymbol{f}_{0t}\boldsymbol{e}_{jt}' \right\| = O_P(T^{-1/2}N^{-1/(2p+\gamma)})=o_P(T^{-1/2}).\]
				Moreover, 
				\begin{multline*}
					\frac{1}{N}\sum_{i=1}^{N}\left\|  \frac{1}{T}\sum_{t=1}^{T}  l_{it}^{(2)} ( \hat{\boldsymbol{f}}_t - \check{\boldsymbol{f}}_{0t}) (\hat{\boldsymbol{f}}_t - \check{\boldsymbol{f}}_{0t})'  \right\| 
					\leq  \left( \max_{1\leq t \leq T}\frac{1}{N}\sum_{i=1}^{N}M(\boldsymbol{x}_{it})  \right)\cdot \left( \frac{1}{T}\sum_{t=1}^{T}\| \hat{\boldsymbol{f}}_t -\check{\boldsymbol{f}}_{0t}\|^2  \right) \\
					= O_P(N^{-1}) = o_P(T^{-1/2}).
				\end{multline*}
				Therefore, 
				\[ \frac{1}{N}\sum_{i=1}^{N}\left\| \frac{1}{T}\sum_{t=1}^{T}  l_{it}^{(2)} \hat{\boldsymbol{f}}_t  (\hat{\boldsymbol{f}}_t - \check{\boldsymbol{f}}_{0t})'\right\| = o_P(T^{-1/2}). \]
				
				Finally, 
				\begin{multline*}
					\frac{1}{N}\sum_{i=1}^{N}\left\| \frac{1}{T}\sum_{t=1}^{T}  l_{it}^{(3)}(\ast)  \check{\boldsymbol{A}}_i^{-1}\hat{\boldsymbol{f}}_t \cdot  \check{\boldsymbol{\lambda}}_{0i}'(\hat{\boldsymbol{f}}_t - \check{\boldsymbol{f}}_{0t}) (\hat{\boldsymbol{f}}_t - \check{\boldsymbol{f}}_{0t})'\check{\boldsymbol{\lambda}}_{0i}\right\| \lesssim \frac{1}{NT}\sum_{i=1}^{N} \sum_{t=1}^{T}  M(\boldsymbol{x}_{it})\|\hat{\boldsymbol{f}}_t - \check{\boldsymbol{f}}_{0t}\|^2 \\
					= O_P(N^{-1}) = o_P(T^{-1/2}).
				\end{multline*}
				Combining all the above results gives the desired conclusion.
		\end{proof}

		\begin{lemma}Under Assumptions 1 to 4, 
			\[\frac{1}{N}\sum_{i=1}^{N}\|\hat{\boldsymbol{\lambda}}_i -\check{\boldsymbol{\lambda}}_{0i}\|^2 = O_P(T^{-1})\]
				and
				\[\frac{1}{N}\sum_{i=1}^{N}\left\|\hat{\boldsymbol{\lambda}}_i -\check{\boldsymbol{\lambda}}_{0i} +  (\hat{\boldsymbol{H}}')^{-1}  \boldsymbol{A}^{-1}_i\cdot \frac{1}{T}\sum_{t=1}^{T}  l_{it}^{(1)}\boldsymbol{f}_{0t}\right\|^2 = o_P(T^{-1}).\]
		\end{lemma}
		\begin{proof}
			By \eqref{A7} it can be shown that
				\begin{multline*}
					\frac{1}{N}\sum_{i=1}^{N}\|\hat{\boldsymbol{\lambda}}_i -\check{\boldsymbol{\lambda}}_{0i}\|^2 \leq \bar{O}_P(\|\hat{ \boldsymbol{\beta}}-\boldsymbol{\beta_{0}}\|^2) 
					+ o_P\left( \frac{1}{N} \sum_{i=1}^{N}\| \hat{\boldsymbol{\lambda}}_i -\check{\boldsymbol{\lambda}}_{0i}\|^2 \right) + 
					O_P(1)\cdot \frac{1}{N}\sum_{i=1}^{N}\left\|\frac{1}{T}\sum_{t=1}^{T}  l_{it}^{(1)}\boldsymbol{f}_{0t}\right\|^2 \\
					+O_P(1)\cdot \frac{1}{N}\sum_{i=1}^{N} \left\| \frac{1}{T}\sum_{t=1}^{T}  l_{it}^{(1)}( \hat{\boldsymbol{f}}_t - \check{\boldsymbol{f}}_{0t})\right\|^2  +  O_P(1)\cdot \frac{1}{N}\sum_{i=1}^{N} \left\| \frac{1}{T}\sum_{t=1}^{T}  l_{it}^{(2)} \hat{\boldsymbol{f}}_t  (\hat{\boldsymbol{f}}_t - \check{\boldsymbol{f}}_{0t})' \right\|^2 \\
					+O_P(1)\cdot \frac{1}{N}\sum_{i=1}^{N} \left\|\frac{1}{T}\sum_{t=1}^{T}  l_{it}^{(3)}(\ast)  \check{\boldsymbol{A}}_i^{-1}\hat{\boldsymbol{f}}_t \cdot  \check{\boldsymbol{\lambda}}_{0i}'(\hat{\boldsymbol{f}}_t - \check{\boldsymbol{f}}_{0t}) (\hat{\boldsymbol{f}}_t - \check{\boldsymbol{f}}_{0t})'\check{\boldsymbol{\lambda}}_{0i}\right\|^2.
				\end{multline*}
				Note that Lemma 5 and 6 now imply $\|\hat{ \boldsymbol{\beta}}-\boldsymbol{\beta_{0}}\| = O_P(T^{-1}) + o_P\left(N^{-1}\sum_{i=1}^{N} \| \hat{\boldsymbol{\lambda}}_i  -\check{\boldsymbol{\lambda}}_{0i}\|\right)=o_P(T^{-1/2})$. Besides, similar to the proof of Lemma 6, it holds that 
				\[\frac{1}{N}\sum_{i=1}^{N}\left\|\frac{1}{T}\sum_{t=1}^{T}  l_{it}^{(1)}\boldsymbol{f}_{0t}\right\|^2 = O_P(T^{-1})\]
				and
				\[\frac{1}{N}\sum_{i=1}^{N} \left\| \frac{1}{T}\sum_{t=1}^{T}  l_{it}^{(1)}( \hat{\boldsymbol{f}}_t - \check{\boldsymbol{f}}_{0t})\right\|^2 = o_P(T^{-1}), \quad \frac{1}{N}\sum_{i=1}^{N} \left\| \frac{1}{T}\sum_{t=1}^{T}  l_{it}^{(2)} \hat{\boldsymbol{f}}_t  (\hat{\boldsymbol{f}}_t - \check{\boldsymbol{f}}_{0t})' \right\|^2 = o_P(T^{-1}).\]
				Moreover, 
				\begin{multline*}
					\frac{1}{N}\sum_{i=1}^{N} \left\|\frac{1}{T}\sum_{t=1}^{T}  l_{it}^{(3)}(\ast)  \check{\boldsymbol{A}}_i^{-1}\hat{\boldsymbol{f}}_t \cdot  \check{\boldsymbol{\lambda}}_{0i}'(\hat{\boldsymbol{f}}_t - \check{\boldsymbol{f}}_{0t}) (\hat{\boldsymbol{f}}_t - \check{\boldsymbol{f}}_{0t})'\check{\boldsymbol{\lambda}}_{0i}\right\|^2 \\
					\lesssim \frac{1}{N}\sum_{i=1}^{N} \left(\frac{1}{T}\sum_{t=1}^{T}  M(\boldsymbol{x}_{it}) \cdot \|\hat{\boldsymbol{f}}_t - \check{\boldsymbol{f}}_{0t}\|^2 \right)^2 \leq \left(\frac{1}{NT}\sum_{i=1}^{N} \sum_{t=1}^{T}  M(\boldsymbol{x}_{it})^2\right)\left(\frac{1}{T}\sum_{t=1}^{T} \|\hat{\boldsymbol{f}}_t - \check{\boldsymbol{f}}_{0t}\|^4 \right) \\
					\lesssim\max_{1\leq t \leq T}\|\hat{\boldsymbol{f}}_t - \check{\boldsymbol{f}}_{0t}\|^2\cdot \frac{1}{T}\sum_{t=1}^{T} \|\hat{\boldsymbol{f}}_t - \check{\boldsymbol{f}}_{0t}\|^2 = o_P(N^{-1}).
				\end{multline*}
				Consequently, the first conclusion of this Lemma holds: $N^{-1}\sum_{i=1}^{N}\|\hat{\boldsymbol{\lambda}}_i -\check{\boldsymbol{\lambda}}_{0i}\|^2 = O_P(T^{-1}).$
				As for the second conclusion, by \eqref{A7} and previous results we have
				\begin{multline*}
					\frac{1}{N}\sum_{i=1}^{N}\left\|\hat{\boldsymbol{\lambda}}_i -\check{\boldsymbol{\lambda}}_{0i} +  (\hat{\boldsymbol{H}}')^{-1}  \boldsymbol{A}^{-1}_i\cdot \frac{1}{T}\sum_{t=1}^{T}  l_{it}^{(1)}\boldsymbol{f}_{0t}\right\|^2 = \bar{O}_P(\|\hat{ \boldsymbol{\beta}}-\boldsymbol{\beta_{0}}\|^2) 
					+ o_P\left( \frac{1}{N} \sum_{i=1}^{N}\| \hat{\boldsymbol{\lambda}}_i -\check{\boldsymbol{\lambda}}_{0i}\|^2 \right) \\
					+O_P(1)\cdot \frac{1}{N}\sum_{i=1}^{N} \left\| \frac{1}{T}\sum_{t=1}^{T}  l_{it}^{(1)}( \hat{\boldsymbol{f}}_t - \check{\boldsymbol{f}}_{0t})\right\|^2  +  O_P(1)\cdot \frac{1}{N}\sum_{i=1}^{N} \left\| \frac{1}{T}\sum_{t=1}^{T}  l_{it}^{(2)} \hat{\boldsymbol{f}}_t  (\hat{\boldsymbol{f}}_t - \check{\boldsymbol{f}}_{0t})' \right\|^2 \\
					+O_P(1)\cdot \frac{1}{N}\sum_{i=1}^{N} \left\|\frac{1}{T}\sum_{t=1}^{T}  l_{it}^{(3)}(\ast)  \check{\boldsymbol{A}}_i^{-1}\hat{\boldsymbol{f}}_t \cdot  \check{\boldsymbol{\lambda}}_{0i}'(\hat{\boldsymbol{f}}_t - \check{\boldsymbol{f}}_{0t}) (\hat{\boldsymbol{f}}_t - \check{\boldsymbol{f}}_{0t})'\check{\boldsymbol{\lambda}}_{0i}\right\|^2 = o_P(T^{-1}).
			\end{multline*}
		\end{proof}

		\vspace{0.5cm}
		
		\begin{lemma}Under Assumptions 1 to 4,  
			\begin{multline*}   \frac{1}{NT}\sum_{i=1}^{N}\sum_{t=1}^{T}  l_{it}^{(3)}(\ast) \cdot (\boldsymbol{x}_{it} -\boldsymbol{B}_i \boldsymbol{A}^{-1}_i \hat{\boldsymbol{H}}^{-1}\hat{\boldsymbol{f}}_t) \cdot   \hat{\boldsymbol{f}}_t' (\hat{\boldsymbol{\lambda}}_i - \check{\boldsymbol{\lambda}}_{0i}) (\hat{\boldsymbol{\lambda}}_i - \check{\boldsymbol{\lambda}}_{0i})'\hat{\boldsymbol{f}}_t \\
				=\frac{1}{T}\cdot \frac{1}{NT}\sum_{i=1}^{N}\sum_{t=1}^{T}  \Ex[ l_{it}^{(3)} \dot{\boldsymbol{x}}_{it} ] \cdot \boldsymbol{f}_{0t}'  \boldsymbol{A}^{-1}_i \boldsymbol{Q}_{i} \boldsymbol{A}^{-1}_i  \boldsymbol{f}_{0t} +o_P(T^{-1}).
			\end{multline*}
		\end{lemma} 
		\begin{proof}
			First, by Assumption 4 and Lemma 7, 
			\begin{align*}
				& \frac{1}{NT}\sum_{i=1}^{N}\sum_{t=1}^{T}  l_{it}^{(3)}(\ast) \cdot (\boldsymbol{x}_{it} -\boldsymbol{B}_i \boldsymbol{A}^{-1}_i \hat{\boldsymbol{H}}^{-1}\hat{\boldsymbol{f}}_t) \cdot   \hat{\boldsymbol{f}}_t' (\hat{\boldsymbol{\lambda}}_i - \check{\boldsymbol{\lambda}}_{0i}) (\hat{\boldsymbol{\lambda}}_i - \check{\boldsymbol{\lambda}}_{0i})'\hat{\boldsymbol{f}}_t \\
				= &  \frac{1}{NT}\sum_{i=1}^{N}\sum_{t=1}^{T}  l_{it}^{(3)} \dot{\boldsymbol{x}}_{it} \cdot   \check{\boldsymbol{f}}_{0t}' (\hat{\boldsymbol{\lambda}}_i - \check{\boldsymbol{\lambda}}_{0i}) (\hat{\boldsymbol{\lambda}}_i - \check{\boldsymbol{\lambda}}_{0i})'\check{\boldsymbol{f}}_{0t} +o_P\left( \frac{1}{N}\sum_{i=1}^{N}\|\hat{\boldsymbol{\lambda}}_i - \check{\boldsymbol{\lambda}}_{0i}\|^2 \right) \\
				= & \frac{1}{T} \cdot \frac{1}{NT}\sum_{i=1}^{N}\sum_{t=1}^{T}  l_{it}^{(3)} \dot{\boldsymbol{x}}_{it} \cdot   \boldsymbol{f}_{0t}' \boldsymbol{A}^{-1}_i  \left( \frac{1}{\sqrt{T}}\sum_{t=1}^{T}  l_{it}^{(1)}\boldsymbol{f}_{0t}  \right) \left( \frac{1}{\sqrt{T}}\sum_{t=1}^{T}  l_{it}^{(1)}\boldsymbol{f}_{0t}  \right)' \boldsymbol{A}^{-1}_i \boldsymbol{f}_{0t} \\ 
				&  + O_P(T^{-1/2})\cdot \left(\frac{1}{N}\sum_{i=1}^{N}\left\|\hat{\boldsymbol{\lambda}}_i -\check{\boldsymbol{\lambda}}_{0i} +  (\hat{\boldsymbol{H}}')^{-1}  \boldsymbol{A}^{-1}_i\cdot \frac{1}{T}\sum_{t=1}^{T}  l_{it}^{(1)}\boldsymbol{f}_{0t}\right\|^2\right)^{1/2}\\
				& + O_P(1)\cdot \frac{1}{N}\sum_{i=1}^{N} \left\| \hat{\boldsymbol{\lambda}}_i -\check{\boldsymbol{\lambda}}_{0i} +  (\hat{\boldsymbol{H}}')^{-1}  \boldsymbol{A}^{-1}_i\cdot \frac{1}{T}\sum_{t=1}^{T}  l_{it}^{(1)}\boldsymbol{f}_{0t} \right\|^2 + o_P\left( \frac{1}{N}\sum_{i=1}^{N}\|\hat{\boldsymbol{\lambda}}_i - \check{\boldsymbol{\lambda}}_{0i}\|^2 \right) \\
				= & \frac{1}{T} \cdot \frac{1}{NT}\sum_{i=1}^{N}\sum_{t=1}^{T}  l_{it}^{(3)} \dot{\boldsymbol{x}}_{it} \cdot   \boldsymbol{f}_{0t}' \boldsymbol{A}^{-1}_i  \left( \frac{1}{\sqrt{T}}\sum_{t=1}^{T}  l_{it}^{(1)}\boldsymbol{f}_{0t}  \right) \left( \frac{1}{\sqrt{T}}\sum_{t=1}^{T}  l_{it}^{(1)}\boldsymbol{f}_{0t}  \right)' \boldsymbol{A}^{-1}_i \boldsymbol{f}_{0t} + o_P(T^{-1}),
			\end{align*}
			where the first equality follows from \begin{multline*}
					\left\| \frac{1}{NT}\sum_{i=1}^{N}\sum_{t=1}^{T}  l_{it}^{(3)}(\ast) \cdot (\boldsymbol{x}_{it} -\boldsymbol{B}_i \boldsymbol{A}^{-1}_i \hat{\boldsymbol{H}}^{-1}\hat{\boldsymbol{f}}_t) \cdot   \hat{\boldsymbol{f}}_t' (\hat{\boldsymbol{\lambda}}_i - \check{\boldsymbol{\lambda}}_{0i}) (\hat{\boldsymbol{\lambda}}_i - \check{\boldsymbol{\lambda}}_{0i})'\hat{\boldsymbol{f}}_t \right. \\
					\left. - \frac{1}{NT}\sum_{i=1}^{N}\sum_{t=1}^{T}  l_{it}^{(3)} \dot{\boldsymbol{x}}_{it} \cdot   \check{\boldsymbol{f}}_{0t}' (\hat{\boldsymbol{\lambda}}_i - \check{\boldsymbol{\lambda}}_{0i}) (\hat{\boldsymbol{\lambda}}_i - \check{\boldsymbol{\lambda}}_{0i})'\check{\boldsymbol{f}}_{0t} \right\| \lesssim \left(\|\hat{ \boldsymbol{\beta}}-\boldsymbol{\beta}_0\|+\max_{1\leq i \leq N}\|\hat{\boldsymbol{\lambda}}_i - \check{\boldsymbol{\lambda}}_{0i}\| + \max_{1\leq t \leq T}\|\hat{\boldsymbol{f}}_t - \check{\boldsymbol{f}}_{0t}\|\right)\\
					\cdot \left(\max_{1\leq i \leq N}\frac{1}{T}\sum_{t=1}^{T}M(\boldsymbol{x}_{it})\right)\cdot \left(\frac{1}{N}\sum_{i=1}^{N}\|\hat{\boldsymbol{\lambda}}_i - \check{\boldsymbol{\lambda}}_{0i}\|^2\right) = o_P\left(\frac{1}{N}\sum_{i=1}^{N}\|\hat{\boldsymbol{\lambda}}_i - \check{\boldsymbol{\lambda}}_{0i}\|^2\right),
				\end{multline*}
				and the second equality can be derived by
				\begin{multline*}
					\left\|\frac{1}{NT}\sum_{i=1}^{N}\sum_{t=1}^{T}  l_{it}^{(3)} \dot{\boldsymbol{x}}_{it}   \check{\boldsymbol{f}}_{0t}' (\hat{\boldsymbol{\lambda}}_i - \check{\boldsymbol{\lambda}}_{0i})(\hat{\boldsymbol{\lambda}}_i - \check{\boldsymbol{\lambda}}_{0i})' \check{\boldsymbol{f}}_{0t}\right. \\
					\left.- \frac{1}{NT}\sum_{i=1}^{N}\sum_{t=1}^{T}  l_{it}^{(3)} \dot{\boldsymbol{x}}_{it}  \boldsymbol{f}_{0t}' \boldsymbol{A}^{-1}_i  \left( \frac{1}{T}\sum_{t=1}^{T}  l_{it}^{(1)}\boldsymbol{f}_{0t}  \right) \left( \frac{1}{T}\sum_{t=1}^{T}  l_{it}^{(1)}\boldsymbol{f}_{0t}  \right)' \boldsymbol{A}^{-1}_i \boldsymbol{f}_{0t}\right\| \\
					\lesssim \frac{1}{N}\sum_{i=1}^{N}\left(\frac{1}{T}\sum_{t=1}^{T}  M(\boldsymbol{x}_{it})\right)\cdot  \left\| \frac{1}{T}\sum_{t=1}^{T}  l_{it}^{(1)}\boldsymbol{f}_{0t} \right\| \cdot \left\| \hat{\boldsymbol{\lambda}}_i -\check{\boldsymbol{\lambda}}_{0i} +  (\hat{\boldsymbol{H}}')^{-1}  \boldsymbol{A}^{-1}_i\cdot \frac{1}{T}\sum_{t=1}^{T}  l_{it}^{(1)}\boldsymbol{f}_{0t} \right\| \\
					+ \frac{1}{N}\sum_{i=1}^{N}\left(\frac{1}{T}\sum_{t=1}^{T}  M(\boldsymbol{x}_{it})\right)\cdot \left\| \hat{\boldsymbol{\lambda}}_i -\check{\boldsymbol{\lambda}}_{0i} +  (\hat{\boldsymbol{H}}')^{-1}  \boldsymbol{A}^{-1}_i\cdot \frac{1}{T}\sum_{t=1}^{T}  l_{it}^{(1)}\boldsymbol{f}_{0t} \right\|^2 \\
					\leq O_p(1)\cdot \left[\frac{1}{N}\sum_{i=1}^{N}  \left\| \frac{1}{T}\sum_{t=1}^{T}  l_{it}^{(1)}\boldsymbol{f}_{0t} \right\|^2\right]^{1/2}\cdot \left[\frac{1}{N}\sum_{i=1}^{N}\left\| \hat{\boldsymbol{\lambda}}_i -\check{\boldsymbol{\lambda}}_{0i} +  (\hat{\boldsymbol{H}}')^{-1}  \boldsymbol{A}^{-1}_i\cdot \frac{1}{T}\sum_{t=1}^{T}  l_{it}^{(1)}\boldsymbol{f}_{0t} \right\|^2\right]^{1/2} \\
					+ O_P(1)\cdot \frac{1}{N}\sum_{i=1}^{N} \left\| \hat{\boldsymbol{\lambda}}_i -\check{\boldsymbol{\lambda}}_{0i} +  (\hat{\boldsymbol{H}}')^{-1}  \boldsymbol{A}^{-1}_i\cdot \frac{1}{T}\sum_{t=1}^{T}  l_{it}^{(1)}\boldsymbol{f}_{0t} \right\|^2.
			\end{multline*}

			Next, it can be shown that:
			\begin{align*}
				&\frac{1}{NT}\sum_{i=1}^{N}\sum_{t=1}^{T}  l_{it}^{(3)} \dot{\boldsymbol{x}}_{it,j} \cdot   \boldsymbol{f}_{0t}' \boldsymbol{A}^{-1}_i  \left( \frac{1}{\sqrt{T}}\sum_{t=1}^{T}  l_{it}^{(1)}\boldsymbol{f}_{0t}  \right) \left( \frac{1}{\sqrt{T}}\sum_{t=1}^{T}  l_{it}^{(1)}\boldsymbol{f}_{0t}  \right)' \boldsymbol{A}^{-1}_i \boldsymbol{f}_{0t} \\
				= &\tr\left[ \frac{1}{N}\sum_{i=1}^{N}\left( \frac{1}{T}\sum_{t=1}^{T}  l_{it}^{(3)} \dot{\boldsymbol{x}}_{it,j} \cdot \boldsymbol{f}_{0t}  \boldsymbol{f}_{0t}' \right) \boldsymbol{A}^{-1}_i  \left( \frac{1}{\sqrt{T}}\sum_{t=1}^{T}  l_{it}^{(1)}\boldsymbol{f}_{0t}  \right) \left( \frac{1}{\sqrt{T}}\sum_{t=1}^{T}  l_{it}^{(1)}\boldsymbol{f}_{0t}  \right)' \boldsymbol{A}^{-1}_i   \right]  \\
				=&  \tr\left[ \frac{1}{N}\sum_{i=1}^{N}\left( \frac{1}{T}\sum_{t=1}^{T}  \Ex[ l_{it}^{(3)} \dot{\boldsymbol{x}}_{it,j} ] \cdot \boldsymbol{f}_{0t}  \boldsymbol{f}_{0t}' \right) \boldsymbol{A}^{-1}_i  \left( \frac{1}{T}\sum_{t=1}^{T}\sum_{s=1}^{T}  \Ex\left[ l_{it}^{(1)}l_{is}^{(1)}\right]  \boldsymbol{f}_{0t}\boldsymbol{f}_{0s}'  \right)  \boldsymbol{A}^{-1}_i \right]  +o_P(1) \\
				=& \frac{1}{NT}\sum_{i=1}^{N}\sum_{t=1}^{T}  \Ex[ l_{it}^{(3)} \dot{\boldsymbol{x}}_{it,j} ] \cdot \boldsymbol{f}_{0t}'  \boldsymbol{A}^{-1}_i \boldsymbol{Q}_{i} \boldsymbol{A}^{-1}_i  \boldsymbol{f}_{0t} +o_P(1).
			\end{align*}
			Let $\bm{R}_{ij}=T^{-1}\sum_{t=1}^{T}  l_{it}^{(3)} \dot{\boldsymbol{x}}_{it,j} \cdot \boldsymbol{f}_{0t}  \boldsymbol{f}_{0t}'.$
			To obtain the second equality in the above equation, note that
			\begin{multline}\label{a17}
				\frac{1}{N}\sum_{i=1}^{N}\bm{R}_{ij} \boldsymbol{A}^{-1}_i  \left( \frac{1}{\sqrt{T}}\sum_{t=1}^{T}  l_{it}^{(1)}\boldsymbol{f}_{0t}  \right) \left( \frac{1}{\sqrt{T}}\sum_{t=1}^{T}  l_{it}^{(1)}\boldsymbol{f}_{0t}'  \right) \boldsymbol{A}^{-1}_i \\
				=  \frac{1}{N}\sum_{i=1}^{N}\left( \bm{R}_{ij} - \Ex[\bm{R}_{ij}] \right) \boldsymbol{A}^{-1}_i  \left( \frac{1}{T}\sum_{t=1}^{T} \sum_{s=1}^{T} l_{it}^{(1)}l_{is}^{(1)}\boldsymbol{f}_{0t}\boldsymbol{f}_{0s}' \right) \boldsymbol{A}^{-1}_i \\
				+ \frac{1}{N}\sum_{i=1}^{N}\Ex[\bm{R}_{ij}]\cdot \boldsymbol{A}^{-1}_i  \left( \frac{1}{T}\sum_{t=1}^{T} \sum_{s=1}^{T} l_{it}^{(1)}l_{is}^{(1)}\boldsymbol{f}_{0t}\boldsymbol{f}_{0s}' \right) \boldsymbol{A}^{-1}_i
			\end{multline}
			First, since
			\[\Ex\left\|\bm{R}_{ij}-\Ex[\bm{R}_{ij}]\right\|^{2p} = \Ex \left\|\frac{1}{T}\sum_{t=1}^{T}  \left( l_{it}^{(3)} \dot{\boldsymbol{x}}_{it,j} -\Ex\left[l_{it}^{(3)} \dot{\boldsymbol{x}}_{it,j}\right] \right) \cdot \boldsymbol{f}_{0t}  \boldsymbol{f}_{0t}'\right\|^{2p}=O(T^{-p}),\]
			it holds that 
			\[\max_{1\leq i \leq N}\left\|\bm{R}_{ij}-\Ex[\bm{R}_{ij}]\right\|=O_P(T^{1/2p-1/2})=o_P(1).\]
			Next, 
			\[\Ex\left\|\frac{1}{T}\sum_{t=1}^{T} \sum_{s=1}^{T} l_{it}^{(1)}l_{is}^{(1)}\boldsymbol{f}_{0t}\boldsymbol{f}_{0s}'\right\|  \leq   C\cdot \frac{1}{T}\mathop{\sum\sum}_{1 \leq t\leq s \leq T}\left|\Ex\left[l_{it}^{(1)}l_{is}^{(1)}\right]\right| = O(1),\]
			where the last equality follows from Lemma 3. 
			Thus, for the first term on the right-hand side of \eqref{a17},
			\begin{multline*}
				\left\|\frac{1}{N}\sum_{i=1}^{N}\left( \bm{R}_{ij} - \Ex[\bm{R}_{ij}] \right) \boldsymbol{A}^{-1}_i  \left( \frac{1}{T}\sum_{t=1}^{T} \sum_{s=1}^{T} l_{it}^{(1)}l_{is}^{(1)}\boldsymbol{f}_{0t}\boldsymbol{f}_{0s}' \right) \boldsymbol{A}^{-1}_i\right\| \\
				\leq C\cdot \max_{1\leq i \leq N}\left\|\bm{R}_{ij}-\Ex[\bm{R}_{ij}]\right\|\cdot \frac{1}{N}\sum_{i=1}^{N}\left\|\frac{1}{T}\sum_{t=1}^{T} \sum_{s=1}^{T} l_{it}^{(1)}l_{is}^{(1)}\bm{f}_{0t}\bm{f}_{0s}'\right\|= o_P(1).
			\end{multline*}
			For the second term, we have
			\begin{multline*}
				\Ex\left[\frac{1}{N}\sum_{i=1}^{N}\Ex[\bm{R}_{ij}]\cdot \boldsymbol{A}^{-1}_i  \left( \frac{1}{T}\sum_{t=1}^{T} \sum_{s=1}^{T} l_{it}^{(1)}l_{is}^{(1)}\boldsymbol{f}_{0t}\boldsymbol{f}_{0s}' \right) \boldsymbol{A}^{-1}_i\right] \\
				= \frac{1}{N}\sum_{i=1}^{N}\Ex[\bm{R}_{ij}]\cdot \boldsymbol{A}^{-1}_i  \left( \frac{1}{T}\sum_{t=1}^{T}\sum_{s=1}^{T}  \Ex\left[ l_{it}^{(1)}l_{is}^{(1)}\right]  \boldsymbol{f}_{0t}\boldsymbol{f}_{0s}'  \right) \boldsymbol{A}^{-1}_i.
			\end{multline*}
			Thus, it remains to show that the variance of each element of the second term of \eqref{a17} is $o(1)$. \\
			Let $w_{ij,ts}$ denote a generic element of $\Ex[\bm{R}_{ij}]\boldsymbol{A}^{-1}_i\boldsymbol{f}_{0t}\boldsymbol{f}_{0s}'\boldsymbol{A}^{-1}_i$, then a generic element of
			\[\frac{1}{N}\sum_{i=1}^{N}\Ex[\bm{R}_{ij}]\cdot \boldsymbol{A}^{-1}_i  \left( \frac{1}{T}\sum_{t=1}^{T} \sum_{s=1}^{T} l_{it}^{(1)}l_{is}^{(1)}\boldsymbol{f}_{0t}\boldsymbol{f}_{0s}' \right) \boldsymbol{A}^{-1}_i\]
			can be written as $(NT)^{-1}\sum_{i=1}^{N}\sum_{t=1}^{T} \sum_{s=1}^{T} l_{it}^{(1)}l_{is}^{(1)}w_{ij,ts}$. By Lemma 3 it can be shown that 
			\[ \text{Var}\left[\frac{1}{NT}\sum_{i=1}^{N}\sum_{t=1}^{T} \sum_{s=1}^{T} l_{it}^{(1)}l_{is}^{(1)}w_{ij,ts}\right]=o(1).\]
			Then the desired result follows.
		\end{proof}

		\vspace{0.5cm}	
		
		\begin{lemma}Under Assumptions 1 to 4,  
			\[ \frac{1}{NT}\sum_{i=1}^{N}\sum_{t=1}^{T}  l_{it}^{(3)}(\ast) \cdot (\boldsymbol{x}_{it} -\boldsymbol{B}_i \boldsymbol{A}^{-1}_i \hat{\boldsymbol{H}}^{-1}\hat{\boldsymbol{f}}_t) \cdot    \check{\boldsymbol{\lambda}}_{0i}'(\hat{\boldsymbol{f}}_t - \check{\boldsymbol{f}}_{0t}) (\hat{\boldsymbol{\lambda}}_i - \check{\boldsymbol{\lambda}}_{0i})'\hat{\boldsymbol{f}}_t
			=o_P(T^{-1}).  \]
		\end{lemma}
		\begin{proof}
			By Lemma 6, 7 and Assumption 4, 
			\begin{align*}
				& \frac{1}{NT}\sum_{i=1}^{N}\sum_{t=1}^{T}  l_{it}^{(3)}(\ast) \cdot (\boldsymbol{x}_{it} -\boldsymbol{B}_i \boldsymbol{A}^{-1}_i \hat{\boldsymbol{H}}^{-1}\hat{\boldsymbol{f}}_t) \cdot    \check{\boldsymbol{\lambda}}_{0i}'(\hat{\boldsymbol{f}}_t - \check{\boldsymbol{f}}_{0t}) (\hat{\boldsymbol{\lambda}}_i - \check{\boldsymbol{\lambda}}_{0i})'\hat{\boldsymbol{f}}_t  \\
				= & \frac{1}{NT}\sum_{i=1}^{N}\sum_{t=1}^{T}  l_{it}^{(3)} \dot{\boldsymbol{x}}_{it} \cdot    \boldsymbol{\lambda}_{0i}' \hat{\boldsymbol{H}}^{-1}(\hat{\boldsymbol{f}}_t - \check{\boldsymbol{f}}_{0t}) (\hat{\boldsymbol{\lambda}}_i - \check{\boldsymbol{\lambda}}_{0i})' \hat{\boldsymbol{H}} \boldsymbol{f}_{0t}   + o_P(T^{-1}) \\
				= & -\frac{1}{NT}\sum_{i=1}^{N}\sum_{t=1}^{T}  l_{it}^{(3)} \dot{\boldsymbol{x}}_{it} \cdot    \boldsymbol{\lambda}_{0i}' \hat{\boldsymbol{H}}^{-1}  \hat{\boldsymbol{\Psi}}'\left( \frac{1}{N}\sum_{j=1}^{N} \boldsymbol{e}_{jt}\right)\left(  \frac{1}{T}\sum_{s=1}^{T}  l_{is}^{(1)}\boldsymbol{f}_{0s}' \right)\boldsymbol{A}_i^{-1} \boldsymbol{f}_{0t} \\
				& +O_P(N^{-1/2})\cdot \left(\frac{1}{N}\sum_{i=1}^{N}\left\|\hat{\boldsymbol{\lambda}}_i -\check{\boldsymbol{\lambda}}_{0i} +  (\hat{\boldsymbol{H}}')^{-1}  \boldsymbol{A}^{-1}_i\cdot \frac{1}{T}\sum_{t=1}^{T}  l_{it}^{(1)}\boldsymbol{f}_{0t}\right\|^2\right)^{1/2}+ o_P(T^{-1})\\
				= &-\frac{1}{NT}\sum_{i=1}^{N}\sum_{t=1}^{T}  l_{it}^{(3)} \dot{\boldsymbol{x}}_{it} \cdot    \boldsymbol{\lambda}_{0i}' \hat{\boldsymbol{H}}^{-1}  \hat{\boldsymbol{\Psi}}'\left( \frac{1}{N}\sum_{j=1}^{N} \boldsymbol{e}_{jt}\right)\left(  \frac{1}{T}\sum_{s=1}^{T}  l_{is}^{(1)}\boldsymbol{f}_{0s}' \right)\boldsymbol{A}_i^{-1} \boldsymbol{f}_{0t} + o_P(T^{-1}).
			\end{align*}
			Note that
			\begin{multline*} 
				\frac{1}{NT}\sum_{i=1}^{N}\sum_{t=1}^{T}  l_{it}^{(3)} \dot{\boldsymbol{x}}_{it,k} \cdot    \boldsymbol{\lambda}_{0i}' \hat{\boldsymbol{H}}^{-1}  \hat{\boldsymbol{\Psi}}'\left( \frac{1}{N}\sum_{j=1}^{N} \boldsymbol{e}_{jt}\right)\left(  \frac{1}{T}\sum_{s=1}^{T}  l_{is}^{(1)}\boldsymbol{f}_{0s}' \right)\boldsymbol{A}_i^{-1} \boldsymbol{f}_{0t} \\
				= \tr \left[  \hat{\boldsymbol{H}}^{-1}  \hat{\boldsymbol{\Psi}}' \cdot \frac{1}{NT}\sum_{i=1}^{N}\sum_{t=1}^{T}  l_{it}^{(3)} \dot{\boldsymbol{x}}_{it,k}   \left( \frac{1}{N}\sum_{j=1}^{N} \boldsymbol{e}_{jt}\right)\boldsymbol{f}_{0t}' \boldsymbol{A}_i^{-1} \left(  \frac{1}{T}\sum_{s=1}^{T}  l_{is}^{(1)}\boldsymbol{f}_{0s} \right) \boldsymbol{\lambda}_{0i}' \right],
			\end{multline*}
			and we can show that 
		\begin{align*}
					& \left\|  \frac{1}{NT}\sum_{i=1}^{N}\sum_{t=1}^{T}  l_{it}^{(3)} \dot{\boldsymbol{x}}_{it,k}   \left( \frac{1}{N}\sum_{j=1}^{N} \boldsymbol{e}_{jt}\right)\boldsymbol{f}_{0t}' \boldsymbol{A}_i^{-1} \left(  \frac{1}{T}\sum_{s=1}^{T}  l_{is}^{(1)}\boldsymbol{f}_{0s} \right) \boldsymbol{\lambda}_{0i}'  \right\| \\
					=&\left\|  \frac{1}{N}\sum_{i=1}^{N}          \left[       \frac{1}{NT}\sum_{j=1}^{N}\sum_{t=1}^{T}    l_{it}^{(3)} \dot{\boldsymbol{x}}_{it,k} \boldsymbol{e}_{jt}  \boldsymbol{f}_{0t}'       \right] \cdot   \left[\boldsymbol{A}_i^{-1} \left(  \frac{1}{T}\sum_{s=1}^{T}  l_{is}^{(1)}\boldsymbol{f}_{0s} \right) \boldsymbol{\lambda}_{0i}' \right]  \right\|  \\
					\lesssim &  \left(\frac{1}{N}\sum_{i=1}^{N}\left\| \frac{1}{NT}\sum_{j=1}^{N}\sum_{t=1}^{T}    l_{it}^{(3)} \dot{\boldsymbol{x}}_{it,k} \boldsymbol{e}_{jt}  \boldsymbol{f}_{0t}'  \right\|^2\right)^{1/2}  \cdot \left(\frac{1}{N}\sum_{i=1}^{N}\left\| \frac{1}{T}\sum_{s=1}^{T}  l_{is}^{(1)}\boldsymbol{f}_{0s} \right\|^2\right)^{1/2} \\
					=& O_P(T^{-1/2}N^{-1/2+\epsilon/[2(2+\epsilon)]}) \cdot O_P( T^{-1/2} ) =o_P(T^{-1})
				\end{align*}
				for any $\epsilon>0$, where the inequality follows from Cauchy-Schwarz inequality, and the second equality follows from $N^{-1}\sum_{i=1}^{N}\Ex\left\| T^{-1}\sum_{s=1}^{T}  l_{is}^{(1)}\boldsymbol{f}_{0s} \right\|^2 = O(T^{-1})$
				and
				\[\frac{1}{N}\sum_{i=1}^{N}\Ex\left\| \frac{1}{NT}\sum_{j=1}^{N}\sum_{t=1}^{T}    l_{it}^{(3)} \dot{\boldsymbol{x}}_{it,k} \boldsymbol{e}_{jt}  \boldsymbol{f}_{0t}'  \right\|^2 = O(T^{-1}N^{-1+\epsilon/(2+\epsilon)}),\]
				which can be derived similarly to the proof of Lemma 6 and Lemma 1. Thus, the desired result follows.
		\end{proof}

		\vspace{0.5cm}	
		
		\begin{lemma}Under Assumptions 1 to 4,  
			\begin{multline}
				\frac{1}{N}\sum_{i=1}^{N} \boldsymbol{B}_i \boldsymbol{A}^{-1}_i\hat{\boldsymbol{H}}^{-1}\left( \frac{1}{T}\sum_{t=1}^{T}  l_{it}^{(2)} \hat{\boldsymbol{f}}_{t} \hat{\boldsymbol{f}}_t ' -  \check{\boldsymbol{A}}_i\right) (\hat{\boldsymbol{\lambda}}_i - \check{\boldsymbol{\lambda}}_{0i} )\\ =- \frac{1}{T}\cdot \frac{1}{NT}\sum_{i=1}^{N}\sum_{t=1}^{T}\sum_{s=1}^{T} \Ex\left[ l_{it}^{(2)}l_{is}^{(1)}\right]\boldsymbol{B}_i \boldsymbol{A}^{-1}_i \boldsymbol{f}_{0t} \cdot  \boldsymbol{f}_{0t} ' \boldsymbol{A}^{-1}_i \boldsymbol{f}_{0s} \cdot  +o_P(T^{-1}).
			\end{multline}
		\end{lemma}
		\begin{proof}
			Note that 
			\begin{align*} 
				& \frac{1}{N}\sum_{i=1}^{N} \boldsymbol{B}_i \boldsymbol{A}^{-1}_i\hat{\boldsymbol{H}}^{-1}\left( \frac{1}{T}\sum_{t=1}^{T}  l_{it}^{(2)} \hat{\boldsymbol{f}}_{t} \hat{\boldsymbol{f}}_t ' -  \check{\boldsymbol{A}}_i\right) (\hat{\boldsymbol{\lambda}}_i - \check{\boldsymbol{\lambda}}_{0i} ) \\
				=& \frac{1}{N}\sum_{i=1}^{N} \boldsymbol{B}_i \boldsymbol{A}^{-1}_i\left( \frac{1}{T}\sum_{t=1}^{T}  l_{it}^{(2)} \boldsymbol{f}_{0t} \boldsymbol{f}_{0t} ' - \boldsymbol{A}_i \right) \hat{\boldsymbol{H}}'(\hat{\boldsymbol{\lambda}}_i - \check{\boldsymbol{\lambda}}_{0i} ) +\\
				&\frac{1}{N}\sum_{i=1}^{N} \boldsymbol{B}_i \boldsymbol{A}^{-1}_i\hat{\boldsymbol{H}}^{-1} \left( \frac{1}{T}\sum_{t=1}^{T}  l_{it}^{(2)} (  \hat{\boldsymbol{f}}_{t} \hat{\boldsymbol{f}}_t ' - \check{\boldsymbol{f}}_{0t} \check{\boldsymbol{f}}_{0t} ' )\right) (\hat{\boldsymbol{\lambda}}_i - \check{\boldsymbol{\lambda}}_{0i} ).   
			\end{align*}
			Similar to the proof of Lemma 9, the second term on the right-hand side of the above equation is $o_P(T^{-1})$, while the first term can be written as 
			\begin{align*}
				&  -\frac{1}{T}\cdot \frac{1}{N}\sum_{i=1}^{N} \boldsymbol{B}_i \boldsymbol{A}^{-1}_i\left( \frac{1}{\sqrt{T}}\sum_{t=1}^{T}  l_{it}^{(2)} \boldsymbol{f}_{0t} \boldsymbol{f}_{0t} ' - \boldsymbol{A}_i \right) \boldsymbol{A}^{-1}_i \left( \frac{1}{\sqrt{T}}\sum_{t=1}^{T}  l_{it}^{(1)}\boldsymbol{f}_{0t}\right)  + o_P(T^{-1})  \\
				=&- \frac{1}{T}\cdot \frac{1}{NT}\sum_{i=1}^{N}\sum_{t=1}^{T}\sum_{s=1}^{T} \Ex\left[ l_{it}^{(2)}l_{is}^{(1)}\right]\boldsymbol{B}_i \boldsymbol{A}^{-1}_i \boldsymbol{f}_{0t} \cdot  \boldsymbol{f}_{0t} ' \boldsymbol{A}^{-1}_i \boldsymbol{f}_{0s} \cdot  +o_P(T^{-1}).
			\end{align*}
			Then the desired result follows.
		\end{proof}

		\vspace{0.5cm}	
		
		\begin{lemma}Under Assumptions 1 to 4,  
			\begin{multline}
				\frac{1}{N}\sum_{i=1}^{N}  \left( \frac{1}{T }\sum_{t=1}^{T}  l_{it}^{(2)} \boldsymbol{x}_{it}\hat{\boldsymbol{f}}_t'  - \boldsymbol{B}_i \hat{\boldsymbol{H}}'\right)  (\hat{\boldsymbol{\lambda}}_i - \check{\boldsymbol{\lambda}}_{0i})=	\\
				- \frac{1}{T}\cdot  \frac{1}{NT}\sum_{i=1}^{N}\sum_{t=1}^{T}\sum_{s=1}^{T}\Ex\left[l_{it}^{(2)}l_{is}^{(1)} \boldsymbol{x}_{it}\right] \boldsymbol{f}_{0t}'\boldsymbol{A}^{-1}_i \boldsymbol{f}_{0s}+o_P(T^{-1}).
			\end{multline}
		\end{lemma}
		\begin{proof}
			Note that 
			\begin{align*} 
				& \frac{1}{N}\sum_{i=1}^{N}  \left( \frac{1}{T }\sum_{t=1}^{T}  l_{it}^{(2)} \boldsymbol{x}_{it}\hat{\boldsymbol{f}}_t'  - \boldsymbol{B}_i \hat{\boldsymbol{H}}'\right)  (\hat{\boldsymbol{\lambda}}_i - \check{\boldsymbol{\lambda}}_{0i})  \\
				=&\frac{1}{N}\sum_{i=1}^{N}  \left( \frac{1}{T }\sum_{t=1}^{T}  l_{it}^{(2)} \boldsymbol{x}_{it}\boldsymbol{f}_{0t}'  - \boldsymbol{B}_i \right) \hat{\boldsymbol{H}}' (\hat{\boldsymbol{\lambda}}_i - \check{\boldsymbol{\lambda}}_{0i}) +  \frac{1}{NT}\sum_{i=1}^{N} \sum_{t=1}^{T}  l_{it}^{(2)} \boldsymbol{x}_{it} (\hat{\boldsymbol{f}}_t  -\check{\boldsymbol{f}}_{0t}) '   (\hat{\boldsymbol{\lambda}}_i - \check{\boldsymbol{\lambda}}_{0i}). 
			\end{align*}
			Similar to the proof of Lemma 9, the second term on the right-hand side of the above equation is $o_P(T^{-1})$, while the first term can be written as 
			\begin{align*}
				& - \frac{1}{T}\cdot  \frac{1}{N}\sum_{i=1}^{N}  \left( \frac{1}{\sqrt{T} }\sum_{t=1}^{T}  l_{it}^{(2)} \boldsymbol{x}_{it}\boldsymbol{f}_{0t}'  - \boldsymbol{B}_i \right)\boldsymbol{A}^{-1}_i \left(\frac{1}{\sqrt{T}}\sum_{t=1}^{T}  l_{it}^{(1)}\boldsymbol{f}_{0t}\right) + o_P(T^{-1}) \\
				=&- \frac{1}{T}\cdot  \frac{1}{NT}\sum_{i=1}^{N}\sum_{t=1}^{T}\sum_{s=1}^{T}\Ex\left[l_{it}^{(2)}l_{is}^{(1)} \boldsymbol{x}_{it}\right] \boldsymbol{f}_{0t}'\boldsymbol{A}^{-1}_i \boldsymbol{f}_{0s}+o_P(T^{-1}),
			\end{align*}
			then the desired result follows.
		\end{proof}

		\vspace{0.5cm}
		
		\noindent{\textbf{Proof of Theorem 1:} }
		\begin{proof}
			From \eqref{A8} to \eqref{A14} and Lemma 8 to Lemma 11, we get
			\begin{multline*}
				\boldsymbol{\Delta} (\hat{\boldsymbol{\beta}} -\boldsymbol{\beta}_0)+o_P(\| \hat{\boldsymbol{\beta}} -\boldsymbol{\beta}_0\|) =- \frac{1}{NT}\sum_{i=1}^{N}\sum_{t=1}^{T}  \left[ l_{it}^{(1)}\dot{\boldsymbol{x}}_{it} +\boldsymbol{C}_t \boldsymbol{H}_0^{-1} \boldsymbol{\Psi}_0' \boldsymbol{e}_{it}\right] \\
				+\frac{1}{N}( \bm{d}^{1}+ \bm{d}^{2}) +\frac{1}{T}( \bm{b}^{1}+ \bm{b}^{2}) +o_P(T^{-1}).
			\end{multline*}
	Let $\bar{\boldsymbol{w}}_{i}=T^{-1/2}\sum_{t=1}^{T}\boldsymbol{w}_{it}$ where $\boldsymbol{w}_{it}$ is defined in Assumption 5, so that by Assumption 4(v) implies
				\[\sqrt{NT}\boldsymbol{\Delta} (\hat{\boldsymbol{\beta}} -\boldsymbol{\beta}_0)+o_P(\sqrt{NT}\| \hat{\boldsymbol{\beta}} -\boldsymbol{\beta}_0\|) = -\frac{1}{\sqrt{N}}\sum_{i=1}^{N}\bar{\boldsymbol{w}}_{i}+\kappa^{-1}\bm{d}+\kappa\bm{b}+o_P(1).\]
				For any $\boldsymbol{a}\in\mathbb{R}^k$, $N^{-1}\sum_{i=1}^{N}\text{Var}(\boldsymbol{a}'\bar{\boldsymbol{w}}_{i}) = \boldsymbol{a}'\cdot N^{-1}\sum_{i=1}^{N} \Ex[\bar{\boldsymbol{w}}_{i}\bar{\boldsymbol{w}}_{i}']\cdot\boldsymbol{a}\rightarrow \boldsymbol{a}'\boldsymbol{\Omega}\boldsymbol{a}$ and
				\[\frac{1}{N^{1+\delta/2}}\sum_{i=1}^{N}\Ex\left|\boldsymbol{a}'\bar{\boldsymbol{w}}_{i}\right|^{2+\delta}\leq \|\boldsymbol{a}\|^{2+\delta}\cdot\frac{1}{N^{1+\delta/2}} \sum_{i=1}^{N}\Ex\left\|\frac{1}{\sqrt{T}}\sum_{t=1}^{T}\boldsymbol{w}_{it}\right\|^{2+\delta}=O(N^{-\delta/2})=o(1)\]
				for any $0<\delta\leq 2p-2$. Thus, Lyapunov’s central limit theorem implies that:
				\[\frac{1}{\sqrt{N}} \sum_{i=1}^{N}\boldsymbol{a}'\bar{\boldsymbol{w}}_{i}  \overset{d}{\rightarrow} \mathcal{N} \left( 0,\boldsymbol{a}'\boldsymbol{\Omega}\boldsymbol{a}   \right),\]
				which leads to Theorem 1 by Cramér–Wold theorem. Then Theorem 1 follows.
		\end{proof}

		\subsection{Proofs of Other Theorems}
		The proofs of Theorem 2 to Theorem 4 are relegated to the Online Appendix to save space.


	\normalsize
	
	\begin{spacing}{1}
		\bibliographystyle{chicago}
		\bibliography{inter_reference}

\begin{thebibliography}{}

\bibitem[\protect\citeauthoryear{Ahn and Horenstein}{Ahn and
  Horenstein}{2013}]{ahn2013eigenvalue}
Ahn, S.~C. and A.~R. Horenstein (2013).
\newblock Eigenvalue ratio test for the number of factors.
\newblock {\em Econometrica\/}~{\em 81\/}(3), 1203--1227.

\bibitem[\protect\citeauthoryear{Andersen}{Andersen}{1970}]{andersen1970asymptotic}
Andersen, E.~B. (1970).
\newblock Asymptotic properties of conditional maximum-likelihood estimators.
\newblock {\em Journal of the Royal Statistical Society: Series B\/}~{\em
  32\/}(2), 283--301.

\bibitem[\protect\citeauthoryear{Ando, Bai, and Li}{Ando
  et~al.}{2022}]{ando2022bayesian}
Ando, T., J.~Bai, and K.~Li (2022).
\newblock Bayesian and maximum likelihood analysis of large-scale panel choice
  models with unobserved heterogeneity.
\newblock {\em Journal of Econometrics\/}~{\em 230\/}(1), 20--38.

\bibitem[\protect\citeauthoryear{Andrews}{Andrews}{2005}]{andrews2005cross}
Andrews, D.~W. (2005).
\newblock Cross-section regression with common shocks.
\newblock {\em Econometrica\/}~{\em 73\/}(5), 1551--1585.

\bibitem[\protect\citeauthoryear{Bai}{Bai}{2003}]{bai2003inferential}
Bai, J. (2003).
\newblock Inferential theory for factor models of large dimensions.
\newblock {\em Econometrica\/}~{\em 71\/}(1), 135--171.

\bibitem[\protect\citeauthoryear{Bai}{Bai}{2009}]{bai2009panel}
Bai, J. (2009).
\newblock Panel data models with interactive fixed effects.
\newblock {\em Econometrica\/}~{\em 77\/}(4), 1229--1279.

\bibitem[\protect\citeauthoryear{Baker, Greenwood, and Wurgler}{Baker
  et~al.}{2003}]{BAKER2003261}
Baker, M., R.~Greenwood, and J.~Wurgler (2003).
\newblock The maturity of debt issues and predictable variation in bond
  returns.
\newblock {\em Journal of Financial Economics\/}~{\em 70\/}(2), 261--291.

\bibitem[\protect\citeauthoryear{Baker and Wurgler}{Baker and
  Wurgler}{2000}]{Baker2000}
Baker, M. and J.~Wurgler (2000).
\newblock The equity share in new issues and aggregate stock returns.
\newblock {\em The Journal of Finance\/}~{\em 55\/}(5), 2219--2257.

\bibitem[\protect\citeauthoryear{Boneva and Linton}{Boneva and
  Linton}{2017}]{boneva2017discrete}
Boneva, L. and O.~Linton (2017).
\newblock A discrete-choice model for large heterogeneous panels with
  interactive fixed effects with an application to the determinants of
  corporate bond issuance.
\newblock {\em Journal of Applied Econometrics\/}~{\em 32\/}(7), 1226--1243.

\bibitem[\protect\citeauthoryear{Bradley}{Bradley}{2005}]{bradley2005basic}
Bradley, R.~C. (2005).
\newblock Basic properties of strong mixing conditions: A survey and some open
  questions.
\newblock {\em Probability Surveys\/}~{\em 2}, 107--144.

\bibitem[\protect\citeauthoryear{Chamberlain}{Chamberlain}{2010}]{chamberlain2010binary}
Chamberlain, G. (2010).
\newblock Binary response models for panel data: Identification and
  information.
\newblock {\em Econometrica\/}~{\em 78\/}(1), 159--168.

\bibitem[\protect\citeauthoryear{Chen}{Chen}{2022}]{chen2022twostep}
Chen, L. (2022).
\newblock Two-step estimation of quantile panel data models with interactive
  fixed effects.
\newblock {\em Econometric Theory\/}, forthcoming.

\bibitem[\protect\citeauthoryear{Chen}{Chen}{2016}]{chen2016estimation}
Chen, M. (2016).
\newblock Estimation of nonlinear panel models with multiple unobserved
  effects.
\newblock Warwick economics research papers series (WERPS) (1120).

\bibitem[\protect\citeauthoryear{Chen, Fern{\'a}ndez-Val, and Weidner}{Chen
  et~al.}{2021}]{chen2020nonlinear}
Chen, M., I.~Fern{\'a}ndez-Val, and M.~Weidner (2021).
\newblock Nonlinear factor models for network and panel data.
\newblock {\em Journal of Econometrics\/}~{\em 220\/}(2), 296--324.

\bibitem[\protect\citeauthoryear{Chudik and Pesaran}{Chudik and
  Pesaran}{2015}]{chudik2015large}
Chudik, A. and H.~Pesaran (2015).
\newblock Large panel data models with cross-sectional dependence.
\newblock In {\em The Oxford Handbook of Panel Data}.

\bibitem[\protect\citeauthoryear{Davezies, D'Haultfoeuille, and
  Mugnier}{Davezies et~al.}{2020}]{davezies2020fixed}
Davezies, L., X.~D'Haultfoeuille, and M.~Mugnier (2020).
\newblock Fixed effects binary choice models with three or more periods.
\newblock {\em arXiv preprint arXiv:2009.08108\/}.

\bibitem[\protect\citeauthoryear{Dhaene and Jochmans}{Dhaene and
  Jochmans}{2015}]{dhaene2015split}
Dhaene, G. and K.~Jochmans (2015).
\newblock Split-panel jackknife estimation of fixed-effect models.
\newblock {\em The Review of Economic Studies\/}~{\em 82\/}(3), 991--1030.

\bibitem[\protect\citeauthoryear{Dong, Hirshleifer, and Teoh}{Dong
  et~al.}{2012}]{DHT2012}
Dong, M., D.~Hirshleifer, and S.~H. Teoh (2012, 12).
\newblock {Overvalued Equity and Financing Decisions}.
\newblock {\em The Review of Financial Studies\/}~{\em 25\/}(12), 3645--3683.

\bibitem[\protect\citeauthoryear{Fernández-Val and Weidner}{Fernández-Val and
  Weidner}{2016}]{fernandez2016individual}
Fernández-Val, I. and M.~Weidner (2016).
\newblock Individual and time effects in nonlinear panel models with large n,
  t.
\newblock {\em Journal of Econometrics\/}~{\em 192\/}(1), 291 -- 312.

\bibitem[\protect\citeauthoryear{Galvao and Kato}{Galvao and
  Kato}{2016}]{galvao2016smoothed}
Galvao, A.~F. and K.~Kato (2016).
\newblock Smoothed quantile regression for panel data.
\newblock {\em Journal of Econometrics\/}~{\em 193\/}(1), 92--112.

\bibitem[\protect\citeauthoryear{Gao, Liu, Peng, and Yan}{Gao
  et~al.}{2023}]{gao2023binary}
Gao, J., F.~Liu, B.~Peng, and Y.~Yan (2023).
\newblock Binary response models for heterogeneous panel data with interactive
  fixed effects.
\newblock {\em Journal of Econometrics\/}, forthcoming.

\bibitem[\protect\citeauthoryear{Hahn and Kuersteiner}{Hahn and
  Kuersteiner}{2011}]{hahn2011bias}
Hahn, J. and G.~Kuersteiner (2011).
\newblock Bias reduction for dynamic nonlinear panel models with fixed effects.
\newblock {\em Econometric Theory\/}~{\em 27\/}(6), 1152--1191.

\bibitem[\protect\citeauthoryear{Hahn and Newey}{Hahn and
  Newey}{2004}]{hahn2004jackknife}
Hahn, J. and W.~Newey (2004).
\newblock Jackknife and analytical bias reduction for nonlinear panel models.
\newblock {\em Econometrica\/}~{\em 72\/}(4), 1295--1319.

\bibitem[\protect\citeauthoryear{Hong, Wang, and Yu}{Hong
  et~al.}{2008}]{HONG2008119}
Hong, H., J.~Wang, and J.~Yu (2008).
\newblock Firms as buyers of last resort.
\newblock {\em Journal of Financial Economics\/}~{\em 88\/}(1), 119--145.

\bibitem[\protect\citeauthoryear{Honor{\'e} and Kyriazidou}{Honor{\'e} and
  Kyriazidou}{2000}]{honore2000panel}
Honor{\'e}, B.~E. and E.~Kyriazidou (2000).
\newblock Panel data discrete choice models with lagged dependent variables.
\newblock {\em Econometrica\/}~{\em 68\/}(4), 839--874.

\bibitem[\protect\citeauthoryear{Honor{\'e} and Lewbel}{Honor{\'e} and
  Lewbel}{2002}]{honore2002semiparametric}
Honor{\'e}, B.~E. and A.~Lewbel (2002).
\newblock Semiparametric binary choice panel data models without strictly
  exogeneous regressors.
\newblock {\em Econometrica\/}~{\em 70\/}(5), 2053--2063.

\bibitem[\protect\citeauthoryear{Honor{\'e} and Weidner}{Honor{\'e} and
  Weidner}{2020}]{honore2020moment}
Honor{\'e}, B.~E. and M.~Weidner (2020).
\newblock Moment conditions for dynamic panel logit models with fixed effects.
\newblock {\em arXiv preprint arXiv:2005.05942\/}.

\bibitem[\protect\citeauthoryear{Karabiyik, Reese, and Westerlund}{Karabiyik
  et~al.}{2017}]{Karabiyik201760}
Karabiyik, H., S.~Reese, and J.~Westerlund (2017).
\newblock On the role of the rank condition in {CCE} estimation of
  factor-augmented panel regressions.
\newblock {\em Journal of Econometrics\/}~{\em 197\/}(1), 60 -- 64.

\bibitem[\protect\citeauthoryear{Khan, Ponomareva, and Tamer}{Khan
  et~al.}{2020}]{khan2020identification}
Khan, S., M.~Ponomareva, and E.~Tamer (2020).
\newblock Identification of dynamic binary response models.
\newblock {\em Working Paper, Harvard University\/}.

\bibitem[\protect\citeauthoryear{Ma}{Ma}{2019}]{ma2019nonfinancial}
Ma, Y. (2019).
\newblock Nonfinancial firms as cross-market arbitrageurs.
\newblock {\em The Journal of Finance\/}~{\em 74\/}(6), 3041--3087.

\bibitem[\protect\citeauthoryear{Manski}{Manski}{1987}]{manski1987semiparametric}
Manski, C.~F. (1987).
\newblock Semiparametric analysis of random effects linear models from binary
  panel data.
\newblock {\em Econometrica\/}~{\em 55\/}(2), 357--362.

\bibitem[\protect\citeauthoryear{Moon and Weidner}{Moon and
  Weidner}{2015}]{moon2015linear}
Moon, H.~R. and M.~Weidner (2015).
\newblock Linear regression for panel with unknown number of factors as
  interactive fixed effects.
\newblock {\em Econometrica\/}~{\em 83\/}(4), 1543--1579.

\bibitem[\protect\citeauthoryear{Newey and West}{Newey and
  West}{1987}]{newey1987}
Newey, W.~K. and K.~D. West (1987).
\newblock A simple, positive semi-definite, heteroskedasticity and
  autocorrelation consistent covariance matrix.
\newblock {\em Econometrica\/}~{\em 55\/}(3), 703--708.

\bibitem[\protect\citeauthoryear{Neyman and Scott}{Neyman and
  Scott}{1948}]{neyman1948consistent}
Neyman, J. and E.~L. Scott (1948).
\newblock Consistent estimates based on partially consistent observations.
\newblock {\em Econometrica\/}~{\em 16\/}(1), 1--32.

\bibitem[\protect\citeauthoryear{Pesaran}{Pesaran}{2006}]{pesaran2006estimation}
Pesaran, M.~H. (2006).
\newblock Estimation and inference in large heterogeneous panels with a
  multifactor error structure.
\newblock {\em Econometrica\/}~{\em 74\/}(4), 967--1012.

\bibitem[\protect\citeauthoryear{Rio}{Rio}{2017}]{rio2017asymptotic}
Rio, E. (2017).
\newblock {\em Asymptotic theory of weakly dependent random processes},
  Volume~80.
\newblock Springer.

\bibitem[\protect\citeauthoryear{Rosenthal}{Rosenthal}{1970}]{rosenthal1970subspaces}
Rosenthal, H.~P. (1970).
\newblock On the subspaces of lp ($p>2$) spanned by sequences of independent
  random variables.
\newblock {\em Israel Journal of Mathematics\/}~{\em 8\/}(3), 273--303.

\bibitem[\protect\citeauthoryear{Wang}{Wang}{2022}]{wang2022maximum}
Wang, F. (2022).
\newblock Maximum likelihood estimation and inference for high dimensional
  generalized factor models with application to factor-augmented regressions.
\newblock {\em Journal of Econometrics\/}~{\em 229\/}(1), 180--200.

\bibitem[\protect\citeauthoryear{Zhu}{Zhu}{2022}]{zhu2022sufficient}
Zhu, Y. (2022).
\newblock A sufficient and necessary condition for identification of binary
  choice models with fixed effects.
\newblock {\em arXiv preprint arXiv:2206.10475\/}.

\end{thebibliography}
	\end{spacing}

\end{document}